\documentclass[10pt,a4paper]{article}
\usepackage[a4paper, left=2cm, right=2cm, top=2cm]{geometry}

\usepackage[utf8]{inputenc}
\usepackage{amsmath, amsthm, amssymb, amsfonts}     
\usepackage{mathrsfs}                               
\usepackage{accents}                                
\usepackage{natbib}                                 
\usepackage{hyperref}                               
\usepackage{doi}                                    
\usepackage{authblk}                                
\usepackage{siunitx}                                
\usepackage{booktabs}                               
\usepackage{ulem}                                   
\usepackage{nomencl}                                
\makenomenclature                                   
\usepackage[linesnumbered,ruled,vlined]{algorithm2e}
\SetCommentSty{mycommfont}
\usepackage{subfigure}                              
\usepackage{soul}                                   

\usepackage{bm}                     

\newcommand{\vect}[1]{\bm{#1}} 
\newcommand{\matr}[1]{\bm{#1}} 
\usepackage{graphicx}
\usepackage{tikz}
\usepackage{pgfplots}

\usetikzlibrary{patterns}

\definecolor{bblue}{HTML}{4F81BD}
\definecolor{rred}{HTML}{C0504D}
\definecolor{ggreen}{HTML}{9BBB59}
\definecolor{ppurple}{HTML}{9F4C7C}

\usetikzlibrary{decorations.pathmorphing,arrows,intersections,arrows.meta,angles,quotes}

\pgfplotsset{
	kurze Legende/.style={
		legend image code/.code={
			\draw[##1,mark repeat=2,line width=0.6pt]
			plot coordinates {
				(0cm,0cm)
				(0.3cm,0cm)
			};
		}
	}
}

\pgfplotsset{
	compat = newest,
	scale only axis, 
	max space between ticks = 50pt,
	ticklabel style = {font=\footnotesize},
	legend style =  {font=\footnotesize},
	grid = major,
	grid style = {dotted},
	legend columns=1, 
	xtick pos=left,
	ytick pos=left
}

\pgfplotsset{select coords between index/.style 2 args={
		x filter/.code={
			\ifnum\coordindex<#1\fi
			\ifnum\coordindex>#2\fi
		}
}}

\definecolor{color1}{HTML}{0060AD} 
\definecolor{color2}{HTML}{FF4500} 
\definecolor{color3}{HTML}{FFA500} 
\definecolor{color4}{HTML}{006400} 
\definecolor{color5}{HTML}{9400D3} 
\definecolor{color6}{HTML}{800000} 
\definecolor{color7}{HTML}{000000} 
\definecolor{color8}{HTML}{0000FF} 
\definecolor{color9}{HTML}{FF0000} 
\definecolor{mycolor_blue}{RGB}{66,124,161}
\definecolor{mycolor_grey}{RGB}{198,198,198} 

\tikzstyle{line1} = [color=color7,semithick] 
\tikzstyle{line2} = [color=color2,densely dotted,semithick]
\tikzstyle{line3} = [color=color1,densely dashed,semithick]
\tikzstyle{line4} = [color=color5,dash dot,semithick]
\tikzstyle{line5} = [color=color4,dash dot dot,semithick]
\tikzstyle{line6} = [color=color6,semithick]

\tikzstyle{mark1} = [color=color7,mark=x,mark size=2pt,mark options=solid,semithick] 
\tikzstyle{mark2} = [color=color2,mark=o,mark size=2pt,mark options=solid,semithick]
\tikzstyle{mark3} = [color=color1,mark=*,mark size=2pt,mark options=solid,semithick]
\tikzstyle{mark4} = [color=color5,mark=triangle,mark size=2pt,mark options=solid,semithick]
\tikzstyle{mark5} = [color=color4,mark=square,mark size=2pt,mark options=solid,semithick]
\tikzstyle{mark6} = [color=color7,mark=o,mark size=2pt,mark options=solid,semithick]
\tikzstyle{mark7} = [color=color7,mark=*,mark size=2pt,mark options=solid,semithick]
\tikzstyle{mark8} = [color=color7,mark=triangle,mark size=2pt,mark options=solid,semithick]

\usetikzlibrary{matrix,decorations.pathreplacing, calc, positioning}
\usepackage{multirow}
\newcommand\mydots{\ifmmode\ldots\else\makebox[0.5em][c]{.\hfil.}\fi}

\usepackage{etoolbox}
\renewcommand\nomgroup[1]{
  \item[\bfseries
  \ifstrequal{#1}{F}{Field Quantities}{
  \ifstrequal{#1}{B}{Boundary Quantities}{
  \ifstrequal{#1}{C}{Constant Quantities}{
  \ifstrequal{#1}{L}{Algebraic Quantities}{
  \ifstrequal{#1}{O}{Further Quantities}{}}}}}
]}
\usepackage{multicol}           

\usetikzlibrary{external}
\title{An Incremental Singular Value Decomposition Approach for Large-Scale Spatially Parallel \& Distributed but Temporally Serial Data -- Applied to Technical Flows}
\author[1,2]{Niklas K\"uhl\thanks{niklas.kuehl@tuhh.de}}
\author[3,4]{Hendrik Fischer}
\author[5]{Michael Hinze}
\author[2]{Thomas Rung}

\affil[1]{Hamburg Ship Model Basin, Bramfelder Strasse 164, D-22305 Hamburg, Germany}
\affil[2]{Hamburg University of Technology, Institute for Fluid Dynamics and Ship Theory, Am Schwarzenberg-Campus 4, D-21075 Hamburg, Germany}
\affil[3]{Leibniz Universität Hannover, Institute of Applied Mathematics, Welfengarten 1, D-30167 Hannover, Germany}
\affil[4]{Universit\'e Paris-Saclay, CentraleSup\'elec, ENS Paris-Saclay,  LMPS - Laboratoire de Mecanique Paris-Saclay, F-91190 Gif-sur-Yvette, France}
\affil[5]{University Koblenz-Landau, Department of Mathematics, Campus Koblenz, Universit\"atsstrasse 1, D-56070 Koblenz}

\begin{document}

\providetoggle{tikzExternal}
\settoggle{tikzExternal}{false}

\maketitle

\begin{abstract}
The paper presents a strategy to construct an incremental Singular Value Decomposition (SVD) for time-evolving, spatially 3D discrete data sets. A low memory access  procedure for reducing and deploying the snapshot data is presented.
Considered examples refer to Computational Fluid Dynamic (CFD) results extracted from unsteady flow simulations, which are computed spatially parallel using domain decomposition strategies and the Message Passing Interface (MPI) inter-processor communication protocol. The framework addresses state of the art PDE-solvers dedicated to practical applications. In particular, the spatial size of the data is assumed to be time-invariant and significantly larger than the temporal size due to the large computational mesh and the number of field quantities. 
Although the approach is applied to technical flows, it is applicable in similar applications under the umbrella of Computational Science and Engineering (CSE).

The research highlights issues associated with large-scale simulations performed by spatially decomposing the data sets on distributed memory systems along the routes of a Single Instruction Multiple Data (SIMD) concept. Aspects relevant to the performance
are scrutinized in detail. To this end, we introduce a bunch matrix that allows the aggregation of multiple time steps and 
 SVD updates, and significantly
increases the computational efficiency.

The incremental SVD strategy is initially verified and validated by simulating the 2D laminar single-phase flow around a circular cylinder.
Subsequent studies analyze the proposed strategy for a 2D submerged hydrofoil located 
 in turbulent two-phase flows.
Attention is directed to the 
accuracy of the SVD-based reconstruction  
based on local and global flow quantities, their physical realizability, the independence of the domain partitioning, and related implementation aspects. Moreover, the influence of lower and (adaptive) upper construction rank thresholds on both the effort and the accuracy are assessed.
Finally, various pseudo-codes support the traceability and reprogramming. 

%
%
%
 The incremental SVD process is applied to analyze and compress the predicted flow field around a Kriso container ship in harmonic head waves at $\mathrm{Fn} = 0.26$ and $\mathrm{Re}_\mathrm{L} = 1.4 \cdot 10^7$. With a numerical overhead of $\mathcal{O}(10\%)$, the snapshot matrix of size $\mathcal{O}( \mathbb{R}^{10E8  \times 10E4 } )$ computed on approximately 3000 processors can be incrementally compressed by $\mathcal{O}(95\%)$. The storage reduction is accompanied by errors in integral force and local wave elevation quantities of $\mathcal{O}(10^{-2}\%)$. 
\end{abstract}

\begin{flushleft}
\small{\textbf{{Keywords:}}} -- Principal Component Analysis, Reduced Order Modelling, Incremental Singular Value Decomposition, Large Spatio/Temporal Data Sets, Navier-Stokes Flow, Computational Fluid Dynamics
\end{flushleft}

\nomenclature[C]{$\rho$}{Fluid's density}
\nomenclature[C]{$\nu$}{Fluid's kinematic viscosity}
\nomenclature[C]{$\varphi^\mathrm{ref}$}{Reference quantity}
\nomenclature[C]{$\varphi^\mathrm{v}$}{Reference Velocity}
\nomenclature[C]{$\varphi^\mathrm{p}$}{Reference Pressure}
\nomenclature[C]{$\varphi^\mathrm{k}$}{Reference turbulent kinetic energy}
\nomenclature[C]{$\varphi^\mathrm{\varepsilon}$}{Reference turbulent kinetic energy's dissipation}
\nomenclature[C]{$\eta^\mathrm{r}$}{Retained energy}
\nomenclature[C]{$\eta^\mathrm{u}$}{Upper energy}
\nomenclature[C]{$\eta^\mathrm{l}$}{Lower energy}
\nomenclature[C]{$\eta^\mathrm{n}$}{Neglected energy}
\nomenclature[C]{$\tilde{\eta}^\mathrm{q}$}{Retained energy}
\nomenclature[C]{$\tilde{\eta}^\mathrm{o}$}{Modified retained energy}
\nomenclature[C]{$A$}{Projected reference area}
\nomenclature[C]{$D$}{Cylinder diameter}
\nomenclature[C]{$G$}{Gravitational acceleration}
\nomenclature[C]{$f$}{Vortex shedding frequency}
\nomenclature[C]{$F_i$}{Flow induced force}
\nomenclature[C]{$V_1$}{Bulk flow velocity}
\nomenclature[C]{$V$}{Volume}
\nomenclature[C]{$V_a$}{Volume of species $a$}
\nomenclature[C]{$y^+$}{Dimensionless wall normal distance}
\nomenclature[C]{T}{Number of time steps}
\nomenclature[C]{P}{Number of spatial partitions}
\nomenclature[C]{L}{Number of spatial degrees of freedom (per partition)}
\nomenclature[C]{S}{Number of states}
\nomenclature[C]{Y}{Length of local state vector}
\nomenclature[C]{$t$}{Time integration counter}
\nomenclature[C]{$p$}{Partition counter}
\nomenclature[C]{$l$}{Spatial degree of freedom counter}
\nomenclature[C]{$s$}{State counter}
\nomenclature[C]{$d$}{Maximum number of general singular values}
\nomenclature[C]{$r$}{Matrix rank, i.e. maximum number of reduced (non-zero) singular values}
\nomenclature[C]{$q$}{Truncation/Construction rank}
\nomenclature[C]{$\tilde{q}$}{Reconstruction rank}
\nomenclature[C]{$l$}{No. of economically computable singular values}
\nomenclature[C]{N}{Global number of rows}
\nomenclature[C]{Np}{Local number of rows on partition $p$}
\nomenclature[C]{$b$}{Bunch matrix width}
\nomenclature[C]{$\tilde{d}$}{Disk usage}
\nomenclature[C]{$\delta_{ik}$}{Kronecker Delta - unit coordinates / matrix}
\nomenclature[C]{$\mathrm{Re}$}{Reynolds number}
\nomenclature[C]{$\mathrm{St}$}{Strouhal number}
\nomenclature[C]{$\mathrm{Fn}$}{Froude number}
\nomenclature[C]{$L^\mathrm{c}$}{Hydrofoil's chord length}
\nomenclature[C]{$L^\mathrm{w}$}{Hydrofoil's incoming wave length}
\nomenclature[C]{$\lambda^\mathrm{c}$}{Hydrofoil's induced wave length}
\nomenclature[C]{$\lambda^\mathrm{w}$}{Ship hull's induced wave length}
\nomenclature[C]{$S^\mathrm{w}$}{Hydrofoil's submergence depth}
\nomenclature[C]{$H^\mathrm{w}$}{Hydrofoil's wave height}
\nomenclature[C]{$\tau_\mathrm{w}$}{Wall friction}
\nomenclature[C]{$c_\mathrm{f}$}{Friction coefficient}
\nomenclature[C]{$u_\mathrm{\tau}$}{Friction velocity}
\nomenclature[C]{$\alpha$}{Smallest re-presentable number}
\nomenclature[C]{$o$}{Minimal rank}
\nomenclature[C]{$L^\mathrm{pp}$}{Length between perpendiculars}
\nomenclature[C]{$d^\mathrm{h}$}{Hull draught}
\nomenclature[C]{$\tilde{t}^\mathrm{full}$}{Consumed wall clock time of full simulation}
\nomenclature[C]{$\tilde{t}^\mathrm{svd}$}{Consumed wall clock time of the itSVD}
\nomenclature[C]{$\tilde{t}^\mathrm{line}$}{Consumed wall clock time of specific algorithmic parts}
\nomenclature[C]{$\tilde{t}^\mathrm{RANS}$}{Consumed wall clock time of the RANS time integration}
\nomenclature[C]{$I$}{Turbulence intensity}
\nomenclature[C]{$\mathrm{T}^\mathrm{w}$}{Wave period}

\nomenclature[F]{$p$}{Fluid pressure}
\nomenclature[F]{$v_i$}{Flow velocity}
\nomenclature[F]{$c$}{Volume concentration}
\nomenclature[F]{$S_{ik}$}{Shear rate tensor}
\nomenclature[F]{$k$}{Turbulent kinetic energy}
\nomenclature[F]{$x_k$}{Spatial coordinate}
\nomenclature[F]{$\varepsilon$}{Turbulent kinetic energy dissipation}
\nomenclature[F]{$\tilde{p}$}{Reconstructed fluid pressure}
\nomenclature[F]{$\tilde{v}_i$}{Reconstructed flow velocity}
\nomenclature[F]{$\tilde{c}$}{Reconstructed volume concentration}
\nomenclature[F]{$\tilde{k}$}{Reconstructed turbulent kinetic energy}
\nomenclature[F]{$\tilde{\varepsilon}$}{Reconstructed turbulent kinetic energy dissipation}
\nomenclature[F]{$D^q$}{Dissipation of the turbulent quantity $q$}
\nomenclature[F]{$P^q$}{Production of the turbulent quantity $q$}
\nomenclature[F]{$R^q$}{Reduction of the turbulent quantity $q$}
\nomenclature[F]{$\varphi$}{Generic Field Quantity}
\nomenclature[F]{$\varphi$}{Reconstructed Generic Field Quantity}

\nomenclature[B]{$n_k$}{Surface normal vector}
\nomenclature[B]{$\Gamma$}{Surface area}
\nomenclature[B]{$y^+$}{Dimensionless first cell layer's height}

\nomenclature[L]{$\vect{\varphi}$}{Single state vector}
\nomenclature[L]{$\tilde{\vect{\varphi}}$}{Approximated single state vector}
\nomenclature[L]{$\vect{a}$}{Generic vector}
\nomenclature[L]{$\matr{A}$}{Generic matrix}
\nomenclature[L]{$\vect{y}$}{Local total state vector}
\nomenclature[L]{$\tilde{\vect{y}}$}{Approximated local total state vector}
\nomenclature[L]{$\matr{Y}$}{Global state matrix}
\nomenclature[L]{$\matr{Y}_q$}{Truncated (rank-$q$) global state matrix}
\nomenclature[L]{$\matr{U}$}{Left singular value matrix}
\nomenclature[L]{$\matr{S}$}{Singular value matrix}
\nomenclature[L]{$\matr{V}$}{Right singular value matrix}
\nomenclature[L]{$\matr{U}_r$}{Reduced (rank-r) left singular value matrix}
\nomenclature[L]{$\matr{S}_r$}{Reduced (rank-r) singular value matrix}
\nomenclature[L]{$\matr{V}_r$}{Reduced (rank-r) right singular value matrix}
\nomenclature[L]{$\matr{U}_q$}{Truncated (rank-$q$) left singular value matrix}
\nomenclature[L]{$\matr{S}_q$}{Truncated (rank-$q$) singular value matrix}
\nomenclature[L]{$\matr{V}_q$}{Truncated (rank-$q$) right singular value matrix}
\nomenclature[L]{$\tilde{\matr{U}}$}{Updated left singular value matrix}
\nomenclature[L]{$\tilde{\matr{S}}$}{Updated singular value matrix}
\nomenclature[L]{$\tilde{\matr{V}}$}{Updated right singular value matrix}
\nomenclature[L]{$\matr{U}^\prime$}{Temporary left singular value matrix}
\nomenclature[L]{$\matr{S}^\prime$}{Temporary singular value matrix}
\nomenclature[L]{$\matr{V}^\prime$}{Temporary right singular value matrix}
\nomenclature[L]{$\matr{K}$}{Temporary rSVD matrix}
\nomenclature[L]{$\matr{Q}_A$}{Q matrix of QR decomposition of $\matr{A}$}
\nomenclature[L]{$\matr{R}_A$}{R matrix of QR decomposition of $\matr{A}$}
\nomenclature[L]{$\matr{B}$}{Arbitrary matrix for additive rSVD modification}
\nomenclature[L]{$\matr{C}$}{Arbitrary matrix for additive rSVD modification}
\nomenclature[L]{$\matr{I}$}{Identity matrix}
\nomenclature[L]{$\matr{F}$}{Matrix with global quantities of interest}
\nomenclature[L]{$\tilde{\matr{F}}$}{Matrix with appr. global quantities of interest}
\nomenclature[L]{$\vect{f}$}{Vector with global modes to be evaluated}
\nomenclature[L]{$\vect{s}$}{Vector containing the singular values}

\nomenclature[O]{$\delta_{ik}$}{Kronecker delta}
\nomenclature[O]{$\Delta t$}{Discrete temporal increment}
\nomenclature[O]{$\Delta x_i$}{Discrete spatial increment}
\nomenclature[O]{$s_i$}{Singular values}
\nomenclature[O]{$y_{ik}$}{Entries of matrix $\matr{Y}$}
\nomenclature[O]{$t$}{Physical time}
\nomenclature[O]{$c_\mathrm{d}$}{Drag coefficient}
\nomenclature[O]{$c_\mathrm{l}$}{Lift coefficient}
\nomenclature[O]{$\tilde{c}_\mathrm{d}$}{Reconstructed drag coefficient}
\nomenclature[O]{$\tilde{c}_\mathrm{l}$}{Reconstructed lift coefficient}
\nomenclature[O]{$\tilde{t}$}{Measured wall clock time for (re)construction}
\nomenclature[O]{$\tilde{d}$}{Allocated memory for (re)construction}

\begin{multicols}{2}
\printnomenclature[5mm]
\end{multicols}

\section{Introduction}
Engineering computational methods are nowadays demanded to deliver more than the mere result of a performance analysis for a device exposed to particular operating conditions. In addition, the simulation results -- which often involve several millions and more degrees of freedom (DoF) --  should also serve investigations of designs modifications, different cost functional, and/or different operating conditions, preferably at a low computational cost.
 
When attention is directed to a few (integral) cost functions, aka. output parameters, 
as a function of a few input parameters,   
data-based machine learning (ML) approaches are nowadays intensively used to provide such input/output relations. A typical example might refer to the dynamic forces on an airfoil in response to its angle of incidence. 
Data-driven ML methods can estimate the cost functional (output) in response to formerly not explicitly investigated input parameters at a virtually negligible computational cost. However, they usually require the availability of a considerable body of data compiled by detailed first-principle simulations, e.g., transient, 3D fluid dynamic 
 simulations. The latter are expensive due to the many degrees of freedom and yield substantial offline or background computing costs. 
As an alternative to ML strategies, physics-based reduced order models (ROMs) are employed to compress the information of detailed first-principle simulations into a rapidly analyzable form, using different variants of principle component analysis (PCA), aka. proper orthogonal decomposition (POD), singular value decomposition (SVD), or Karhunen-Lo\`eve Expansion-based projections, cf. \cite{benner2015survey, mainini2015surrogate}. ROM aims at extracting the physically dominant features or modes of the simulated fields, which are frequently limited to a few ten contributions.
Linearity assumptions somewhat restrict the use of ROM strategies. Nevertheless, the exponential convergence of ROM approximations is often recovered also in nonlinear fluid problems, e.g. \cite{lassila2014model},  due to the rapid decay in their Kolmogorov N-width, cf. \cite{kolmogoroff1936uber}. Hence, the solution manifold can  often be well approximated by a low-rank subspace, and
 POD (\cite{sirovich1987turbulence, kunisch2002galerkin, willcox2002balanced}) is still an appreciated tool in model-order reduction of unsteady nonlinear systems. An example refers to fluid dynamic Navier-Stokes systems, where the framework was successfully applied to laminar (\cite{caiazzo2014numerical, stabile2018finite, grassle2019pod}) as well as turbulent problems (\cite{wang2012proper, lorenzi2016pod, hijazi2020data}). Additionally, clear links  between ROM and ML approaches using convolutional neural network (CNN) methods exist for linear problems (\cite{murata2020nonlinear, agostini2020exploration}), which relate the reduced space of CNN auto-encoders to the modes of a ROM strategy. Therefore the application of hybrid ROM/ML strategies that blend data-based and physics-based information is currently receiving increasing attention \cite{benner2015survey, mainini2015surrogate, hesthaven2018non, swischuk2019projection, pache2022data}.
 Many of these techniques require an SVD, emphasizing the need for an efficient, ideally incremental parallel SVD applicable to large-scale  applications.
 
Typical reduced-space applications refer to near-to-real-time extraction of output parameters entering decision support systems (\cite{walton2013reduced, mainini2015surrogate, pache2022data}) or field reconstructions for transient optimization studies (\cite{stoll2015low, dolgov2017low, vezyris2019incremental, buenger2020low, benner2020low, li2021towards, margetis2021lossy, margetis2022reducing, kodakkal2022risk, nobis2023modal}). Superficially, data-/order-reduction strategies require the availability of the complete data. However, the size of the spatio-temporal data sets generated by, for example, fluid dynamic simulation methods might involve O($10^8$) grid points and O($10^5$) time steps, and data handling can therefore be an issue. In such situations, data compression with modest computational overhead is of significance, i.e., using  existing domain parallel High-Performance Computing (HPC) fragments and, at the same time  incrementally preparing the reduction in transient frameworks.
%

\subsubsection*{Objectives and Outline of the Paper}
The paper tries to convey a strategy to perform incremental order reduction methods for time-evolving, spatially partitioned 3D data sets. 
A low memory access procedure for reducing and deploying
the snapshot data in parallel is presented. Considered examples refer to Computational Fluid Dynamic (CFD) results extracted from unsteady flow simulations,
%
 which are computed spatially parallel using domain decomposition strategies and the Message Passing Interface (MPI) inter-processor communication protocol (\cite{mpi}). The spatial size of the data is assumed to be time-invariant and significantly larger than the temporal size due to the large computational mesh and the number of field quantities involved in practical simulations. 
 The suggested procedure is grid type/structure independent.
Some aspects of this research have been highlighted previously, e.g., PDE solution tracking (\cite{fareed2018incremental, fareed2019note, fareed2020error}), low-rank PDE approximations (\cite{stoll2015low, buenger2020low, benner2020low}) or large-scale, partially distributed systems/networks (\cite{mastronardi2010fast, iwen2016distributed, lin2021low}). 
A very recent paper of \cite{li2022enhanced} shares the motivation of the present paper and discusses similar aspects. As opposed to the present contributions, the authors focus on building a POD basis instead of compressing all snapshot data in a memory-efficient manner. 
%
Similar to the present study, they incrementally build a global SVD on data distributed across processors, where they utilize the 
libROM library \cite{librom}, but without describing in-depth details of the parallelization strategy, which we among other things provide in the present work.
The above-referenced studies did not address combinations of large engineering systems and complex physics. The combination poses particular challenges in incrementally constructing the reduced-order model, particularly on the computational efficiency, as well as subsequent evaluation of the ROM regarding the physical realizability of reconstructed values.
Examples included in the present study refer to positive values of phase concentration and turbulent kinetic energy. 
Unlike other studies, we employ snapshot bunching in the parallel algorithm, which aims to significantly reduce the computational overheads of the online SVD and thereby enables to process of even larger data sets. Moreover, we complement this feature with an adaptive rank determination criterion based on the energy content of the SVD modes. In contrast to previous works, we compute this quantity exactly at a scalar product's complexity, thus eliminating the necessity of elaborate energy bounds. The potential of the rank adaptive parallel incremental SVD using snapshot bunching is reflected in a complex engineering application for 3D two-phase flows at Re 14$\times 10^6$ and Fn =0.26, involving around 30 million control volumes and 10 thousand time steps on 2880 processors.

The paper is organized as follows: Sec. \ref{sec:starting_point} outlines the algorithmic starting point and 
the theoretical framework. Section \ref{sec:parallelization_aspects} 
is devoted to a spatially distributed, temporally serial implementation, whereby several pseudo-codes should increase the comprehensibility of the 
parallelization aspects. Additionally, an exemplary Matlab$^\copyright$ code is provided, cf. \cite{matlabitSVD}. Sec. \ref{sec:governing_equations} provides the governing equations of investigated flow fields, as well as a brief description of the employed solver. Verification, validation and parallelization, as well as accuracy and efficiency aspects are reported for a 2D generic laminar flow around a circular cylinder in Sec. \ref{sec:verification_validation}. Subsequently, practical  implementation aspects and physical realizability as well as process adaptivity issues are investigated for a 2D turbulent two-phase flow  
in Sec. \ref{sec:practical_issues}. Section \ref{sec:application} displays the performance for the two-phase flow 
around a 3D container ship benchmark exposed to regular head waves. Final conclusions are drawn in Sec. \ref{sec:conclusion_outlook}.

In the remainder of the paper, 
 field quantities are defined regarding Cartesian coordinates denoted by Latin subscripts, and Einstein's summation convention is applied to repeated subscripts. 
%
Moreover, we employ lower and upper case bold letters to denote vectors and tensors, respectively. 
%
All numerical experiments have been conducted on the NHR Göttingen and Berlin, cf. \href{https://www.hlrn.de}{www.hlrn.de}.

\section{Algorithmic Starting Point}
\label{sec:starting_point}
We consider the flow field to consist of $s = [1,...,\mathrm{S}]$ individual states, e.g. the pressure, density, longitudinal velocity etc.. Moreover, we assume the snapshots to co-exist in all $l = [1,...,\mathrm{L}]$ spatial degrees of 
freedom, e.g. the number of control volumes of a finite volume scheme, of a partition (or processor, thread) $p \in [1, ..., \mathrm{P}]$.

The concatenation of all $s = [1,...,\mathrm{S}]$ state vectors $\vect{\varphi}$ of length $l = [1,...,\mathrm{L}]$ allows for the construction of a local state vector $\vect{y}$ of length ${\mathrm{Y}=}  (\mathrm{S} \, \mathrm{L})$ at time $t \in [1, ..., \mathrm{T}]$ on a partition $p \in [1, ..., \mathrm{P}]$, viz.
\begin{align}
    \vect{y}^{(t,p)} =
    \frac{1}{\mathrm{P} \, \mathrm{L}}
    \begin{bmatrix}
        \vect{\varphi}^\mathrm{(1,1:L,t,p)} \\
        \vdots \\
        \vect{\varphi}^\mathrm{(S,1:L,t,p)}
	\end{bmatrix}
	\qquad \qquad \mathrm{with} \qquad \qquad
    \vect{\varphi}^\mathrm{(s,1:L,t,p)} =
    \frac{1}{\varphi^\mathrm{ref}}
    \begin{bmatrix}  
		\varphi^\mathrm{(s,1,t,p)} \\
    	\vdots \\
	    \varphi^\mathrm{(s,L,t,p)} 
	\end{bmatrix} \, . \label{equ:SVD_start}
\end{align}
Here $\varphi^\mathrm{ref}$ refers to a state-specific reference value  that serves the non-dimensionalization of the involved states and supports   their combination. To this end it acts as a preconditioner that numerically harmonizes the field variables which may occur in strongly varying magnitudes. 
%
In addition, all state vectors are divided by the constant number of discrete global DoF ($= \mathrm{P} \, \mathrm{L}$) to bound 
subsequent global matrix-vector products in a range that is
 supported by double-precision arithmetics.
%
The partition specific state snapshot vectors 
 provide the following global state snapshot matrix 
 $\matr{Y} \in \mathbb{R}^{\mathrm{N} \, \times \mathrm{T}}$ that evolves in time
\begin{align}
    \matr{Y} = 
    \begin{bmatrix}
            \vect{y}^\mathrm{(1,1)} & \cdots    & \vect{y}^\mathrm{(T,1)}   \\
            \vdots                  &           & \vdots                    \\
            \vect{y}^\mathrm{(1,P)} & \cdots    & \vect{y}^\mathrm{(T,P)}
    \end{bmatrix} \, , \label{equ:state_matrix}
\end{align}
 where 
 $\mathrm{N} =  (\mathrm{S} \, \mathrm{L}) \, \mathrm{P} >> \mathrm{T}$ typically holds, i.e., 
 the global state matrix features much more rows than columns.
 %
%
For the application discussed in Sec. \ref{sec:application}, 
$\mathrm{S} = 7$ state variables are involved on a grid with approximately $\mathrm{L} \approx \SI{10000}{}$ control volumes per partition, using $\mathrm{P} = 2880$ partitions.
%
%
The flow field is analyzed over $\mathrm{T} = \SI{10 000}{}$ time steps and yields a global state matrix of approximately $\matr{Y} \in \mathbb{R}^{\SI{201 600 000}{} \times \SI{10 000}{}}$.
Managing a matrix of this size is cumbersome, if not impossible, which is why resource-efficient singular value decomposition (SVD) techniques are appreciated.
%
The following subsection refers to standard information and is reproduced for reasons of completeness and, above all, for the sake of understanding; their proofs are assumed to be known. 

\subsection{Singular Value Decomposition}
\label{subsec:svd}
Let 
$\matr{Y} \in \mathbb{R}^{\mathrm{N} \times \mathrm{T}}$, then there exist the orthogonal matrices $\matr{U} = \left( \vect{u}_1, ..., \vect{u}_\mathrm{N} \right) \in \mathbb{R}^{\mathrm{N} \times \mathrm{N}}$ and $\matr{V} = \left( \vect{v}_1, ..., \vect{v}_\mathrm{T} \right) \in \mathbb{R}^{\mathrm{T} \times \mathrm{T}}$ and a rectangular diagonal matrix $\matr{S} = \mathrm{diag}(s_\mathrm{i}) \in \mathbb{R}^{\mathrm{N} \times \mathrm{T}}$ such that
\begin{align}
    \matr{Y} = \matr{U} \matr{S} \matr{V}^T = \sum_{i=1}^d s_i \vect{u}_i \vect{v}_i^{T} \label{equ:general_svd}
\end{align}
with unique singular values $s_1 \geq ... \geq s_\mathrm{d} \geq 0$, singular vectors $\vect{u}_\mathrm{i}, \vect{v}_\mathrm{j}$ and $d = \mathrm{min}(\mathrm{N}, \mathrm{T})$.
Note that the theoretical equality $\matr{Y} = \matr{U} \matr{S} \matr{V}^T$  only holds in the absence of numerical errors.
The latter motivates the introduction of two norms which will be utilized below  to adapt the SVD, viz.
\begin{align}
    ||\matr{Y}||_2 = s_1
    \qquad \qquad
    \mathrm{and}
    \qquad \qquad
    ||\matr{Y}||_\mathrm{F} = \sqrt{\sum_{i=1}^\mathrm{N} \sum_{i=1}^\mathrm{T} |y_\mathrm{ij}|^2} = \sqrt{\sum_{i=1}^d s_\mathrm{i}^2} \label{equ:norms} \, .
\end{align}
In Eqn. (\ref{equ:norms})  $||\cdot||_2$ and $||\cdot||_\mathrm{F}$ refer to the L2 and Frobenius norm, respectively. Additionally, we define $r = \mathrm{max}(k \in \mathbb{N}: s_\mathrm{k} \neq 0)$ as the amount of non-zero singular values which is equivalent to  $\mathrm{rank}(\matr{Y}) = r$. Hence, only the first $r$ singular vectors are required to determine the matrix $\matr{Y}$ from Eqn. (\ref{equ:general_svd})
This allows for the definition of a reduced SVD (rSVD), i.e., let $\mathrm{rank}(\matr{Y}) = r \leq \mathrm{min}(\mathrm{N},\mathrm{T})$ such that
\begin{align}
    \matr{Y} = \matr{U}_\mathrm{r} \matr{S}_\mathrm{r} \matr{V}_\mathrm{r}^\mathrm{T} = \sum_{i=1}^\mathrm{r} s_\mathrm{i} \vect{u}_\mathrm{i} \vect{v}_\mathrm{i}^\mathrm{T} \label{equ:reduced_svd}
\end{align}
where the rSVD matrices read $\matr{U}_\mathrm{r} = \left( \vect{u}_1, ..., \vect{u}_\mathrm{r} \right) \in \mathbb{R}^{\mathrm{N} \times \mathrm{r}}$ and $\matr{V}_\mathrm{r} = \left( \vect{v}_1, ..., \vect{v}_\mathrm{r} \right) \in \mathbb{R}^{\mathrm{T} \times \mathrm{r}}$ as well as $\matr{S}_\mathrm{r} = \mathrm{diag}(s_\mathrm{1}, ..., s_\mathrm{r}) \in \mathbb{R}^{\mathrm{r} \times \mathrm{r}}$. 
Instead of $\matr{Y}$, the matrices $\matr{U}_\mathrm{r}$ and $\matr{V}_\mathrm{r}$ and the vector of diagonal $\matr{S}$-entries $\vect{s} \in \mathbb{R}^\mathrm{r}$ have to be stored. In practice, this results in a storage overhead when the matrix $\matr{Y}$ is of high rank, although a steep decay in singular values is observed. 
%
Consequently, further reduction of the efforts are sought. 

Due to its broad applicability and versatility, 
a POD-based model-order reduction is often pursued. The approach aims at a low-rank approximation while also preserving the overall solution behavior by
truncating the smallest non-zero singular values up to a specific truncation criterion. In line with the assumption of a fast decay in Kolmogorov N-width (\cite{kolmogoroff1936uber, benner2020model}), 
the POD modes corresponding to small singular values are deemed to have an insignificant impact on the approximation quality, which leads to the  truncated SVD (tSVD). The rank-$q$ tSVD approximation of $\matr{Y}$ with $1 \leq q < r$ is defined as
\begin{align}
    \matr{Y}_\mathrm{q} = \matr{U}_\mathrm{q} \matr{S}_\mathrm{q} \matr{V}_\mathrm{q}^\mathrm{T} = \sum_\mathrm{i=1}^\mathrm{q} s_\mathrm{i} \vect{u}_\mathrm{i} \vect{v}_\mathrm{i}^\mathrm{T} \label{equ:truncated_svd}
     \; . 
\end{align}
 The tSVD matrices read $\matr{U}_\mathrm{q} = \left( \vect{u}_1, ..., \vect{u}_\mathrm{q} \right) \in \mathbb{R}^{\mathrm{N} \times \mathrm{q}}$, $\matr{V}_\mathrm{q} = \left( \vect{v}_1, ..., \vect{v}_\mathrm{q} \right) \in \mathbb{R}^{\mathrm{T} \times \mathrm{q}}$ and $\matr{S}_\mathrm{q} = \mathrm{diag}(s_1, ..., s_\mathrm{q}) \in \mathbb{R}^{\mathrm{q} \times \mathrm{q}}$. $\matr{Y}_\mathrm{q} \in \mathbb{R}^{\mathrm{N} \times \mathrm{T}}$ denotes the matrix obtained 
 from the $q$ largest singular values of $\matr{Y}$ in 
 (\ref{equ:reduced_svd}).
%
There is no clear route to determine $q$ a-priori.
Generally, the goal is 
to choose  $q$ as small as possible and as large as necessary so that $\matr{Y}_\mathrm{q} \approx \matr{Y}$. 
Although a rank-$q$ approximation is optimal in the least-squared sense (\cite{eckart1936approximation}), this does not yield a clear path for rank determination. A widely used criterion to judge the quality of the 
tSVD  refers to \textit{retained energy or information content $\eta^\mathrm{q}$} heuristics, cf. \cite{grassle2018pod, gubisch2017proper, lassila2014model}. The latter has no rigid physical meaning, but
is typically defined by the ratio of squared Frobenius norms, cf. Eqn. (\ref{equ:norms}), 
viz. 
\begin{align}
    \eta^\mathrm{q}= \frac{||\matr{Y}_\mathrm{q}||_\mathrm{F}^2}{||\matr{Y}||_\mathrm{F}^2} = \frac{\sum_\mathrm{i=1}^\mathrm{q} s_\mathrm{i}^2}{\sum_\mathrm{i=1}^\mathrm{r} s_\mathrm{i}^2} = \frac{\sum_\mathrm{i=1}^\mathrm{q} s_\mathrm{i}^2}{\sum_\mathrm{i=1}^\mathrm{T} ||\vect{y}_\mathrm{i}||^2} \label{equ:retained_energy} \, .
\end{align}
Equation (\ref{equ:retained_energy}) requires the totality of all $r$ singular values, which is impractical for large matrices of technical applications. However, owing to Eqn. (\ref{equ:norms}) and $d \to r$, one can 
compute $||\matr{Y}||_\mathrm{F}^2$ directly if the state snapshot vectors are at least once available during the construction. This is the case in the time-advancing framework of this paper, i.e., $\sum_\mathrm{i=1}^\mathrm{T} ||\vect{y}_\mathrm{i}||^2$ can be computed 
without any truncation error in an incremental fashion as described in the subsequent section \ref{sec:additive_svd}.
%
\subsection*{Adaptation Heuristics}
Applications considered in this paper reveal that a minimal truncation rank $o$ is advantageous for adaptivity aspects. The reason 
is the frequently observed 
dominance of the first singular values that tend to drive the determination of the energetic bounds in (\ref{equ:retained_energy}) towards machine accuracy.
Therefore, expression (\ref{equ:retained_energy}) is heuristically adjusted in combination with a two step approach. The present heuristics neglects the first $o$ singular values to judge and adapt the quality of the SVD, viz.
\begin{align}
    \eta^\mathrm{o} = \frac{\sum_\mathrm{i=o}^\mathrm{q} s_\mathrm{i}^2}{\sum_\mathrm{i=o}^\mathrm{r} s_\mathrm{i}^2} = \frac{\sum_\mathrm{i=o}^\mathrm{q} s_\mathrm{i}^2}{\sum_\mathrm{i=1}^\mathrm{T} ||\vect{y}_\mathrm{i}||^2 - \sum_\mathrm{i=1}^\mathrm{o-1} s_\mathrm{i}^2} \label{equ:adaptive_energy} \, .
\end{align}
Note that equation (\ref{equ:retained_energy}) is recovered for $o = 1$. 
Since this adaptivity criterion contains the exact reference energy in the denominator, the latter is subtracted by the --always available-- first singular values.
The parameter $o$ serves as a starting point of the adaptation (step 1). Applications included in this paper typically employ $o=\mathcal{O}(\textrm{T}/100)$. Though it can not be a priory guaranteed, the first 0.5\%-1\% modes usually retain a fair amount of the matrix energy, i.e., at least 50\%.  Subsequently (step 2), the desired $\eta^o$ should be chosen to initiate the rank adaptation. Since $\eta^o$ describes the share of the initially missed matrix energy that should be recovered throughout the adaptation, this choice
hinges on step 1. Assuming step 1 to retain 50\% [60\%, 70\%, 80\%, 90\%] of the matrix energy, the choice of $\eta^o = 0.8$ will finally provide 90\% [92\%, 94\%, 96\%, 98\%] of the energy. Similarly $\eta^o = 0.9$ results in 95\% [96\%, 97\%, 98\%, 99\%] of the energy. These exemplary numbers indicate that $\eta^o = [0.8,0.9]$ is a fair choice to return sufficient amounts of total energy, and $\eta^o = 0.7$ might become critical since less than 90\% of the matrix energy are retained if step 1 falls below 66\%. 

From the preceding it is deduced that 
one could approximate the global state matrix via a (r,t)SVD according to Eqns. (\ref{equ:general_svd}), (\ref{equ:reduced_svd}) or (\ref{equ:truncated_svd}). However, the global matrix must be available for a one-shot computation, which 
 seems unfeasible for large scale applications and 
motivates an on-the-fly construction. 

\subsection{Additive Modification of a Singular Value Decomposition}
\label{sec:additive_svd}
This section describes an algorithm that gradually adapts an already existing SVD without any additional knowledge about 
the underlying system matrix. The method can be used to update the SVD incrementally. For this purpose, a general approach of an additive rank-$b$ modification of the rSVD, mainly developed by \cite{brand2002incremental, brand2006fast}, is initially summarized in brief.

For a given matrix $\matr{Y} \in \mathbb{R}^{\mathrm{N} \times \mathrm{T}}$, let $\matr{Y} = \matr{U} \matr{S} \matr{V}^\mathrm{T}$ be its rank-r rSVD. Further, let $\matr{B} \in \mathbb{R}^{\mathrm{N} \times \mathrm{b}}$ and $\matr{C} \in \mathbb{R}^{\mathrm{T} \times \mathrm{b}}$ be arbitrary matrices of rank $b$ that describe a desired modification of the system matrix $\matr{Y} \to \matr{Y} + \matr{B} \matr{C}^\mathrm{T}$. Then the rSVD of $\matr{Y} + \matr{B} \matr{C}^\mathrm{T} = \tilde{\matr{U}} \tilde{\matr{S}} \tilde{\matr{V}}^\mathrm{T}$ is given by the three updated SVD matrices
\begin{align}
    \tilde{\matr{V}} &= 
    \begin{bmatrix}
            \matr{V} & \matr{Q}_\mathrm{C}
    \end{bmatrix} \matr{V}^\prime \label{equ:update_V_full} \\
    \tilde{\matr{S}} &= \matr{S}^\prime \label{equ:update_S_full} \\
    \tilde{\matr{U}} &= 
    \begin{bmatrix}
            \matr{U} & \matr{Q}_\mathrm{B}
    \end{bmatrix} \matr{U}^\prime \label{equ:update_U_full} 
\end{align}
where $\matr{Q}_\mathrm{B} \matr{R}_\mathrm{B} = (\matr{I} - \matr{U}\matr{U}^\mathrm{T})\matr{B}$ and $ \matr{Q}_\mathrm{C} \matr{R}_\mathrm{C} = (\matr{I} - \matr{V}\matr{V}^\mathrm{T})\matr{C}$ follow from two QR decompositions, respectively. Additionally, $\matr{U}^\prime \matr{S}^\prime {\matr{V}^\prime}^\mathrm{T}$ follows from the rSVD of a matrix $\matr{K} \in \mathbb{R}^{\mathrm{(r+b)} \times \mathrm{(r+b)}}$ that reads
\begin{align}
    \matr{K} = 
    \begin{bmatrix}
            \matr{S} & \matr{0} \\ \matr{0} & \matr{0}
    \end{bmatrix}
    +
    \begin{bmatrix}
            \matr{U}^\mathrm{T} \matr{B} \\ \matr{R}_\mathrm{B}
    \end{bmatrix}
    \begin{bmatrix}
            \matr{V}^\mathrm{T} \matr{C} \\ \matr{R}_\mathrm{C}
    \end{bmatrix}^\mathrm{T} 
    \label{equ:matrix_K_general} \, .
\end{align}
The approach outlined above provides a variety of possibilities to modify the system matrix. 
In addition to updating and down-dating, individual values can be modified, or rows and columns can be exchanged, cf. \cite{brand2002incremental, brand2006fast}. The necessary QR decomposition employs, e.g., a modified Gram-Schmidt procedure, and the decomposition's resulting matrices have dimensions of $\matr{Q}_\mathrm{B} \in \mathbb{R}^{\mathrm{N} \times \mathrm{b}}, \matr{Q}_\mathrm{C} \in \mathbb{R}^{\mathrm{T} \times \mathrm{b}}$ and $\matr{R}_\mathrm{B}, \matr{R}_\mathrm{C} \in \mathbb{R}^{\mathrm{b} \times \mathrm{b}}$.
The modification presented above can result in an increased rank,
cf. matrix $\matr{K} \in \mathbb{R}^{\mathrm{(r+b)} \times \mathrm{(r+b)}}$ and, thus, a higher memory consumption. Since $\tilde{\matr{S}} = \matr{S}^\prime$ in Eqn. (\ref{equ:update_S_full}), the rank-$q$ truncation of an additive rank-$b$ modification can be achieved by truncating $\matr{K}$ in Eqn. (\ref{equ:matrix_K_general}), i.e.,
\begin{align}
    \tilde{\matr{V}}_\mathrm{q} &= 
    \begin{bmatrix}
            \matr{V} & \matr{Q}_\mathrm{C}
    \end{bmatrix} \matr{V}^\prime(:,1:q) \label{equ:update_V_truncated} \\
    \tilde{\matr{S}}_\mathrm{q} &= \matr{S}^\prime(1:q, 1:q) \label{equ:update_S_truncated} \\
    \tilde{\matr{U}}_\mathrm{q} &= 
    \begin{bmatrix}
            \matr{U} & \matr{Q}_\mathrm{B}
    \end{bmatrix} \matr{U}^\prime(:,1:q) \label{equ:update_U_truncated} \, .
\end{align}
The focal point of this paper is on temporal updates, i.e., column extensions of the system matrix $\matr{Y}$.
 Assigning $\matr{C} = \begin{bmatrix} \matr{0} & \matr{I} \end{bmatrix}^\mathrm{T}$, a column update via the matrix $\matr{B}$ is achieved by extending $\matr{Y}$ and $\matr{V}$ with $b$ additional zero columns such that $\begin{bmatrix} \matr{Y} & \matr{0} \end{bmatrix} = \matr{U} \matr{S} \begin{bmatrix} \matr{V} & \matr{0} \end{bmatrix}^\mathrm{T} \in \mathbb{R}^{\mathrm{N} \times (\mathrm{T+b})}$. 
Since $\matr{C}$ refers to an augmented identity matrix, its QR decomposition follows a certain structure and supports simplifying the former rank-$b$ modification, i.e., $\matr{Q}_\mathrm{C} = \matr{I}$ and $\matr{R}_\mathrm{C} = \matr{C}$. Additionally, the zero entries in $\matr{C} = \begin{bmatrix} \matr{0} & \matr{I} \end{bmatrix}^\mathrm{T}$ yield $\begin{bmatrix} \matr{V} & \matr{0} \end{bmatrix}^\mathrm{T} \matr{C} = \matr{0}$, so that a $q$SVD 
of 
 $\matr{Y} + \matr{B} \matr{C}^\mathrm{T} \approx \tilde{\matr{U}}_\mathrm{q} \tilde{\matr{S}}_\mathrm{q} \tilde{\matr{V}}^\mathrm{T}_\mathrm{q}$ 
 from (\ref{equ:update_V_truncated})-(\ref{equ:update_U_truncated}) 
 is now given by
\begin{align}
    \tilde{\matr{V}}_\mathrm{q} &= 
    \begin{bmatrix}
            \matr{V} & \matr{0} \\
            \matr{0} & \matr{I}
    \end{bmatrix} \matr{V}^\prime(:,1:q) \label{equ:update_V_column} \\
    \tilde{\matr{S}}_\mathrm{q} &= \matr{S}^\prime(1:q,1:q) \label{equ:update_S_column} \\
    \tilde{\matr{U}}_\mathrm{q} &= 
    \begin{bmatrix}
            \matr{U} & \matr{Q}_\mathrm{B}
    \end{bmatrix} \matr{U}^\prime(:,1:q) \label{equ:update_U_column} \, ,
\end{align}
where $\matr{U}^\prime \matr{S}^\prime {\matr{V}^\prime}^\mathrm{T}$ denotes the rSVD of $\matr{K}$ with
\begin{align}
    \matr{K} = 
    \begin{bmatrix}
            \matr{S} & \matr{U}^\mathrm{T} \matr{B} \\ \matr{0} & \matr{R}_\mathrm{B}
    \end{bmatrix}
    \label{equ:matrix_K_column} \, .
\end{align}
Eqns. (\ref{equ:update_V_column})-(\ref{equ:matrix_K_column}) serve as the basis for the on-the-fly or incrementally computed tSVD (itSVD) in this paper.

\paragraph{Important interim conclusion:} At each time step, the instantaneous state snapshot vector $\vect{y}$ is aligned as an additional column of the so-called bunch-matrix $\matr{B}$, and every $b$ time steps when $\matr{B}$ is filled, the latter is appended to the previous itSVD based on (\ref{equ:update_V_column}) - (\ref{equ:matrix_K_column}). Subsequently, $\matr{B}$ is cleared, and the process restarts. Hence, a trade-off between the bunch-size or column-size $b$ of $\matr{B}$, the overall itSVD construction time, and the required  memory arises. Two limit cases are conceivable: 
\begin{itemize} 
\item [(a)] The number of columns of the bunch matrix reaches the maximum number of time steps ($b \to \mathrm{T}$), 
and the 
incremental  tSVD
 merges into the 
classical one-shot SVD performed at the end of the simulation. 
\item [(b)] In contrast, one may expand the itSVD every time step and 
transform the matrix $\matr{B}$ into a column vector ($b \to 1$). 
\end{itemize}
%
While option (a) is mainly memory demanding, option (b) increases the simulation time.
 Analogous to the truncation value $q$, 
the bunch matrix width $b$ should be chosen as large as possible and as small as necessary.
The aspect is primarily of technical nature but nevertheless crucial for the attainable efficiency. It is mainly influenced by the utilized  hardware --e.g., maximum (random access) memory sizes or simulation times, etc.-- and will be discussed later in Sec. \ref{sec:verification_validation}.

\subsection{Evaluation of the parallel itSVD}
Once the itSVD has been completed, e.g., at the end of the time horizon of interest, each column --and thus an approximation of the reconstructed state vector $\tilde{\vect{\varphi}}(:) \approx \vect{\varphi}(:)$-- follows from the subsequent evaluation of the itSVD. For this purpose, relation (\ref{equ:SVD_start}) is twisted, offering an expression for the approximated local instantaneous snapshot vector $\tilde{\vect{\varphi}}^\mathrm{(s,1:L,t,p)}$ of state $s$ and length $\mathrm{L}$ at time $t$ on partition $p$, viz.
\begin{align}
    \begin{bmatrix}  
		\tilde{\varphi}^\mathrm{(s,1,t,p)} \\
    	\vdots \\
	    \tilde{\varphi}^\mathrm{(s,L,t,p)} 
	\end{bmatrix}
	= \varphi^\mathrm{ref} \tilde{\vect{\varphi}}^\mathrm{(s,1:L,t,p)}
	\quad\mathrm{with} \quad
    \begin{bmatrix}
        \tilde{\vect{\varphi}}^\mathrm{(1,1:L,t,p)} \\
        \vdots \\
        \tilde{\vect{\varphi}}^\mathrm{(S,1:L,t,p)}
	\end{bmatrix} 
	= \left( \mathrm{P} \, \mathrm{L}\right)  \tilde{\vect{y}}^{(t,p)}(:) = \tilde{\matr{U}}_q(:,1:\tilde{q}) \tilde{\matr{S}}_q(1:\tilde{q},1:\tilde{q}) \tilde{\matr{V}}_q^\mathrm{T}(t,1:\tilde{q})  \, . \label{equ:SVD_end}
\end{align}
Note that the reference quantities applied during the reconstruction (\ref{equ:SVD_end}) process should match those of the construction (\ref{equ:SVD_start}) phase. The same holds for the constant factor concerning the overall DoF $\mathrm{P} \, \mathrm{L}$. Furthermore, another parameter $\tilde{q} \leq q$ was introduced, which determines the reconstruction rank and should match the construction rank $\tilde{q} \to q$ for an approximation as accurate as possible in terms of the itSVD. Nevertheless, in some situations, one may be interested in information of lower rank, e.g., if only the first $\tilde{q}$, particularly energy-rich modes, are of interest. In this case, the numerical costs of the matrix-vector multiplications to be performed in Eqn. (\ref{equ:SVD_end}) decrease. The choice of $\tilde{q}$ concerning the reconstruction quality (which varies from case to case) is studied in detail in Sec. \ref{sec:verification_validation}.

To simplify the notation and to increase readability, the paper's remaining part consistently assumes the construction of a rank-$q$ itSVD so that the tilde marking and subscripts $\tilde{\matr{U}}_q \tilde{\matr{S}}_q \tilde{\matr{V}}_q^\mathrm{T} \to \matr{U} \matr{S} \matr{V}^\mathrm{T}$ in (\ref{equ:update_V_column})-(\ref{equ:update_U_column}) are no longer explicitly stated.

Finally, physical plausibility should be ensured by the itSVD evaluation. However, numerical inaccuracies or even inaccurate approximations, e.g., via a too-small construction/truncation rank $q$/$\tilde{q}$, may cause the reconstructed field not to satisfy fundamental principles and circumstances. These may include, e.g., boundary, far-field, or coupling conditions that are no longer met. In the case of inhomogeneous Dirichlet boundary conditions, e.g., the description of varying inflow conditions, a potential defect can be addressed by the introduction of lifting functions, cf. \cite{ballarin2016pod, nonino2021monolithic}. Therein, the non-homogeneous boundary values are subtracted from the single-state snapshots to provide homogenized companions. This naturally guarantees the fulfillment of the Dirichlet boundary conditions of the reconstructed single-state. However, issues concerning boundary conditions were of minor importance for the present research. Significantly more crucial are physically unrealistic reconstructed fields that might inhere, e.g., negative energies or volume concentrations, as shown in the application part of the paper, cf. Sec. \ref{sec:practical_issues}. 
Additionally, when the itSVD is composed of divergence free snapshots originating, e.g., from a continuity constraint, any reconstructed reduced solution is also divergence free, i.e., the continuity equation is automatically fulfilled , cf. \cite{quarteroni2014reduced, caiazzo2014numerical, ballarin2015supremizer}.

\section{Parallelization Aspects}
\label{sec:parallelization_aspects}
\begin{figure}[!ht]
    \centering
    \iftoggle{tikzExternal}{
    \input{tikz/general/reduced_svd.tikz}
    }{
    \includegraphics{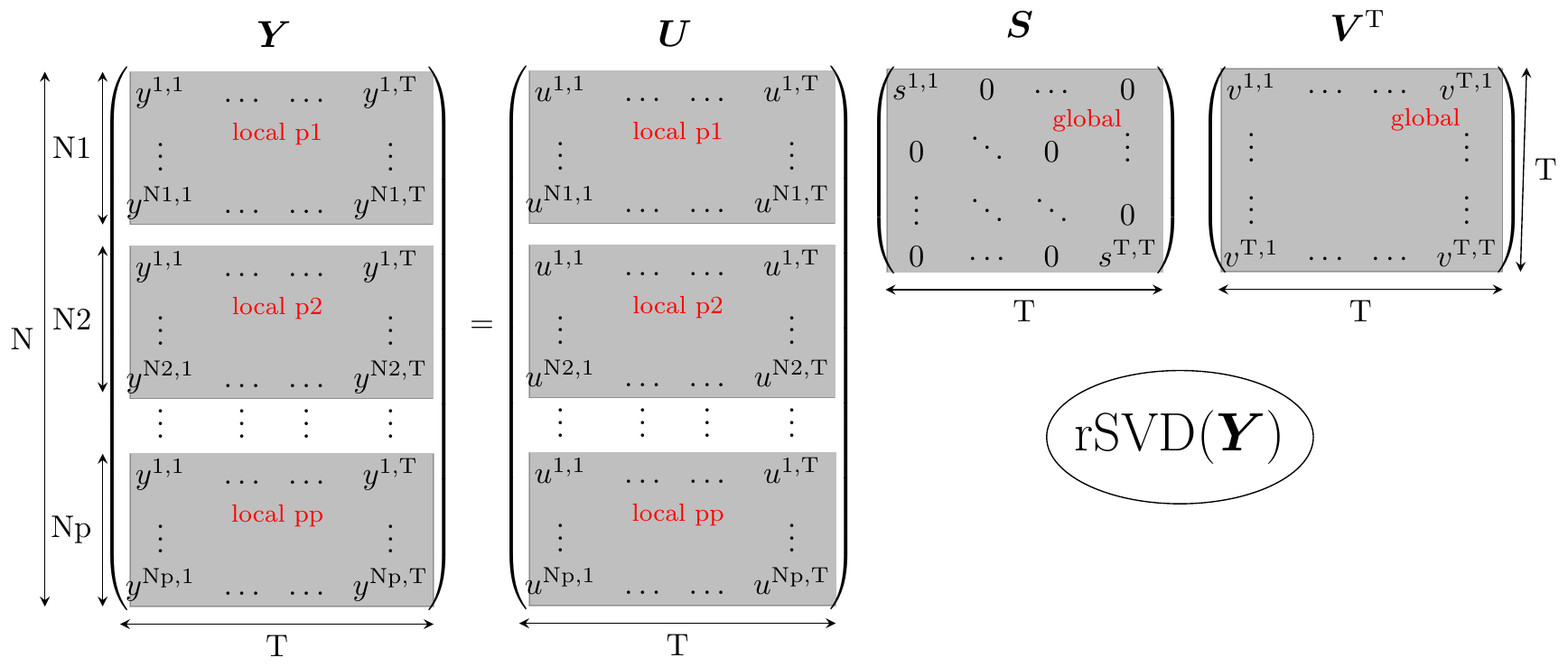}
    }
    \caption{Schematic representation of the reduced, spatially (rows, N) parallel / temporally (columns, M) serial Singular Value Decomposition of the global total state matrix $\matr{Y}$ ($\mathrm{N} >> \mathrm{T}$) based on spatially (or procedural) distributed local state sub-matrices into equally distributed local sub-matrices $\matr{U}$ and two global matrices $\matr{S}$ and $\matr{V}$.}
    \label{fig:reduced_parallel_rpts_svd}
\end{figure}

The overall goal of this paper refers to the incremental construction of a truncated SVD for Spatially Parallel but Temporally Serial (SPTS) data sets, i.e., a procedural largely distributed global state matrix $\matr{Y}$, cf. Eqn. (\ref{equ:state_matrix}). A schematic representation of the global snapshot matrix's parallel construction is shown in Fig. \ref{fig:reduced_parallel_rpts_svd}, assuming that
\begin{enumerate}
    \item the matrix rows are block-wise procedural distributed with different local lengths (i.e., number of rows, spatial DoF per processor/partition/thread times the number of states): $\mathrm{N1} = \mathrm{S} \cdot \mathrm{L1} = \mathrm{Y1}, \mathrm{N2} = \mathrm{S} \cdot \mathrm{L2} = \mathrm{Y2}, ..., \mathrm{Np} = \mathrm{S} \cdot \mathrm{Lp} = \mathrm{Yp}$ but always identical width (i.e. number of columns/temporal DoF/time steps) $\mathrm{T1} = \mathrm{T2} = ... = \mathrm{Tp} = \mathrm{T}$ and
    \item the number of columns $\mathrm{T}$ is significantly less than the number of rows, i.e. $\mathrm{N1} + \mathrm{N2} + ... + \mathrm{Np} = \mathrm{N} >> \mathrm{T}$ and therefore $\mathrm{d} = \mathrm{T}$ [$\mathrm{r} = \mathrm{T}$] in Eqn. (\ref{equ:general_svd}) [(\ref{equ:reduced_svd})] holds for a general [reduced] SVD in all scenarios considered herein. Hence, the number of time steps dictates the itSVD's rank.
\end{enumerate}
Figure \ref{fig:reduced_parallel_rpts_svd} suggests, in particular, that in the outlined SPTS case, the matrices $\matr{S} \in \mathbb{R}^{\mathrm{T} \times \mathrm{T}}$ and $\matr{V} \in \mathbb{R}^{\mathrm{T} \times \mathrm{T}}$ have a global character, i.e., the singular values must coincide on all partitions, and only the matrix $\matr{U} \in \mathbb{R}^{\mathrm{N} \times \mathrm{T}}$ can be locally differently populated. 

\begin{figure}[!ht]
    \centering
    \iftoggle{tikzExternal}{
    \input{tikz/general/truncated_svd.tikz}
    }{
    \includegraphics{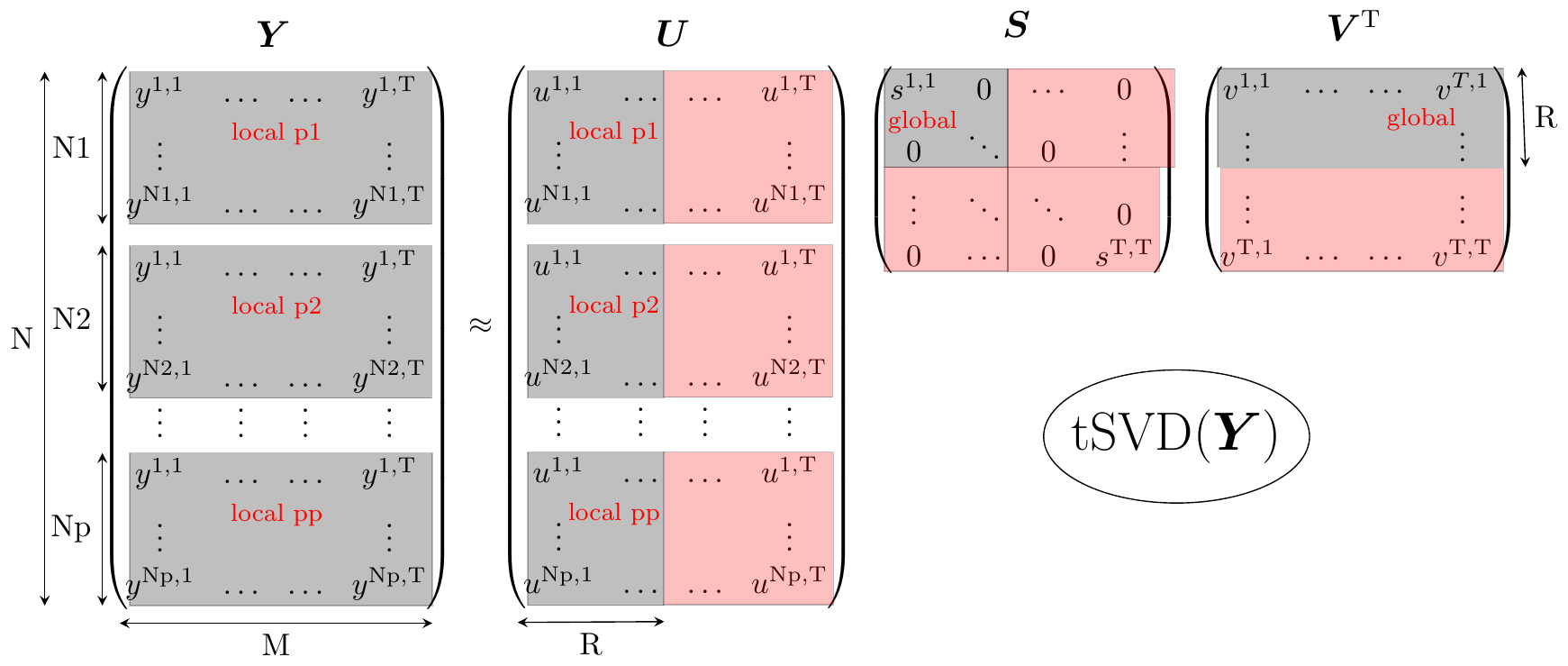}
    }
    \caption{Schematic representation of the truncated, spatially (rows, N) parallel / temporally (columns, M) serial Singular Value Decomposition of the global total state matrix $\matr{Y}$ ($\mathrm{N} >> \mathrm{T}$) based on spatially (or procedural) distributed local state sub-matrices into equally distributed local sub-matrices $\matr{U}$ and two global matrices $\matr{S}$ and $\matr{V}$.}
    \label{fig:truncated_parallel_rpts_svd}
\end{figure}

This section aims at a local$\to$global algorithm for distributed data that allows the modification of an already existing global tSVD without knowledge of the underlying global system matrix. This methodology can then be used to incrementally update a rank-$q$ tSVD as described in the previous section. A recipe for the discrete implementation of the proposed SPTS itSVD strategy is sketched in terms of several pseudo-codes in Algs. \ref{alg:time_integration}-\ref{alg:SVD_adaptive}.
\begin{algorithm}[!ht]
\SetKwInput{KwInput}{Input}
\SetKwInput{KwOutput}{Output}
\SetKwInput{KwOutputOpt}{Optional Output}
\DontPrintSemicolon

\KwInput{$b, q, o\in \mathbb{N}$\tcp*{Bunch size, truncation and minimum truncation rank}}
\KwOutput{$\matr{U} \in \mathbb{R}^\mathrm{Y \times q}$, $\matr{S} \in \mathbb{R}^\mathrm{q}$, $\matr{V} \in \mathbb{R}^\mathrm{T, q}$ \tcp*{SVD matrices/vectors that we are looking for}}
\KwOutputOpt{$\matr{F} \in \mathbb{R}^{\mathrm{T},h}$\tcp*{h quantities of interest (e.g. drag/lift coefficient) over time}}

\SetKwFunction{FMain}{construct{\_}SVD}
\SetKwFunction{FUpdate}{itSVD{\_}update}
 
\SetKwProg{Fn}{Module}{:}{\KwRet}
\Fn{\FMain{}}{
    \textbf{PUBLIC} allocate($\matr{U}$(Y,1),$\vect{s}$(1),$\matr{V}$(1,1),$\matr{B}$(Y,$b$)) \tcp*{itSVD-relevant matrices per processor}
    \textbf{PUBLIC} $e=0$ \tcp*{Matrix energy value incrementally updated}
    \textbf{PRIVATE} $k = 0$ \tcp*{Bunch matrix counter}
    \For{t=1,T} 
    {
        Acquire $\vect{\varphi}(:)$\tcp*{Solve for $1=[1,...,\mathrm{S}]$ instantaneous state variables}
    	$\vect{y}(:) \leftarrow \vect{\varphi}(:) / (\mathrm{P} \, \mathrm{L} \, \varphi^\mathrm{ref})$) \tcp*{Construct local, non-dimensionalized state vector, cf. Eqn. (\ref{equ:SVD_start}}
    	determine $\matr{F}(t,:)$\tcp*{Determine quantities of interest (i.e. drag/lift coefficient)}
    	$e \leftarrow \mathrm{parallel{\_}sum}(\vect{y}(:)^\mathrm{T}\vect{y}(:))$\tcp*{Update the global matrix energy, cf. Eqns. \ref{equ:retained_energy}-\ref{equ:adaptive_energy}}
    	\eIf{$t=1$} 
        {
            $\matr{V}(1,1) = 1$\tcp*{First r.h.s. itSVD vector}
            $\vect{s}(1) = \vect{y}(:)^\mathrm{T} \vect{y}(:)$
            $\mathrm{parallel{\_}sum}(\vect{s}(1)$),
            $\vect{s}(1) = \sqrt{\vect{s}(1)}$ \tcp*{First singular value}
            $\matr{U}(:,1) = \vect{y}(:)/\vect{s}(1)$\tcp*{First l.h.s. itSVD vector}
        }{
            $k++$\tcp*{Increase the bunch matrix counter}
            $\matr{B}(:,k) \leftarrow \vect{y}(:)$ \tcp*{Update the bunch matrix}
            \If{(k=b) \textbf{or} (t=T)}
            {
                itSVD{\_}update()\tcp*{Update the itSVD}
                $\matr{B}(:,:) = 0$, $k = 0$ \tcp*{clear the bunch matrix and its counter}
            }
        }
    }
    \If{procID=0}
    {
        save $\matr{V}$, $\vect{s}$, $\matr{F}$\tcp*{Store global data in one file only, e.g. S{\_}svd{\_}0.h5}
    }
    save $\matr{U}$\tcp*{Store local data for every processor, e.g. U{\_}svd{\_}<procID>.h5}
    \KwRet 0\;
}

\caption{Construction algorithm for a Spatially Parallel / Temporal Serial incrementally updated truncated rank-$q$ Singular Value Decomposition, where $\mathrm{P}$, $\mathrm{L}$, $\mathrm{Y}$, $\mathrm{T}$ are known from the underlying simulation framework, cf. Sec. \ref{sec:starting_point}.}
\label{alg:time_integration}
\end{algorithm}

\begin{algorithm}[!ht]
\SetKwProg{Fn}{Subroutine}{:}{}
\Fn{\FUpdate{}}{
    $u = \mathrm{size}(\matr{U},2)$, $v = \mathrm{size}(\matr{V},1)$, $c=u+b$, $d=v+b$, $l = \mathrm{min}(q,c)$ \tcp*{Required sizes}
    $\matr{M} = \matr{U}^\mathrm{T} \matr{B}$, parallel{\_}sum($\matr{M}$)\tcp*{1) $\matr{M} \in \mathbb{R}^{u \times b}$ 2) Local to Global Operation. cf. Fig. \ref{fig:global_inner_product}}
    $\matr{P} = \matr{B} - \matr{U} \matr{M}$\tcp*{$\matr{P} \in \mathbb{R}^{\mathrm{Y} \times b}$ around Eqn. \ref{equ:matrix_K_general}}
    $\begin{bmatrix}\matr{Q}_\mathrm{P}, \matr{R}_\mathrm{P} \end{bmatrix} = \mathrm{parallel{\_}gram{\_}schmidt}(\matr{P})$\tcp*{Global QR Decomposition}
    \If{procID=0}
    {
        $\matr{K} = \begin{bmatrix}
        \matr{S} & \matr{M} \\
        \matr{0} & \matr{R}_\mathrm{P}
        \end{bmatrix}$\tcp*{Prepare the SVD input matrix $\matr{K} \in \mathbb{R}^{c \times c}$, cf. Eqn. \ref{equ:matrix_K_column}}
        $\begin{bmatrix}
        \matr{U}^\prime, \vect{s}^\prime, \matr{V}^\prime
        \end{bmatrix} = \mathrm{svd}(\matr{K})$\tcp*{Local SVD yields $\matr{U}^\prime, \matr{V}^\prime \in \mathbb{R}^{c \times c}$ and $\vect{s}^\prime \in \mathbb{R}^{c}$ }
        \If{adaptive}
        {
            q = adaptive{\_}truncation($\vect{s}^\prime$)\tcp*{Adapt the truncation rank}
        }
        parallel{\_}bcast($\matr{U}^\prime, \vect{s}^\prime, \matr{V}^\prime, q$)\tcp*{Globalize the results}
    }
    $\matr{R} = \begin{bmatrix} 
    \matr{V} & \matr{0} \\
    \matr{0} & \matr{I}
    \end{bmatrix}$\tcp*{Prepare $\matr{V}$ Update via $\matr{R} \in \mathbb{R}^{d \times l}$, cf. Eqn. \ref{equ:update_V_truncated}}
    $\matr{Q} = \begin{bmatrix}
    \matr{U} & \matr{Q}_\mathrm{P}
    \end{bmatrix}$\tcp*{Prepare $\matr{U}$ Update via $\matr{Q} \in \mathbb{R}^{\mathrm{Y} \times l}$, cf. Eqn. \ref{equ:update_U_truncated}}
    $\begin{bmatrix}\matr{Q}_\mathrm{Q}, \cdot \end{bmatrix} = \mathrm{parallel{\_}gram{\_}schmidt}(\matr{Q})$\tcp*{Global QR Decomposition}
    de-allocate($\matr{U},\vect{s},\matr{V}$), allocate($\matr{U}(\mathrm{Y},l),\vect{s}(l),\matr{V}(d,l))$\tcp*{Increase the $\matr{U}$, $\matr{S}$, and $\matr{V}$ sizes}
    $\matr{V}$ = $\matr{R}$ $\matr{V}^\prime$(:,$1$:$l$)\tcp*{Update $\matr{V}$, cf. Eqn. \ref{equ:update_V_truncated}}
    $\vect{s}$ = $\vect{s}^\prime$(1:l)\tcp*{Update $\vect{s}$, cf. Eqn. \ref{equ:update_S_truncated}}
    $\matr{U}$ = $\matr{Q}_Q$ $\matr{U}^\prime($:$,1$:$l)$\tcp*{Update $\matr{U}$, cf. Eqn. \ref{equ:update_U_truncated}}
}

\caption{Incremental rank $b$ update of a Spatially Parallel / Temporal Serial truncated Singular Value Decomposition, where $\mathrm{Y}$ is known from the underlying simulation framework, cf. Sec. \ref{sec:starting_point} 
}
\label{alg:SVD_construction}
\end{algorithm}

\begin{figure}[!ht]
    \centering
    \iftoggle{tikzExternal}{
    \input{tikz/general/utranspose_times_b.tikz}
    }{
    \includegraphics{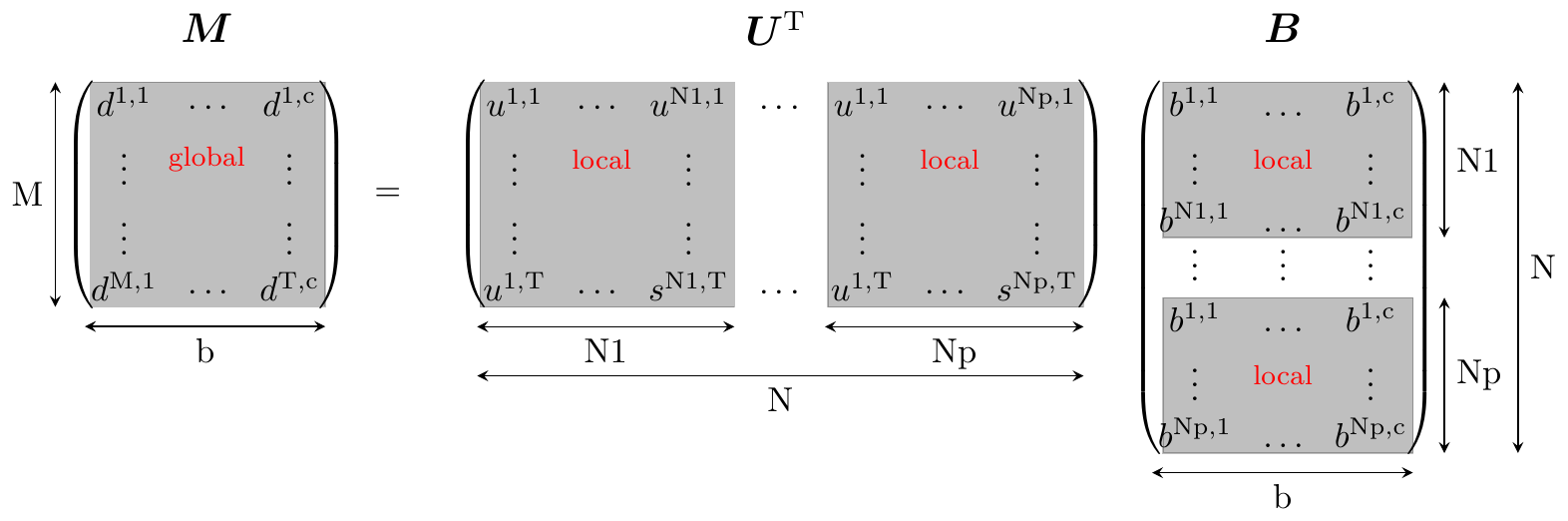}
    }
    \caption{Schematic representation of the product between two distributed matrices $\matr{U}^\mathrm{T} \matr{B}$ towards a globally constant matrix $\matr{M}$, cf. Alg. \ref{alg:SVD_construction} line 3.}
    \label{fig:global_inner_product}
\end{figure}

For the moment, it is assumed that the simulated past has already been reduced in a corresponding itSVD. The bunch matrix $\matr{B}$ has been filled up so far that now an update is to be carried out according to the explanations from Sec. \ref{sec:additive_svd}. Hence, the product of the two global matrices $\matr{U}^\mathrm{T} \matr{B}$ is necessary at two places, once in the QR-decomposition of $(\matr{I} - \matr{U} \matr{U}^\mathrm{T}) \matr{B}$, as well as in the assembly of the upper right part of $\matr{K}$, cf. Eqn. (\ref{equ:matrix_K_column}). The inner product of both actual global matrices can, however, first be computed locally on each thread and then communicated beyond the processor boundaries in the spirit of a local$\to$global operation, cf. Alg. \ref{alg:auxilary_routines} and the schematic sketch in Fig. \ref{fig:global_inner_product}. In this paper, the MPI routine \textit{ALLREDUCE} (\cite{mpi}) is employed, which first combines (i.e., sums up) the $\matr{U}^\mathrm{T} \matr{B}$ values of all processes and then sends them back to all senders so that the global equality of $\matr{U}^\mathrm{T} \matr{B}$ is ensured. Subsequently, the QR decomposition can be assembled and executed locally, cf. Alg. \ref{alg:auxilary_routines}. A modified parallelized Gram-Schmidt method is used for orthonormalization, where the required global inner products are again performed based on the \textit{MPI-ALLREDUCE} instruction. Once the globally determined $\matr{Q}_\mathrm{P}$ and $\matr{R}_\mathrm{P}$ matrices are locally available, they can be inserted into Eqns. (\ref{equ:update_U_column}) and (\ref{equ:matrix_K_column}). Subsequently, the matrix $\matr{K}$ is assembled, and the determination of its rSVD according to Eqn. (\ref{equ:reduced_svd}) can be performed. However, note that $\matr{K}$ consists mainly of the previously determined singular values $\matr{S}$, which should coincide on all partitions. Moreover, the globally invariant sub-matrices $\matr{M} = \matr{U}^\mathrm{T} \matr{B}$ and $\matr{R}_\mathrm{P}$ are added so that $\matr{K}$ actually refers to a global matrix. Therefore, the rSVD of $\matr{K}$ can be computed on all threads simultaneously. Still, identical results can be expected only in the absence of numerical rounding errors, which is why it is recommended --and also practiced in this research-- to perform the rSVD of $\matr{K}$ on one partition for safety reasons and then communicate the results to all remaining threads, for example using the \textit{MPI-BCAST} command, cf. Alg. \ref{alg:auxilary_routines}. In this paper, the local rSVD of $\matr{K}$ is determined using the \textit{LAPACK} library, precisely the built-in \textit{DGESVD} routine, cf. \cite{lapack99}. Once the $\matr{U}^\prime \matr{S}^\prime \matr{V}^\prime$ decomposition of $\matr{K}$ is known on all processors, the $\matr{U} \matr{S} \matr{V}$ matrices can be updated, cf. Eqns. (\ref{equ:update_V_column})-(\ref{equ:update_U_column}), storing only the diagonal of $\matr{S} \to \vect{s}$ in a one-dimensional array for efficiency.
An additional technical observation: For small bunch matrix widths $b$, the itSVD algorithm is invoked frequently, and algebraic subspace rotations involved possibly do not preserve orthogonality. Hence, a numerically induced loss of orthogonality of the matrix $\begin{bmatrix} \matr{U} & \matr{Q}_\mathrm{P} \end{bmatrix}$, and thus the updated $\matr{U}$ may occur. This issue was recognized by other authors (\cite{brand2006fast, fareed2018incremental, bach2019randomized, fareed2019note, fareed2020error, zhang2022answer}), which is why an additional orthonormalization step, acting globally on $\matr{Q}$, is recommended. Afterward, the bunch matrix $\matr{B}$ is emptied, and the time integration is continued for $b$ time steps until the next itSVD update is performed.

The computational effort is in the following three operations. 
Firstly, the QR decompositions to obtain $\matr{Q}_\mathrm{P}$, $ \matr{R}_\mathrm{P}$ and $\matr{Q}_\mathrm{Q}$ take $\mathcal{O}(N \, q^2)$ and $\mathcal{O}(N \,(u+b)^2)$, respectively. Secondly, the SVD of $\matr{K}$ in Eqn. (\ref{equ:matrix_K_column}) has a complexity of $\mathcal{O}((u + b)^3)$. Thirdly, the rotations of the subspace in Eqns. (\ref{equ:update_V_truncated}), (\ref{equ:update_U_truncated}) need $\mathcal{O}((N + T)(u + b)q)$ operations (\cite{brand2002incremental}). The additional computational effort due to the itSVD Alg. \ref{alg:SVD_construction} is studied during later appliation studies in Sec. \ref{sec:application} 

A special case arises when the previous simulation has no rSVD information and the considered time history is limited to one time step. In this case, the global rank-1 rSVD simply reads $\vect{y} = \vect{u} \, s \, v$, where $\vect{u} = \vect{y}/s$, $s=||\vect{y}||_2$, and $v=1$ can be obtained quickly using the \textit{ALLREDUCE} routine mentioned earlier. Although not necessary in principle, the itSVD algorithm presented in this manuscript always starts with a rank-1 rSVD after the first time step. Subsequently, the bunch matrix $\matr{B}$ can be built up to an arbitrarily large rank $b$. The latter can be adjusted to the underlying simulation hardware described in Sec. \ref{sec:practical_issues}.

In the case of a distributed memory environment, an enormous amount of memory can be saved if each thread stores its local matrix $\matr{U}$, while only one partition stores the global vector $\matr{S} \to \vect{s}$ and the matrix $\matr{V}$. This is essential for massively parallel applications with several hundred to thousand partitions and a considerable reduction of the total memory overhead can be achieved. In the context of this paper, the HDF5 (\cite{hdf5}) library is used to compress and store the itSVD data and, for benchmarking and comparison purposes, all instantaneous snapshot vectors.

\begin{algorithm}[!ht]
\DontPrintSemicolon

\SetKwFunction{FGS}{parallel{\_}gram{\_}schmidt}
\SetKwFunction{FSum}{parallel{\_}sum}
\SetKwFunction{FBroadcast}{parallel{\_}broadcast}

\SetKwProg{Fn}{Function}{:}{}
\Fn{\FGS{A}}{
    \For{j=1,size($\matr{A}$,2)}
    {
    $\vect{v}(:) = \matr{A}(:,j)$\;
    \For{i=1,j-1}
        {
        $\matr{R}(i,j)=\matr{Q}(:,i)^\mathrm{T} \vect{v}(:)$,
        parallel{\_}sum($\matr{R}(i,j)$)\;
        $\vect{v}(:) = \vect{v}(:) - \matr{R}(i,j) \matr{Q}(:,i)$\;
        }
    $\matr{R}(j,j) = \vect{v}(:)^\mathrm{T} \vect{v}(:)$,
    parallel{\_}sum($\matr{R}(j,j)$),
    $\matr{R}(j,j) = \sqrt{\matr{R}(j,j)}$\;
    $\matr{Q}(:,j) = \vect{v}(:)/\matr{R}(j,j)$\;
    }
\KwRet $\matr{Q}, \matr{R}$\;
}\;

\SetKwProg{Fn}{Subroutine}{:}{}
\Fn{\FSum{vIn}}{
    mpi{\_}allreduce(vIn)\;
}\;

\SetKwProg{Fn}{Subroutine}{:}{}
\Fn{\FBroadcast{vIn}}{
    mpi{\_}bcast(vIn)\;
}
\caption{Auxiliary functions and routines for: 1) Matrix Orthonormalization via a Gram-Schmidt-based QR-decompostion for distributed row data, 2) summation from different local to the same global value, and 3) broadcast value(s) from one processor to all others.}
\label{alg:auxilary_routines}
\end{algorithm}

\begin{algorithm}[!ht]
\SetKwInput{KwInput}{Input}
\SetKwInput{KwOutputOpt}{Optional Output}
\DontPrintSemicolon

\KwInput{$\matr{U} \in \mathbb{R}^\mathrm{Y \times q}$, $\vect{s} \in \mathbb{R}^\mathrm{q}$, $\matr{V} \in \mathbb{R}^\mathrm{T, q}, \tilde{q} \in \mathbb{N}$\tcp*{itSVD matrices/vectors and evaluation rank (note: $\tilde{q} \leq q$)}}
\KwOutputOpt{$\tilde{\matr{F}} \in \mathbb{R}^{\mathrm{T}, h}$\tcp*{h quantities of interest (e.g. drag/lift coefficient) over time}}

\SetKwFunction{FMain}{evaluate{\_}SVD}
 
\SetKwProg{Fn}{Function}{:}{\KwRet}
\Fn{\FMain{}}{
    \If{procID=0}
    {
        read $\matr{V}, \vect{s}$\tcp*{Read global data in one file only, e.g. S{\_}svd{\_}0.h5}
        parallel{\_}broadcast($\matr{V}, \vect{s}$)\tcp*{Broadcast to all processors}
    }
    read $\matr{U}$\tcp*{Read local data for every processor, e.g. U{\_}svd{\_}<procID>.h5}
    \For{t=1,T} 
    {
        $\vect{f} = \matr{V}(t,1:\tilde{q}) \vect{s}(1:\tilde{q})$\tcp*{$\vect{f} \in \mathbb{R}^{\tilde{q}}$ first $\tilde{q}$ global modes}
        $\tilde{\vect{y}}(:) = \matr{U}(:,1:\tilde{q}) \vect{f}(1:\tilde{q})$\tcp*{Compute the (approximated) local, non-dimensional state vector}
        $\tilde{\vect{\varphi}}(:) \leftarrow (\mathrm{P} \, \mathrm{L} \, \varphi^\mathrm{ref}) \tilde{\vect{y}}(:)$ \tcp*{Extract the $s=[1,...,\mathrm{S}]$ instantaneous state variables, cf. Eqn. (\ref{equ:SVD_end})}
        ensure physical feasibility\\
        determine $\tilde{\matr{F}}(t,:)$\tcp*{Determine (approximated) quantities of interest}
    }
    save $\tilde{\matr{F}}$\tcp*{Store quantities of interest}
    \KwRet 0
}

\caption{Rank $\tilde{q}$ evaluation ($\tilde{q} \leq q$) of the Spatially Parallel / Temporal Serial truncated Singular Value Decomposition algorithm, where $\mathrm{P}$, $\mathrm{L}$ and $\mathrm{Y}$ are known from the underlying simulation framework, cf. Sec. \ref{sec:starting_point}}
\label{alg:SVD_evaluation}
\end{algorithm}

\begin{algorithm}[!ht]
\SetKwInput{KwInput}{Input}
\SetKwInput{KwOutput}{Output}
\DontPrintSemicolon


\SetKwFunction{FMain}{adaptive{\_}truncation}
 
\SetKwProg{Fn}{Function}{:}{\KwRet}
\Fn{\FMain{$\vect{s}$}}{
    \eIf{length($\vect{s}$) $\leq$ o}
    {
        $q = \mathrm{length}(\vect{s})$\tcp*{Use all singular values}
    }{
        \For{q=o,length($\vect{s}$)}
        {
            \If{$ [ {\vect{s}}^\mathrm{T}(o:q)\vect{s}(o:q) ] / [ e - {\vect{s}}^\mathrm{T}(1:o-1)\vect{s}(1:o-1) ] \geq \eta^\mathrm{o}$}{exit\tcp*{Increase $q$ and leave when the energetic bound is met, cf. Eqn. \ref{equ:adaptive_energy}}}
        }
    }
    \KwRet $q$
}
\caption{Adaptive truncation rank estimation based on a minimal rank $o$ and a maximum modified retained energy $\eta^\mathrm{o}$.}
\label{alg:SVD_adaptive}
\end{algorithm}

The procedure outlined in Alg. \ref{alg:SVD_construction} might trigger numerical instabilities in case of more time steps than discrete degrees of freedom (i.e., $\mathrm{N}<\mathrm{T}$, which --however-- does not occur in this research) due to rounding errors within the itSVD update. Therefore, the procedure is extended by an additional matrix multiplication to regularize the $\matr{K} \to \matr{R}_Q \matr{K}$ matrix. The adapted pseudo-code is shown in the appendix in Alg. \ref{alg:SVD_construction_improved}, with differences to Alg. \ref{alg:SVD_construction} highlighted. Additionally, an exemplary Matlab$^\copyright$ code is provided, cf. \cite{matlabitSVD}. The efficiency of the algorithms could be increased by suitable necessity checking of the second QR decomposition on $\matr{Q}$, cf. \cite{zhang2022answer, li2022enhanced}.

\section{Governing Physical Equations \& Numerical Framework}
\label{sec:governing_equations}
The paper's numerical studies consider the flow of two immiscible, inert fluids ($a,b$) featuring constant bulk densities ($\rho^\mathrm{a}, \rho^\mathrm{b}$) and bulk viscosities ($\mu^\mathrm{a}, \mu^\mathrm{b}$), where $b$ is to be distinguished from the bunch matrix size. Fluid $a$ is referred to as foreground fluid and fluid $b$ as background fluid. Single-phase simulations of the upcoming verification \& validation Sec. \ref{sec:verification_validation} drop fluid a. Throughout the paper, the foreground fluid refers to air and the background fluid to water. Both fluids are assumed to share the kinematic field along the Volume-of-Fluid (VoF) approach suggested by \cite{hirt1981volume}. An Eulerian concentration field describes the spatial distribution of the fluids, where $c = c^\mathrm{a} = V^\mathrm{a}/V$ denotes the volume concentration of the foreground fluid, and the volume fraction occupied by the background fluid refers to $c^\mathrm{b} = V^\mathrm{b}/V = (V-V^\mathrm{a})/V = (1-c)$.

The governing equations primarily refer to the mixture's momentum and continuity equation and additional transport equations for the volume concentration of the foreground phase and turbulence quantities. The following set of equations needs to be solved for the pressure $p$, the fluid velocity $v_i$, the concentration $c$, the Turbulent Kinetic Energy (TKE) $k$, and its dissipation rate $\varepsilon$ where $\mathrm{D}^{(\cdot)}$, $\mathrm{P}^({\cdot)}$ as well as $\mathrm{R}^{(\cdot)}$, refer to their diffusion, production, and dissipation, respectively, viz.
\begin{alignat}{3}
&\mathrm{R}^\mathrm{p}=&&  \frac{\partial v_\mathrm{k}}{\partial x_\mathrm{k}} && = 0 \label{equ:rans_mass} \\
&\mathrm{R}^\mathrm{c}=&&  \frac{\partial c}{\partial t} + \frac{\partial v_\mathrm{k}  c}{\partial  x_\mathrm{k}} &&= 0 \label{equ:rans_conce} \\
&\mathrm{R}_\mathrm{i}^\mathrm{v_\mathrm{i}}=&& \frac{\partial \rho  v_\mathrm{i}}{\partial t} +  \frac{\partial v_\mathrm{k} \rho  v_\mathrm{i}}{\partial  x_\mathrm{k}} + \frac{\partial }{\partial  x_\mathrm{k}} \bigg[p^\mathrm{eff}   \delta_\mathrm{ik} - 2  \mu^\mathrm{eff}  S_\mathrm{ik} \bigg] - \rho  g_\mathrm{i} &&=0 \label{equ:rans_mome} \\
&\mathrm{R}^\mathrm{k}=&& \frac{\partial \rho  k}{\partial t} + \frac{\partial v_\mathrm{k} \rho  k}{\partial  x_\mathrm{k}} - \mathrm{D}^\mathrm{k} - \mathrm{P}^\mathrm{k} + \mathrm{R}^\mathrm{k} &&= 0  \label{equ:rans_tke} \\
&\mathrm{R}^\mathrm{\varepsilon}=&&  \frac{\partial \rho  \varepsilon}{\partial t} + \frac{\partial v_\mathrm{k} \rho  \varepsilon}{\partial  x_\mathrm{k}} - \mathrm{D}^\mathrm{\varepsilon} - \mathrm{P}^\mathrm{\varepsilon} + \mathrm{R}^\mathrm{\varepsilon} &&= 0  \label{equ:rans_epsilon} \, .
\end{alignat}
Therein, $\mu^\mathrm{ eff} = \mu + \mu^\mathrm{ t}$ and $p^\mathrm{ eff} = p + p^\mathrm{ t}$ represent effective viscosity and pressure.
 The unit coordinates and the strain rate tensor are denoted by the Kronecker Delta $\delta_\mathrm{ik}$ and $S_\mathrm{ik} = 1/2 (\nabla_\mathrm{k} v_\mathrm{i} + \nabla_\mathrm{i} v_\mathrm{k})$.
The sources and sinks of the turbulent balance equations are model-dependent functions of the local material properties, velocity gradients, and turbulent quantities, e.g., $P^\mathrm{k} := P^\mathrm{k}(\rho,\nabla_\mathrm{k} v_\mathrm{i},k,q)$ for the TKE production, and detailed information can be found in the corresponding literature, e.g., \cite{wilcox1998turbulence}. A foreground phase concentration $c \in [c^\mathrm{a}, c^\mathrm{b}]$ is used to measure phase and fluid properties. The concentration ranges from a foreground state ($c^\mathrm{a}, \rho^\mathrm{a}, \mu^\mathrm{a}$) to a background state ($c^\mathrm{b}, \rho^\mathrm{b}, \mu^\mathrm{b}$) and local material properties within Eqns. (\ref{equ:rans_mome})-(\ref{equ:rans_epsilon}) follow from two Equations of State (EoS), viz.
\begin{align}
\rho = m \rho^\mathrm{\Delta} + \rho^\mathrm{b} , 
\qquad \qquad \text{and} \qquad \qquad
\mu = m \mu^\mathrm{\Delta} + \mu^\mathrm{b} , \label{equ:density_interpolation}
\end{align}
where $\rho^\mathrm{\Delta} = \rho^\mathrm{a} - \rho^\mathrm{b}$ and $\mu^\mathrm{\Delta} = \mu^\mathrm{a} - \mu^\mathrm{b}$ mark the difference of the bulk densities as well as viscosities. The standard EoS corresponds to a simple linear interpolation between the limit states, i.e., $m=c$, cf. \cite{kroger2018adjoint, kuhl2021cahn}

The utilized numerical framework to approximate the partial differential Eqns. (\ref{equ:rans_mass})-(\ref{equ:rans_epsilon}) follows a conventional, pressure-based, second-order accurate Finite-Volume (FV) scheme. Unstructured grids, based on arbitrary polyhedral cells with possibly hanging nodes, can be used, cf. \cite{rung2009challenges}. The sequential procedure uses the strong conservation form and employs a cell-centered, co-located storage arrangement for all transport properties. The framework is dedicated to Single Instruction Multiple Data (SIMD) implementations on a distributed-memory parallel CPU machine for several thousand processes using a domain decomposition method based on the (par)METIS algorithm (\cite{metis, parmetis}) and the MPI communication protocol (\cite{yakubov2013hybrid, mpi}). Algorithms employed by the in-house procedure FresCo$^+$ are described in \cite{rung2009challenges}, and \cite{yakubov2013hybrid, kuhl2021phd}. They ground on the integral form of a generic Eulerian transport equation with residuum $\mathrm{R}^{\varphi}$, Eqns. (\ref{equ:rans_mass})-(\ref{equ:rans_epsilon}), for a scalar field $\varphi(x_i, t)$ exposed to the influence of a possibly non-linear source term $\mathcal{S}^\varphi$ in addition to a modeled (non-linear) gradient diffusion term and its diffusivity $\Gamma^{\mathrm{eff}}$ in a control volume $\Omega$ bounded by the Surface $\Gamma$, viz. 
\begin{align}
	\int \mathrm{R}^\varphi \mathrm{d} \Omega = 0 \, , \qquad \to \qquad \int \left[ \frac{\partial \varphi}{\partial t} - \mathcal{S}^\varphi \right] \mathrm{d} \Omega + \oint \left[ u_\mathrm{i} \varphi - \Gamma^{\mathrm{eff}} \frac{\partial \varphi}{\partial x_\mathrm{i}} \right] n_\mathrm{i} \, \mathrm{d} \Gamma = 0 \; . \label{equ:momentum_numerics}
\end{align}
Here, $x_\mathrm{i}$ refers to the Cartesian spatial coordinates and $n_\mathrm{i}$ as well as $t$ denote the surface normal vector components and physical time, respectively. Hence, the employed FV approximation yields a discrete system of size $\mathrm{P} \, \mathrm{L} \times \mathrm{P} \, \mathrm{L}$, where $\mathrm{P} \, \mathrm{L}$ represents the number of Control-Volumes (CV) or discrete DoF, i.e., the length of one state snapshot vector (cf. Sec. \ref{sec:starting_point}). Each line corresponds to a balance of the particular $\mathrm{CV}$, i.e.,
\begin{equation}
    A^\mathrm{\varphi, CV} \varphi^\mathrm{CV} - \sum_\mathrm{NB(CV)} A^\mathrm{\varphi, NB} \varphi^\mathrm{NB} = S^\mathrm{\varphi, CV} \label{equ:discrete_generic_transport_equation} \, ,
\end{equation}
where $A^\mathrm{\varphi, CV}$, $A^\mathrm{\varphi, NB}$ and $S^\mathrm{\varphi, CV}$ refer to the main diagonal coefficient, its adjacent neighboring entries as well as right-hand side in terms of a generic source-term. The solution is iterated to convergence using an enhanced pressure-correction scheme. The odd-even decoupling between the pressure and the velocities components is suppressed via an appropriate Rhie-Chow interpolation scheme, cf. \cite{rhie1983numerical, yakubov2015experience, kuhl2022discrete}. The numerical integration employs a second-order mid-point rule, where discretized diffusive fluxes follow from second-order accurate central differencing and discrete convective fluxes employ higher-order upwind biased interpolation formulae. Time derivatives are approximated by an Implicit Three-Time Level (ITTL) scheme, embedded in a sub-cycling strategy to efficiently meet stability criteria based on maximum Courant number conditions of $\mathrm{Co} = \mathcal{O}(0.5)$ along the refined free-surface regions, cf. \cite{manzke2013sub, kuhl2021adjoint}. The convective transport of momentum and turbulence parameters follows the Quadratic Upwind Interpolation of Convective Kinematics (QUICK) scheme. In the case of two-phase flows, a compressive High-Resolution Interface Capturing (HRIC) procedure convects the volume fraction field. The latter is extended by an Explicit Interface Sharpening (EIS) procedure to ensure the sharpest possible representation of the fluid-gas interface over a maximum of 1-2 cells (\cite{manzke2018development}). Potential spatiotemporal wave boundary conditions are imposed along the outer boundaries towards the interior inside a buffer region by manipulating the equation system for the concentration field $c$ based on linear (Airy) wave theory, i.e., $A^{c, \mathrm{P}}$, $A^{c, \mathrm{NB}}$ and $S^{c, \mathrm{P}}$ in Eqn. (\ref{equ:discrete_generic_transport_equation}), i.e, the domain's nonlinear Navier-Stokes solution is blended by the linear wave theory towards the boundary, cf. \cite{woeckner2010efficient, luo2017computation, luo2019numerical}. In all cases considered herein, the buffer zone extends one wavelength into the domain, and its manipulation intensity decays quadratically.
Since the data structure is generally unstructured, suitable preconditioned iterative sparse-matrix solvers for symmetric and non-symmetric systems, e.g., GMRES, BiCG, QMR, CGS, or BiCGStab, are used.

\section{Verification \& Validation}
\label{sec:verification_validation}
\begin{figure}[!ht]
\centering
\iftoggle{tikzExternal}{
\input{./tikz/2D/Cylinder/cylinder_sketch.tikz}
}{
\includegraphics{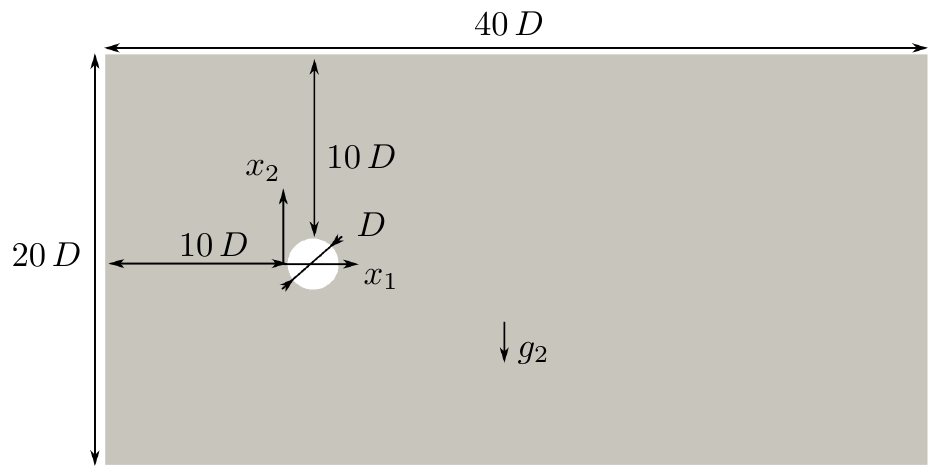}
}
\caption{Schematic drawing of the submerged cylinder case configuration ($\mathrm{Re}_\mathrm{D} = 200$, $\mathrm{Fn}=0.2828$, $\mathrm{St} = 0.2$).}
\label{fig:cylinder_sketch}
\end{figure}

\begin{figure}[!ht]
\centering
\subfigure[]{
\includegraphics[height=0.35\textwidth]{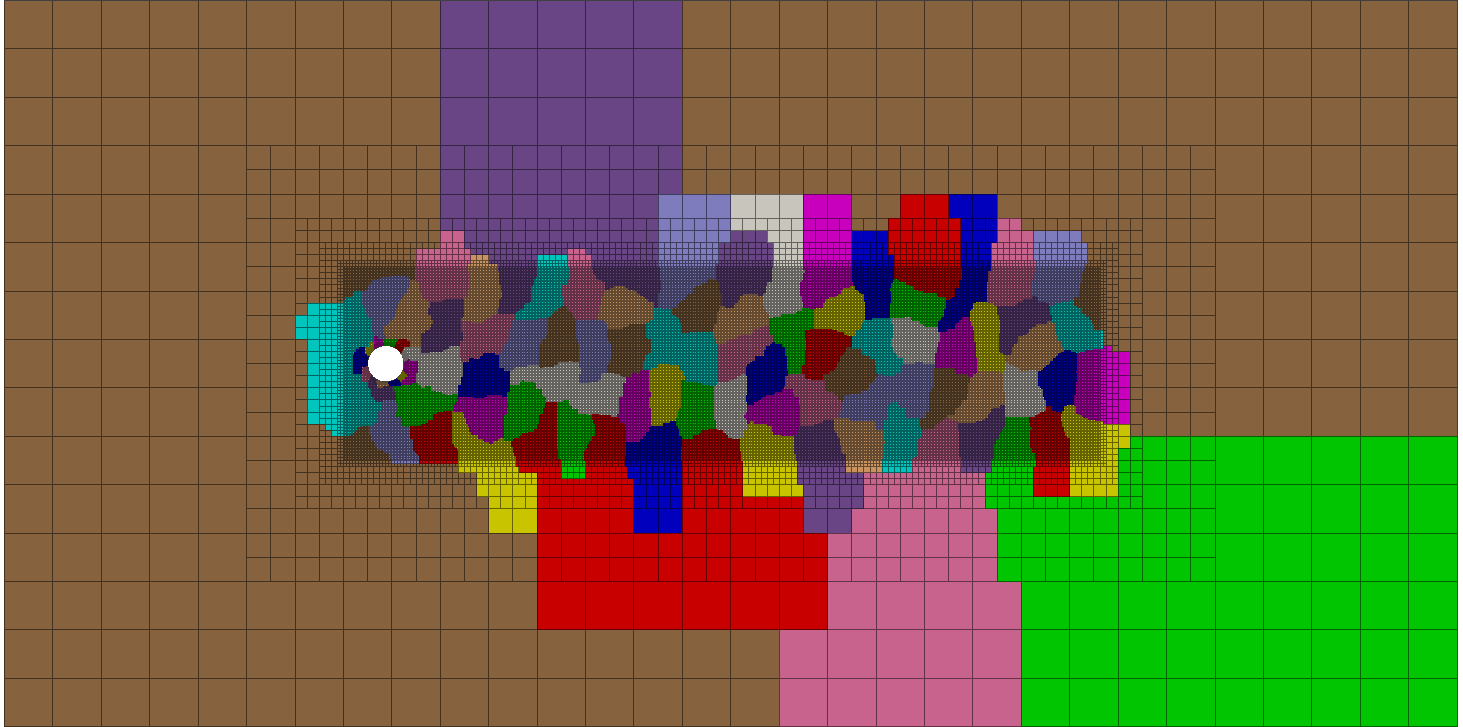}
}
\subfigure[]{
\includegraphics[height=0.35\textwidth]{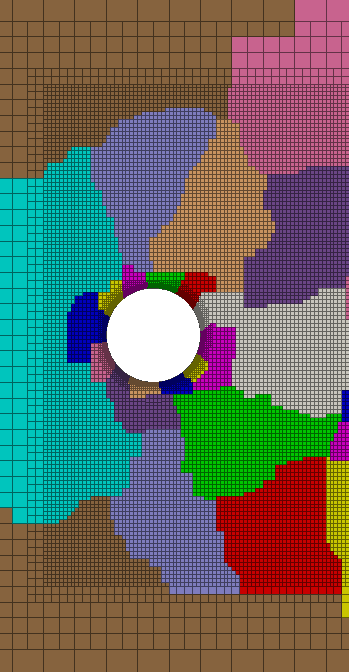}
}
\caption{(a)  Employed unstructured numerical grid of the submerged cylinder case ($\mathrm{Re}_\mathrm{D} = 200$, $\mathrm{Fn}=0.2828$, $\mathrm{St} = 0.2$)  with $\mathrm{P} = 96$ partitions  and (b)  close up of the cylinders's vicinity.}
\label{fig:cylinder_single_phase_grid}
\end{figure}

This section assesses and verifies the credibility of the presented itSVD SPTS approach. The investigations 
refer to a laminar single-phase flow around a two-dimensional submerged circular cylinder at rest. The studies consider local quality measures, e.g. $(\tilde{\varphi} - \varphi) / \varphi^\mathrm{ref}$ as well as global integral forces by means of the drag and lift coefficients, viz.
\begin{align}
    c_\mathrm{d} = \frac{2 F_i \delta_{i1}}{\rho A V_1}
    \quad
    \quad
    \mathrm{and}
    \quad
    \quad
    c_\mathrm{l} = \frac{2 F_i \delta_{i2}}{\rho A V_1}
    \qquad
    \qquad
    \mathrm{with}
    \qquad
    \qquad
    F_i = \int_{\Gamma^\mathrm{C}} \left[ p \delta_\mathrm{ik} - \mu 2 S_{ik} \right] n_\mathrm{k} \, \mathrm{d} \Gamma \, .  \label{equ:drag_lift-coefficient}
\end{align}
Here $\Gamma^\mathrm{C}$ and $n_\mathrm{k}$ denote to the boundary and its normal, and $A$ as well as $V_1$ refer to a reference area and the bulk inlet velocity, respectively. Gravitational forces are active and the transient studies are performed at $\mathrm{Re}_\mathrm{D} = \rho^\mathrm{b} V_\mathrm{1} D / \mu^\mathrm{b} = \SI{200}{}$, $\mathrm{Fn} = V_\mathrm{1}/\sqrt{G \, D} = \SI{0.2828}{}$ and $\mathrm{St} = D f / V_1 \approx 0.2$, based on the gravitational acceleration $G$, the kinematic viscosity $\nu$ and an expected vortex shedding frequency $f \approx 0.2$. The employed two-dimensional domain is depicted in Fig. \ref{fig:cylinder_sketch}. It features a length and a height of $40 \, D$ and $20 \, D$, where the inlet [bottom] boundary is located 10 [9.5] diameters away from the cylinder's origin. At the inlet, a homogeneous unidirectional (horizontal) bulk flow $v_\mathrm{i} = V_1 \delta_{i1}$ is imposed. Slip walls are used along the top and bottom boundaries, and a hydrostatic pressure boundary is employed along with the downstream located outlet. Initial values follow from a hydrostatic pressure field as well as a homogeneous, horizontal flow.
The utilized unstructured numerical grid is displayed in Fig. \ref{fig:cylinder_single_phase_grid} together with an indication of $\mathrm{P} = 96$ employed partitions. It consists of approximately $\SI{75000}{}$ control volumes where the cylinder shape is discretized with 770 surface elements along the circumference. The non-dimensional wall-normal distance of the first grid layer reads $y^+ \approx \SI{0.01}{}$. The grid is refined inside a box of size [$20 \, D \times 5 \, D$] with a spacing of $\Delta x_\mathrm{1} / D = \Delta x_\mathrm{2} / D  = 1 / 20$ to capture the vortex street. The time integration utilizes a constant time increment of $\Delta t \, f = \mathrm{St} / 50$ or $\Delta t / (D / V_1) = 1 / 50$, i.e., one cylinder passage time is discretized with 50 time steps.

\subsection{Parallel vs. Serial Execution} 
\label{sec:accurate}
We first verify that the presented algorithm is independent of the parallelization P and the bunch size $b$. 
For this purpose, a time interval of $\Delta T / (D / V_1) = 100$ cylinder passage times is simulated from scratch to 
establish a periodic flow that features a vortex street.  
The flow field consists of three state variables $\mathrm{S}$, i.e., pressure $p$, horizontal $v_1$ and vertical $v_2$ velocity and reference quantities in Eqns. (\ref{equ:SVD_start}),(\ref{equ:SVD_end}) are set to $\varphi^v = V_1$ and $\varphi^p = \rho^\mathrm{b} \, G \, D$. Since the unpartitioned domain is discretized with approximately $\mathrm{L} \approx \SI{75 000}{}$ cells, the spatial size (length) of the global system matrix is $\mathrm{N} 
\approx \SI{225 000}{}$. 
The number of time steps describes the width which reads 
$\mathrm{T} 
= \SI{5 000}{}$ of this matrix, i.e.,  $\matr{Y} \in \mathbb{R}^{\SI{225 000}{} \times \SI{5 000}{}} $.
 To assess the decomposition influences, we employ 
 $q = \mathrm{T} / 10 =  500$ time steps, referring to ten equivalent flow passings.  
%

 The parallelization of Alg. \ref{alg:time_integration}-\ref{alg:SVD_evaluation} is assessed by examining 20 different scenarios. To this end, ten different spatial partitioning approaches $\mathrm{P} = [1, 2, 3, 4, 5, 6, 12, 24, 48, 96]$ are tested with (a) a fully parallel and (b) a  serial itSVD construction. In the latter case, the parallelization-relevant lines in Algs. \ref{alg:time_integration} (lines 9, 12, 20), \ref{alg:SVD_construction} (lines 3,  6, 11), and \ref{alg:auxilary_routines} (lines 5, 7) of the construction process, as well as Alg. \ref{alg:SVD_evaluation} (lines 2, 4) for the evaluation are neglected, leading to P 
isolated itSVD constructions for each scenario (b). 
In line with Eqn. (\ref{equ:reduced_svd}), no differences between the parallel and serial approaches are expected 
due to the characteristic equality $\matr{A} = \matr{U}_r \matr{S}_r \matr{V}_r^\mathrm{T}$ for a complete, non-truncated rSVD, as confirmed by the schematics in Figs. \ref{fig:reduced_parallel_rpts_svd}, \ref{fig:truncated_parallel_rpts_svd}. However, if the truncation rank is strongly reduced (i.e., $\tilde{q} << q$), the continuity of re-constructed field variables along the partition's boundaries is no longer guaranteed since the retained matrix energy $\eta^\mathrm{q}$  might locally differ considerably, cf. Eqn. (\ref{equ:retained_energy}).
Two partitioning scenarios are examined in more detail for the parallel and the serial approach in Figs. \ref{fig:parallelization_v1} and \ref{fig:parallelization_v1_difference}, i.e. a coarse grained (P=12; left) and a fine grained partitioning (P=96; right).

\subsection*{Accuracy}
\begin{figure}[!ht]
\centering
\iftoggle{tikzExternal}{
\input{./tikz/2D/Cylinder/cylinder_parallelization_v1.tikz}
}{
\includegraphics{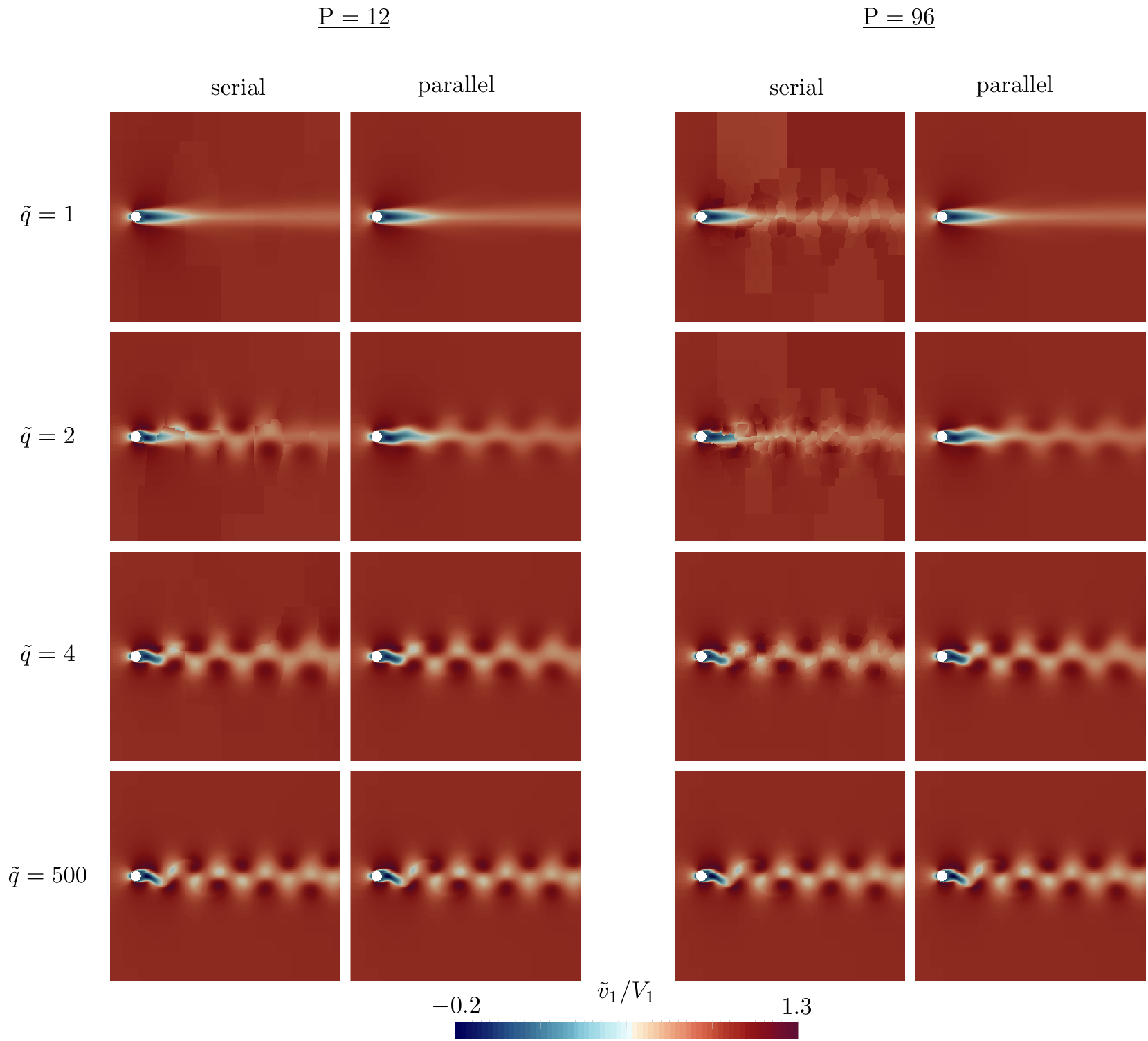}
}
\caption{
Reconstructed primal velocity $\tilde{v}_1/V_1$ for a single phase cylinder flow with $q/\mathrm{T} = 1/10$ ($\mathrm{Re}_\mathrm{D} = 200$, $\mathrm{Fn}=0.2828$, $\mathrm{St} = 0.2, \mathrm{T}=5000$).
%
Reconstructed field at $t/\mathrm{T} = 0.8$ based on Alg. \ref{alg:SVD_evaluation} for different reconstruction ranks (rows) and two partitionings (columns). Each study is performed with a parallel and serial approach that neglects all parallel statements in Algs. \ref{alg:time_integration} (lines 9, 12, 20), \ref{alg:SVD_construction} (lines 3, 5, 6, 11), and Alg. \ref{alg:auxilary_routines} (lines 5, 7) of the construction process, as well as Alg. \ref{alg:SVD_evaluation} (lines 2, 4) for evaluation.}
\label{fig:parallelization_v1}
\end{figure}

\begin{figure}[!ht]
\centering
\iftoggle{tikzExternal}{
\input{./tikz/2D/Cylinder/cylinder_parallelization_v1_difference.tikz}
}{
\includegraphics{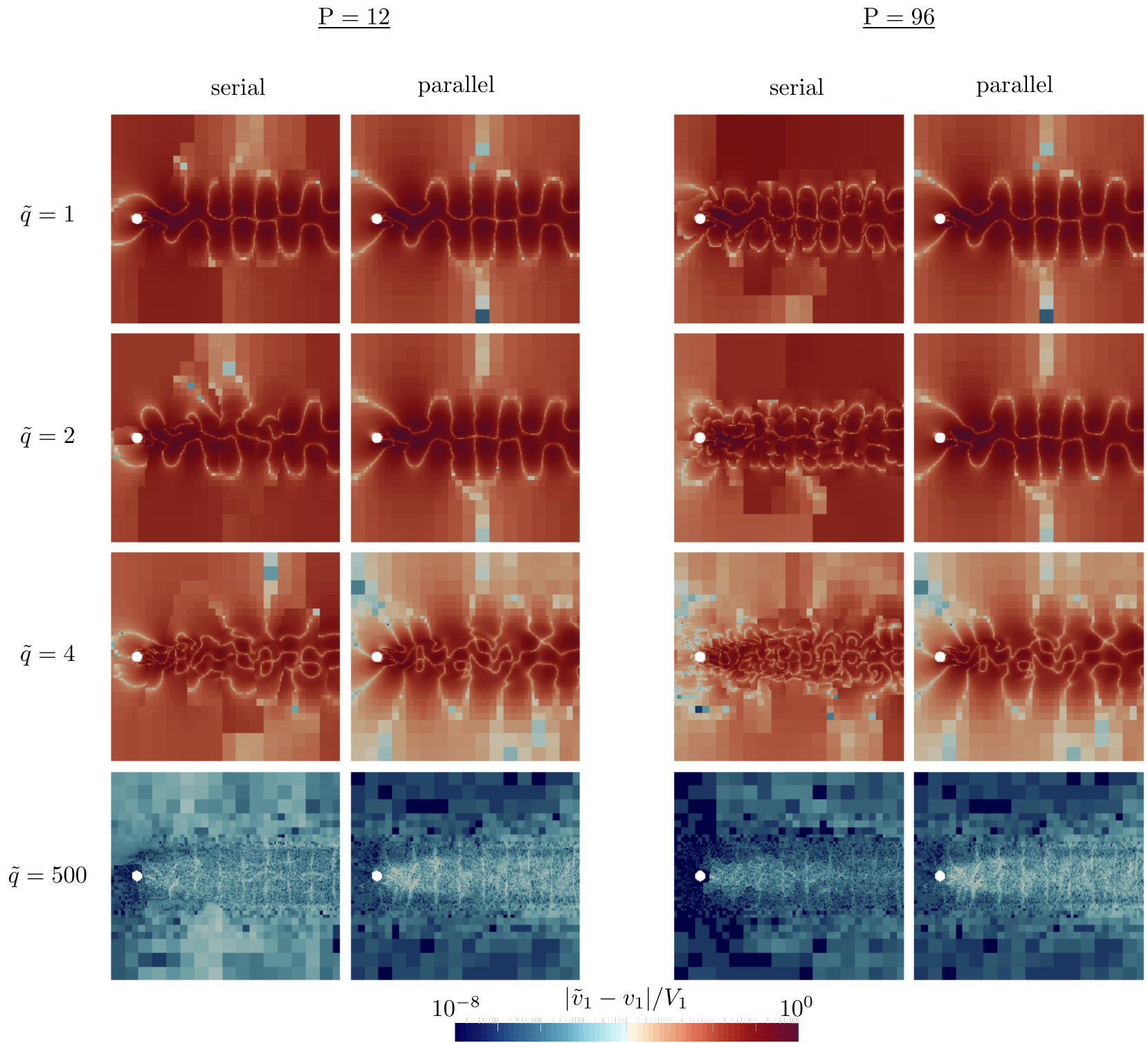}
}
\caption{Reconstruction error $|\tilde{v}_1 - v_1|/V_1$ for a single phase cylinder flow with $q/\mathrm{T} = 1/10$ ($\mathrm{Re}_\mathrm{D} = 200$, $\mathrm{Fn}=0.2828$, $\mathrm{St} = 0.2, \mathrm{T}=5000$). Relative error at $t/\mathrm{T} = 0.8$ based on Alg. \ref{alg:SVD_evaluation} for 4 different reconstruction ranks (rows) and two partitionings (columns). Each study is performed with a parallel and a serial approach that neglects all parallel statements in Algs. \ref{alg:time_integration} (lines 9, 12, 20), \ref{alg:SVD_construction} (lines 3, 5, 6, 11), and Alg. \ref{alg:auxilary_routines} (lines 5, 7) of the construction process, as well as Alg. \ref{alg:SVD_evaluation} (lines 2, 4) for the evaluation.}
\label{fig:parallelization_v1_difference}
\end{figure}

\begin{figure}[!ht]
\centering
\iftoggle{tikzExternal}{
\input{./tikz/2D/Cylinder/cylinder_sv_energy_data.tikz}
}{
\includegraphics{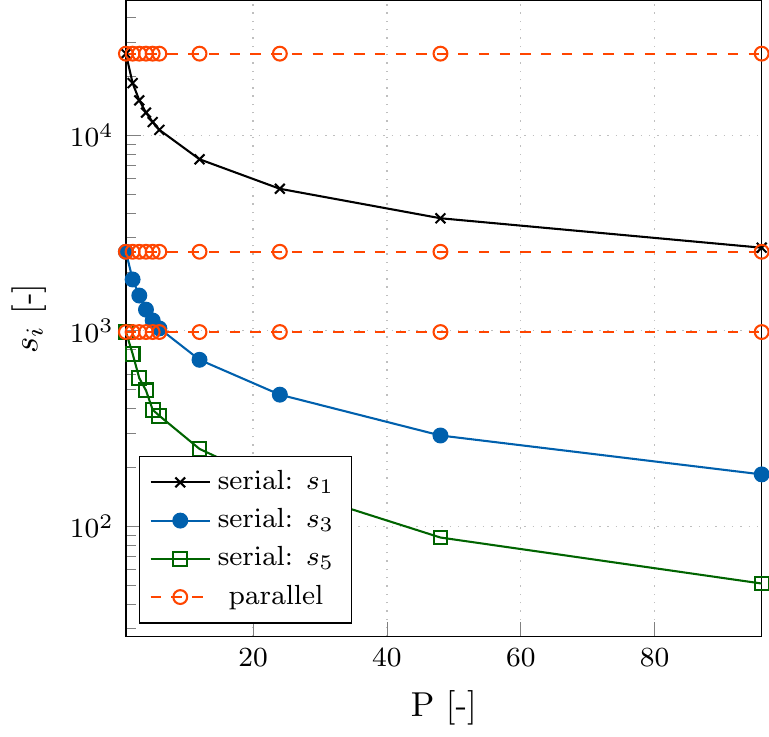}
\includegraphics{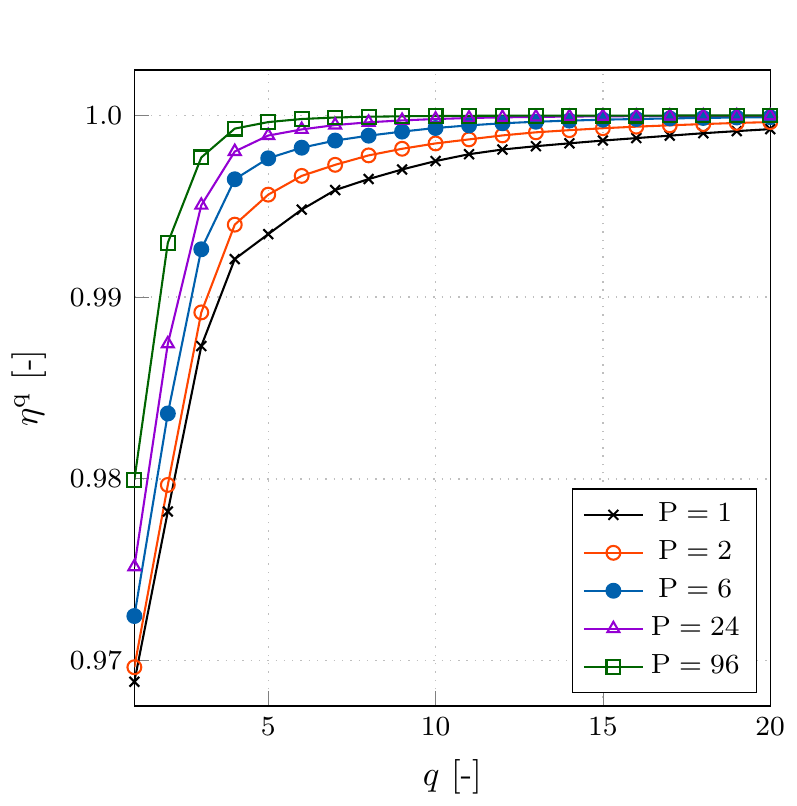}
}
\caption{Single phase cylinder flow with $q / \mathrm{T} = 1 / 10$
($\mathrm{Re}_\mathrm{D} = 200$, $\mathrm{Fn}=0.2828$, $\mathrm{St} = 0.2, \mathrm{T}=5000$). Left: Representation of the first (black), third (blue), and fifth (green) serially determined singular values for different spatial partitionings. The corresponding singular values of the parallel approach are partitioning independent as indicated by the horizontal orange lines. Right:  Retained energy $\eta^\mathrm{q}$ from Eqn. (\ref{equ:retained_energy}) of the first 20 modes on partition $\#0$ for different spatial partitions.}
\label{fig:parallelization_data}
\end{figure}

Figure \ref{fig:parallelization_v1} shows several horizontal velocity fields at time instant $t/\mathrm{T} = 0.8$ for the (a) parallel as well as (b) serial approach. The figure displays  $\tilde{q} =  [1, 2, 4, 500]$ reconstruction ranks that 
increase from top to bottom. 
In all parallel studies, the reconstructed fields are smooth independent of the partitioning. The  energetically most intense bulk mode ($\tilde{q} = 1$) seems to form a steady symmetrical 
 flow. A vortex street follows when adding the subsequent three singular values during the reconstruction. The difference displayed comparing $\tilde{q} =  4$ and $\tilde q  = 500$ is generally smaller compared to the changes introduced between $\tilde{q} =  2$ and $\tilde{q} =  4$,  and will be further investigated by adaptivity studies reported in Sec. \ref{sec:practical_issues}.
In contrast to the parallel framework, the serial approaches reveal a  dependence on the partitioning for small evaluation ranks ($\tilde{q} \leq 4$), cf. Fig. \ref{fig:cylinder_single_phase_grid}. The differences along the partition boundaries are particularly obvious for $\tilde{q} = 2$. 
If the evaluation rank and the retained energy are   increased, the  alignment (smoothness)  of the reconstructed fields improves. 
For $\tilde{q} =  500$, smooth velocities
are also observed in the serial framework, which seem to agree with the parallel approach. 
 Figure 
\ref{fig:parallelization_v1_difference} depicts the local error magnitude of the reconstructed primal velocity $|\tilde{v}_1 - v_1|/V_1$  for all 16 cases.
The figure also outlines partitioning influences for small $\tilde{q}$ when using a serial approach.
%
%
In addition to the visual comparison of the reconstructed primal velocity, the singular values --particularly relevant for the reconstruction-- are displayed in more detail. Figure \ref{fig:parallelization_data} (left) shows the first (black), third (blue), and fifth (green) serially determined singular value of partition $\# 0$ indicated by solid lines. They are compared to 
the corresponding  singular values returned by the parallel approach, indicated by orange symbols and dashed lines. The abscissa of the figure refers to the partitioning.
It is noticed that the singular values obtained from the parallel approach are partitioning independent. As opposed to this, the singular values in all serial cases increase [decrease] for smaller [larger] partitioning sizes.
The decrease in serially determined singular values for fine grained partitioning is expected as the number of spatial degrees of freedom --or the length of the state snapshot vector per thread and thus its magnitude-- decreases, cf. Alg. \ref{alg:time_integration} line 10.
Fig. \ref{fig:parallelization_data} (right) depicts the retained energy $\eta^\mathrm{q}$ on partition $\# 0$ for a range of spatial decompositions, as predicted by Eqn. (\ref{equ:retained_energy}). The abscissa of this figure refers to the truncation rank $q$.
Independent of the partitioning, all curves converge to $\eta^\mathrm{q} \to 1$. However, the convergence starts later [earlier] for low [high] partitioning. The latter is due to the smaller spatial area covered by the serial approach. Mind that the number of cells per partition increases for coarser portionings and finally 
agrees with the parallel SVD 
for $\mathrm{P}=1$, cf. the agreement of serial singular values in the left subfigure. Furthermore, this supports the previous observation that
the retained matrix energy is  significantly partitioning dependent for small ranks $\tilde q$.  However, the latter disappears noticeably for $\tilde{q} \geq 20$. 
%

\subsection*{Storage} 
\label{sec:store}
Next, the occupied disk space $\tilde{d}$ at the end of the respective itSVD construction process is discussed.
To this end, the memory requirement of Alg. \ref{alg:time_integration} (line 20-22) is measured, where the serial itSVD approaches omit line 20 
and 
store the $\matr{V}$ matrix and the $\vect{s}$ vector in addition to the local matrix $\matr{U}$ for each partition. 
The storage requirements are outlined in Tab. \ref{tab:parallelization_memory}. The tabulated data is normalized by the full storage approach $\tilde{d}^\mathrm{\, full}$ that stores the state vector after every time step. The parallel framework always  requires 7.5\% of the full storage disc usage, i.e. $\tilde{d}^\mathrm{\, svd, \, par.} / \tilde{d}^\mathrm{\, full} \approx \SI{0.075}{}$. In contrast, the memory needed in the serial cases $\tilde{d}^\mathrm{\, svd, \, ser.} / \tilde{d}^\mathrm{\, full}$ increases when increasing the amount of partitions. The table also indicates the ratio of the serial to the parallel memory overhead, that is equal to one in the  serial case ($\mathrm{P} = 1$) and approximately triples for  $\mathrm{P} = 96$ partitions.
\begin{table}[!ht]
\caption{Single phase cylinder flow with $q/\mathrm{T} = 1 / 10$ 
($\mathrm{Re}_\mathrm{D} = 200$, $\mathrm{Fn}=0.2828$, $\mathrm{St} = 0.2, \mathrm{T}=5000$). Memory efforts $\tilde{d}$ for the serial and parallel itSVD, normalized with the effort of a full storage approach $\tilde{d}^\mathrm{\, full}$ that stores the state vector after every time step.}
\begin{center}
\begin{tabular}{cllllllll}
\toprule
Partitions P [-] & 96 & 48 & 24 & 12 & 6 & 4 & 2 & 1\\ \midrule 
$\tilde{d}^\mathrm{\, svd, \, ser.} / \tilde{d}^\mathrm{\, full}$ & \SI{0.2293}{} & \SI{0.1512}{} & \SI{0.1127}{} & \SI{0.0931}{} & \SI{0.0832}{}  & \SI{0.0799}{} & \SI{0.0767}{} & \SI{0.0751}{} \\ 
$\tilde{d}^\mathrm{\, svd, \, par.} / \tilde{d}^\mathrm{\, full}$ & \SI{0.0751}{} & \SI{0.0751}{} & \SI{0.0751}{} & \SI{0.0751}{} & \SI{0.0751}{} & \SI{0.0751}{} & \SI{0.0751}{} & \SI{0.0751}{} \\ \midrule \midrule
ratio [-] & \SI{3.0581}{} & \SI{2.0183}{} & \SI{1.498}{} & \SI{1.238}{} & \SI{1.108}{} & \SI{1.065}{} & \SI{1.021}{} & \SI{1.0}{} \\
\bottomrule
\end{tabular}
\end{center}
\label{tab:parallelization_memory}
\end{table}

\subsection*{Scaling}
\begin{figure}[!ht]
\centering
\iftoggle{tikzExternal}{
\input{./tikz/2D/Cylinder/cylinder_strong_scaling.tikz}
}{
\includegraphics{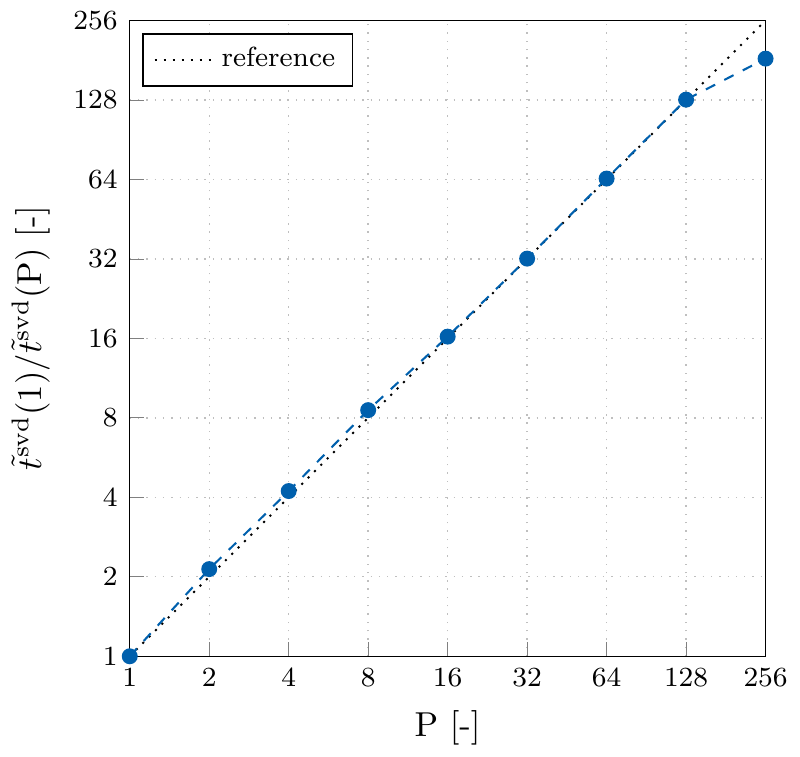}
\includegraphics{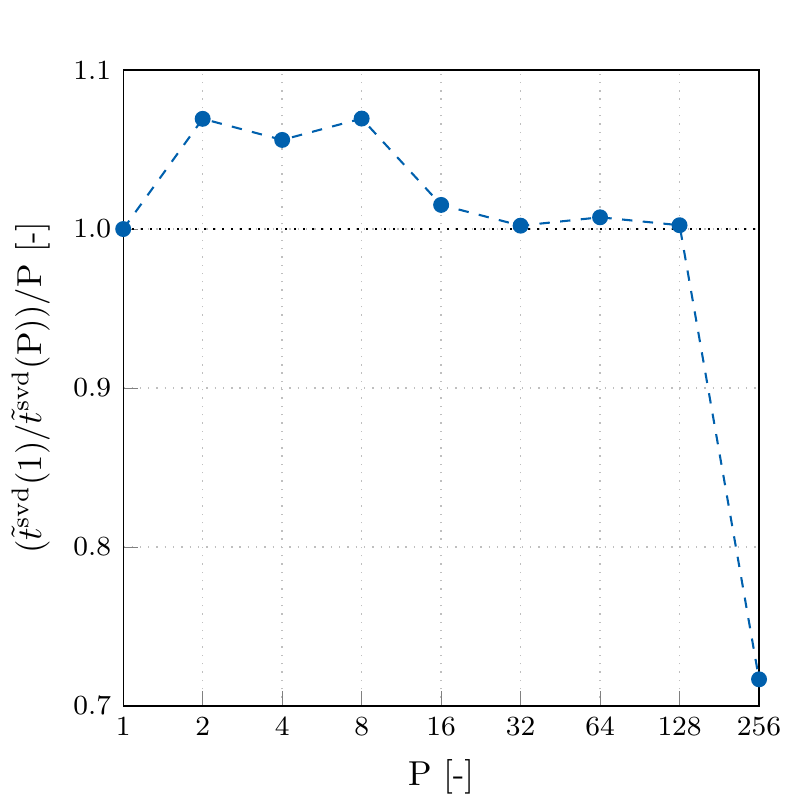}
}
\caption{Scalability test of the parallel itSVD method for the single phase cylinder flow with $q/\mathrm{T} = 1$ and $b/q=0.2$ 
($\mathrm{Re}_\mathrm{D} = 200$, $\mathrm{Fn}=0.2828$, $\mathrm{St} = 0.2, \mathrm{T}=500$). Speedup 
(left) and efficiency (right) of the itSVD algorithm for 
a fine mesh with $\mathrm{N} \approx \SI{3 000 000}{}$.
Given numbers correspond to averages obtained from three runs per scenario.}
\label{fig:cylinder_strong_scaling}
\end{figure}

The scalability of the parallel itSVD algorithm is assessed from a strong scaling test. For this purpose a finer grid with $\mathrm{P L} \approx \SI{1 000 000}{}$ control volumes is employed which is approximately four times finer in each spatial dimension than the grid illustrated in Fig. \ref{fig:cylinder_single_phase_grid}.
%
The scalability tests ground on $\mathrm{T}=500$ time steps using $q/\mathrm{T} =1$ and $b/q=0.2$.
The number of employed processors is consecutively doubled, i.e., $\mathrm{P}=[1,2,4,8,16,32,64,128,256]$ and 
the speedup $\tilde{t}^\mathrm{svd}(1)/\tilde{t}^\mathrm{svd}(P)$ is determined from the wall-clock time required by Algs. \ref{alg:SVD_construction}-\ref{alg:auxilary_routines}, where $\tilde{t}^\mathrm{svd}(1)$ corresponds to the overhead of a serial itSVD construction. To eliminate potential read/write latencies from the results, data w.r.t. wall-clock measurements is computed from the average of three numerical experiments per scenario.
The installed hardware refers to 96 Intel Cascade Lake$^\copyright$ Platinum 9242 processors per node connected with Intel Omni-Path$^\copyright$ on the NHR Göttingen, cf. \href{https://www.hlrn.de}{www.hlrn.de}.
Figure \ref{fig:cylinder_strong_scaling} outlines measured speedup (left) and efficiency (right) results.
%
%
The results  
 initially indicate superlinear scaling, which might be due to cache effects,   
 until $\mathrm{P}=16$ and a fair linear scaling until $\mathrm{P}=128$. 
They also reveal a processor load limit of about $\mathrm{N}/\mathrm{P}=\mathrm{S}\mathrm{L} \approx 15000$, below which the proposed algorithm reaches its scaling limit.
Similar results were obtained for different bunching sizes $b/q$.

\subsection{Overheads and Accuracy of the Parallel itSVD}
The remainder of the paper is restricted to the parallel itSVD version, and all cases of this subsection utilize the fine grained partitioning $\mathrm{P} = 96$ illustrated in Fig. \ref{fig:cylinder_single_phase_grid}.
%
While the storage overhead discussed in Sec. \ref{sec:store} is measured by the required hard-disk usage $\tilde{d}^\mathrm{\, svd}$, the CPU effort is characterized by the additional CPU wall clock time $\tilde{t}^\mathrm{\, svd}$. 
The latter is also normalized with the corresponding data af a full-storage approach which stores the state snapshot vector after each time step during the construction phase and yields the normalization parameter
$\tilde{t}^\mathrm{\, full}$. Note that --to increase efficiency-- the 
output of 
the full storage approach could 
also be 
bunched in line with the procedure employed for the itSVD construction. However, this is 
deliberately not considered as a reference.
%
%
For a maximum construction rank $q / \mathrm{T} = 1$, the error between the computed and reconstructed state should vanish $|\tilde{\varphi} - \varphi|/\varphi^\mathrm{ref} \to 0$ when $\tilde{q} \to q = \mathrm{T}$. 
 To verify this, the construction rank $q = 500$  of the previous subsection is continuously increased, 
i.e., 
$\tilde q / q = 1$ with $q/\mathrm{T} = [1/50, 1/10, 1/5, 1/2, 1]$. In addition, five different bunch sizes 
$b / q = [1, 1/5, 1/10, 1/25, 1/50]$ are examined in each case, resulting in a total of 25 studies.
The untruncated, one-shot rSVD refers to $b/q = 1$ as well as $q/\mathrm{T} = 1$ and can still be performed 
with moderate effort for the present 2D cylinder flow.

\begin{figure}[!ht]
\centering
\iftoggle{tikzExternal}{
\input{./tikz/2D/Cylinder/cylinder_memory_q_5000_cd.tikz}
}{
\includegraphics{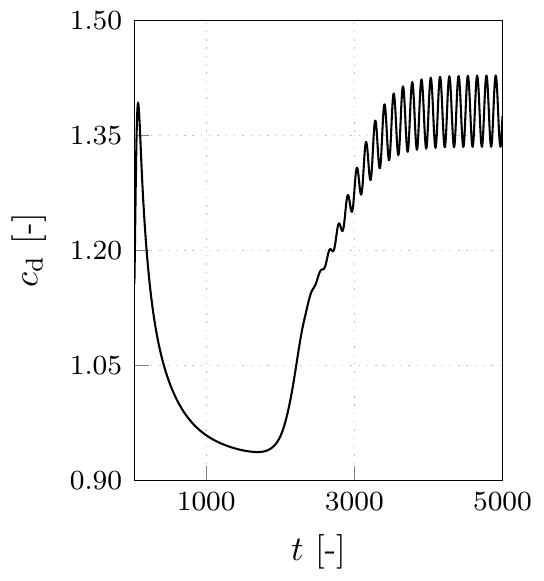}
\includegraphics{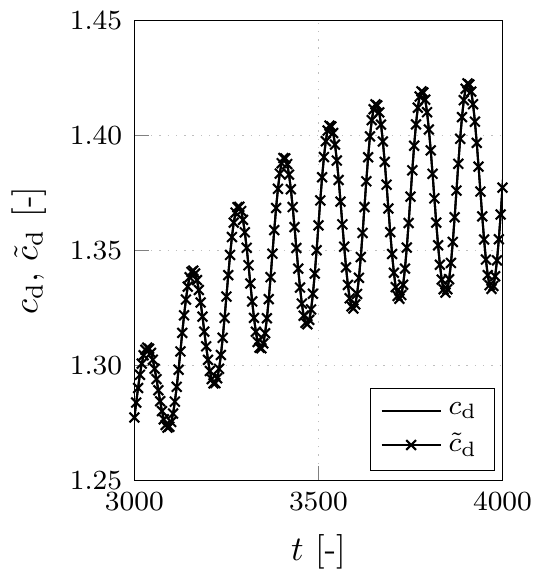}
\includegraphics{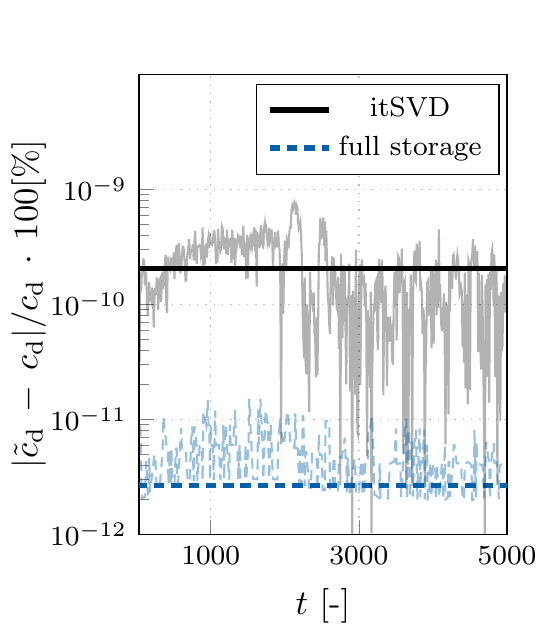}
}
\caption{Single phase cylinder flow for $q/\mathrm{T} = 1 = \tilde{q}/\mathrm{T}$ 
($\mathrm{Re}_\mathrm{D} = 200$, $\mathrm{Fn}=0.2828$, $\mathrm{St} = 0.2, \mathrm{T}=5000$). 
Left: Time evolution of the computed drag coefficient $c_\mathrm{d}$.  
Center: Enlargement of the time step interval $3000 \leq t \leq 4000$ supplemented with the drag coefficient $\tilde{c}_\mathrm{d}$ 
 obtained from the  reconstructed field.
 Right: Relative error between both quantities
 indicated in black, supplemented by the error returned by a full storage approach in blue and respective averages (horizontal lines). All simulations performed in parallel using P=96 partitions. 
 }
\label{fig:memory_data_cd}
\end{figure}

\begin{figure}[!ht]
\centering
\iftoggle{tikzExternal}{
\input{./tikz/2D/Cylinder/cylinder_memory_q_5000_cl.tikz}
}{
\includegraphics{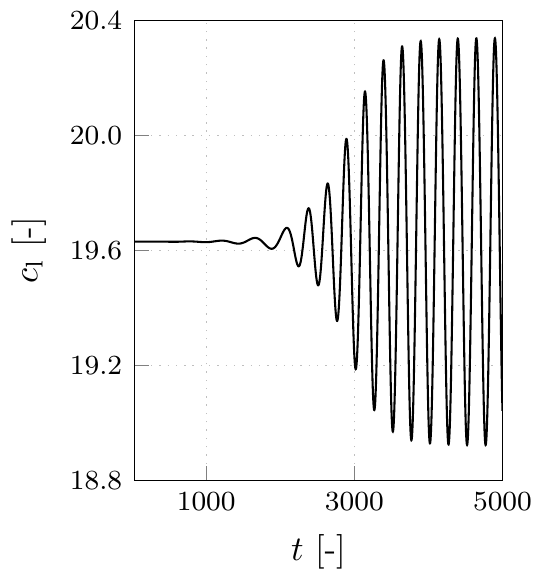}
\includegraphics{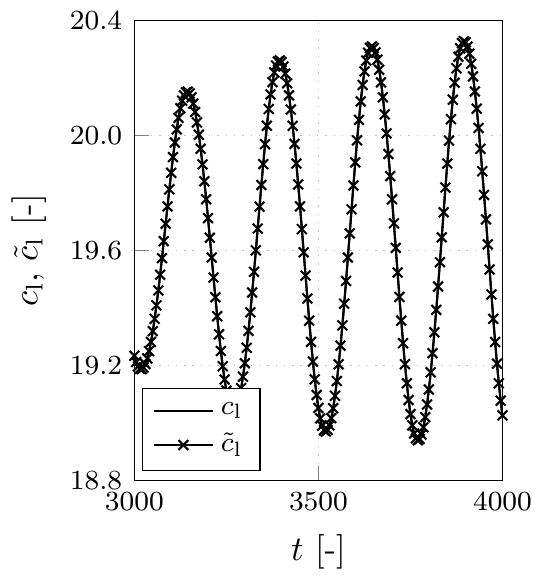}
\includegraphics{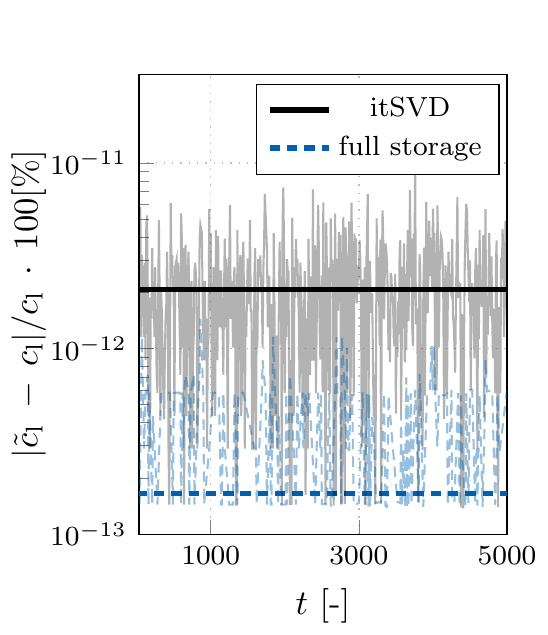}
}
\caption{Single phase cylinder flow with $q/\mathrm{T} = 1 =\tilde{q}/\mathrm{T}$ 
($\mathrm{Re}_\mathrm{D} = 200$, $\mathrm{Fn}=0.2828$, $\mathrm{St} = 0.2, \mathrm{T}=5000$). Left: Time evolution of the lift coefficient $c_\mathrm{l}$. Center: Enlargement of the time step interval $3000 \leq t \leq 4000$ supplemented with the lift coefficient $\tilde{c}_\mathrm{l}$  obtained from the reconstructed field. Right:  Relative error between both quantities
indicated in black, supplemented by the error returned by the full storage approach in blue and respective averages (horizontal lines). All simulations performed in parallel using P=96 partitions. }
\label{fig:memory_data_cl}
\end{figure}

The history of the computed drag ($c_\mathrm{d}$)  and lift ($c_\mathrm{l}$) force coefficients 
is displayed on the left side of Figs. \ref{fig:memory_data_cd} and \ref{fig:memory_data_cl}, respectively.
After approximately 4000 time steps $t$, a periodic flow is established.
The center graphs of Figs. \ref{fig:memory_data_cd} and \ref{fig:memory_data_cl} zoom into the time interval $3000 \leq t \leq 4000$ and compare force data obtained from the reconstructed (symbols, $\tilde c$) and the  original (line, $c$) flow.  Both
 force 
 coefficients $\tilde{c}_\mathrm{d}, \tilde{c}_\mathrm{l}$ are obtained from the reconstructed flow field  using a bunch-size of $b/q = 1$ 
 and $\tilde{q}/\mathrm{T} = 1$. The reconstructed forces  agree with both the original data and their full storage companion. The agreement is supported by the right graphs of Figs. \ref{fig:memory_data_cd} and \ref{fig:memory_data_cl} that outline 
 the error histories and their respective time mean value indicated by horizontal lines.
Errors for the full storage approach are given in blue. 
All errors are, without exception, fairly small. They fall below $10^{-9}$ for the drag coefficient and below $10^{-11}$ for the lift coefficient. 

%

 The accuracy displayed in Figs. \ref{fig:memory_data_cd} and \ref{fig:memory_data_cl} does not change with the  bunch size in case of $q/\mathrm{T} = 1 = \tilde{q}/\mathrm{T}$, i.e. $b/q = [1/5, 1/10, 1/25, 1/50]$ yield virtually identical result.
 While the bunch size $b$ does not influence the quality of the results and the storage effort, 
 it significantly affects the itSVD construction time $\tilde{t}^\mathrm{\, svd}$. 
 Tab. \ref{tab:timing} provides the normalized 
  wall-clock effort ($\tilde{t}^\mathrm{\, svd}$/$\tilde{t}^\mathrm{\, full}$)
 for all considered cases.
 Generally, the itSVD construction time increases when reducing $b$, regardless of the construction rank. The latter is expected since small values of $b$ require  more itSVD updates, cf. Alg. \ref{alg:time_integration}. Combining the maximum rank $q/\mathrm{T} = 1$ with the minimum $b/q = 1/50$, the increase is particularly drastic ($\tilde{t}^{\, svd} / \tilde{t}^\mathrm{\, full} \approx 94$), that was also observed by other authors for similar applications, cf. \cite{vezyris2019incremental, margetis2021lossy, margetis2022reducing}. At the same time, the itSVD construction time naturally decreases with decreasing $q/\mathrm{T}$ and reduces down to about 
 132\%  of the full-storage effort for $q/\mathrm{T} = 1/10$. 
 %
 %
 For the considered cylinder flow, the itSVD only out-performs the full-storage approach 
 if the truncation rank is significantly reduced ($q/\mathrm{T} > 1/10$) and at the same time large bunch sizes are employed ($b/q \le 1 /10$).
\begin{table}[!ht]
\caption{Single phase cylinder flow
($\mathrm{Re}_\mathrm{D} = 200$, $\mathrm{Fn}=0.2828$, $\mathrm{St} = 0.2, \mathrm{T}=5000$). Measured normalized wall-clock time $\tilde{t}^\mathrm{\, svd}$ for different reconstruction ranks $q/\mathrm{T}=[1, 1/2, 1/5, 1/10, 1/50]$ and bunching sizes $b/q = [1, 1/5, 1/10, 1/25, 1/50]$ (reference data refers to full storage wall-clock time $\tilde{t}^\mathrm{\, full}$). Given numbers correspond to averages obtained from three runs per scenario for parallel computations using P=96 partitions.}
\begin{center}
\begin{tabular}{lccccc}
\toprule
$q/\mathrm{T}$ & $\frac{\tilde{t}^\mathrm{\, svd}}{\tilde{t}^\mathrm{\, full}} \left( \frac{b}{q} = 1 \right) $ & $\frac{\tilde{t}^\mathrm{\, svd}}{\tilde{t}^\mathrm{\, full}} \left( \frac{b}{q} = \frac{1}{5} \right) $  & $\frac{\tilde{t}^\mathrm{\, svd}}{\tilde{t}^\mathrm{\, full}} \left( \frac{b}{q} = \frac{1}{10} \right)$ & $\frac{\tilde{t}^\mathrm{\, svd}}{\tilde{t}^\mathrm{\, full}} \left( \frac{b}{q} = \frac{1}{25} \right)$ & $\frac{\tilde{t}^\mathrm{\, svd}}{\tilde{t}^\mathrm{\, full}}\left( \frac{b}{q} = \frac{1}{50} \right)$ \\ \midrule
1 & 10.941 & 13.796 & 22.369 & 49.3725 & 94.0163 \\
1/2 & 9.0980 & 11.487 & 17.400 & 38.0972 & 75.3642 \\
1/5 & 3.0029 & 4.1213 & 6.6093 & 14.5674 & 27.7240 \\
1/10 & 1.3224 & 1.9041 & 3.0843 & 6.70971 & 12.9726 \\
1/50 & 0.2047 & 0.2715 & 0.4372 & 0.93983 & 1.83124\\
\bottomrule
\end{tabular}
\end{center}
\label{tab:timing}
\end{table}

Table \ref{tab:memory} outlines the normalized disk space utilized by the 5 considered construction ranks which is independent of the employed bunch size $b$. As anticipated, the required memory approximately scales with the rank of the itSVD, e.g. for a tenfold reduction 
$q/\mathrm{T} = 1/10$ about 
7.5\% of the full storage memory is required though the results agree well with the full storage data for this rank, cf. bottom row of Figs. \ref{fig:parallelization_v1} and \ref{fig:parallelization_v1_difference}. 
\begin{table}[!ht]
\caption{Single phase cylinder flow 
($\mathrm{Re}_\mathrm{D} = 200$, $\mathrm{Fn}=0.2828$, $\mathrm{St} = 0.2, \mathrm{T}=5000$). Scaling of the normalized memory effort  $\tilde{d}^\mathrm{\, svd}$ with the reconstruction rank for $q/\mathrm{T}=[1, 1/2, 1/5, 1/10, 1/50]$ (reference data refers to disk usage of a full storage approach $\tilde{d}^\mathrm{\, full}$). 
}
\begin{center}
\begin{tabular}{lccccc}
\toprule
$q/\mathrm{T}$ & 1 & 1/2 & 1/5 & 1/10  & 1/50 \\ \midrule
$\tilde{d}^\mathrm{\, svd}/\tilde{d}^\mathrm{\, full}$ & 0.998 & 0.375 & 0.152 & 0.0751 & 0.0169 \\
\bottomrule
\end{tabular}
\end{center}
\label{tab:memory}
\end{table}

\begin{figure}[!ht]
\centering
\iftoggle{tikzExternal}{
\input{./tikz/2D/Cylinder/cylinder_memory_cd_cl_errors.tikz}
}{
\includegraphics{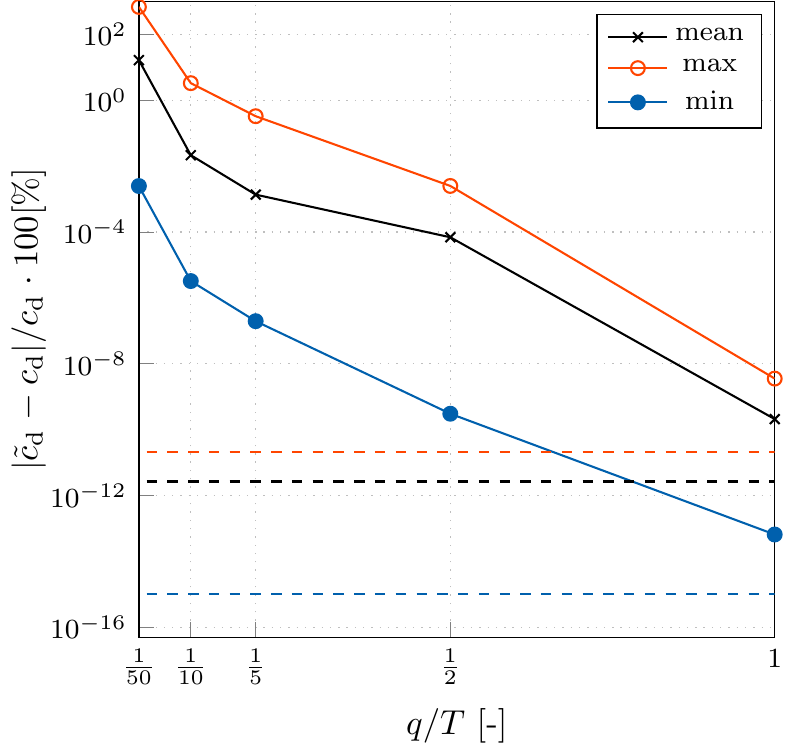}
\includegraphics{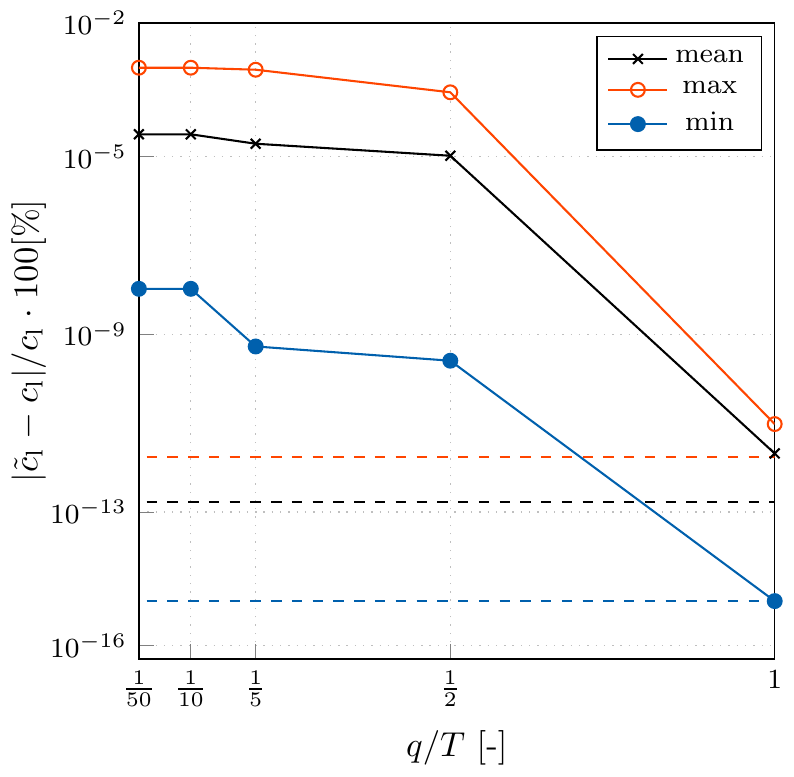}
}
\caption{Single phase cylinder flow for $q/\mathrm{T} = [0.02, 0.1, 0.2, 0.5, 1]$ 
($\mathrm{Re}_\mathrm{D} = 200$, $\mathrm{Fn}=0.2828$, $\mathrm{St} = 0.2, \mathrm{T}=5000$). Left [Right]: mean (black), maximum (orange), minimum (blue) relative error of the reconstructed drag [lift] coefficient $\tilde{c}_\mathrm{d}$ [$\tilde{c}_\mathrm{l}$] for itSVD reductions ($1 \leq t \leq \mathrm{T}$).}
\label{fig:error_over_accuracy}
\end{figure}

\begin{figure}[!ht]
\centering
\iftoggle{tikzExternal}{
\input{./tikz/2D/Cylinder/cylinder_energy_init_vortex_street.tikz}
}{
\includegraphics{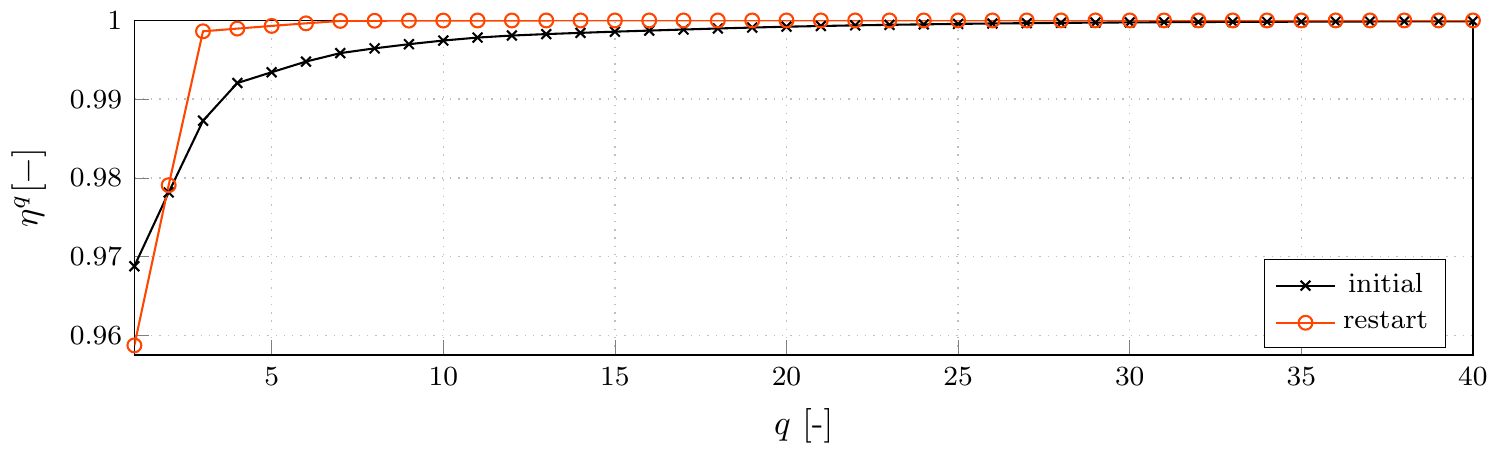}
\includegraphics{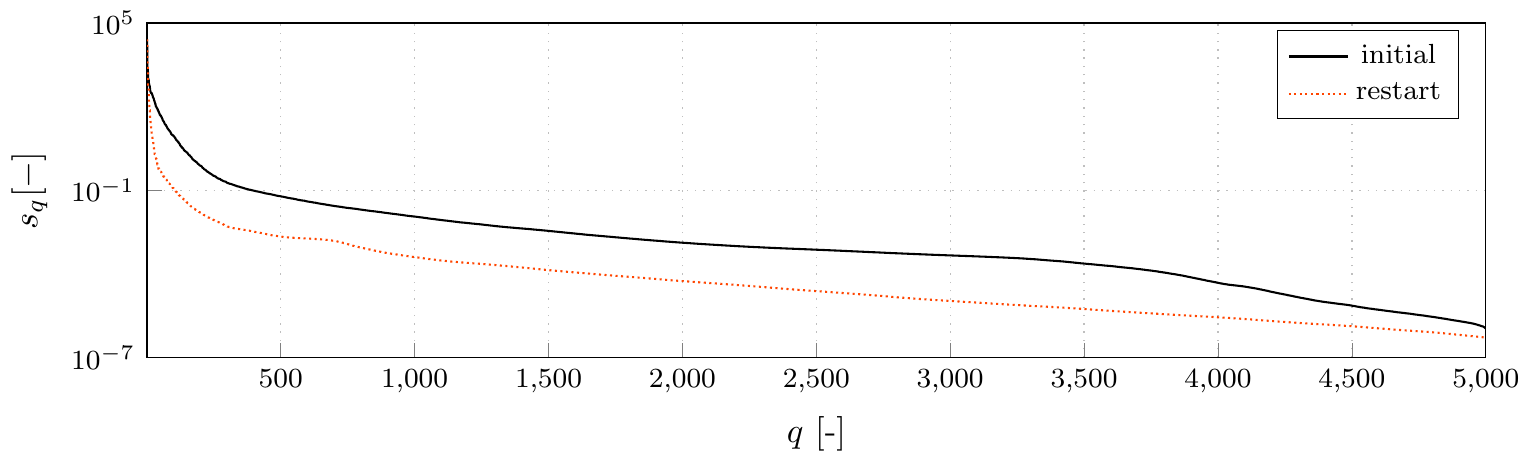}
}
\caption{Single phase cylinder flow with $q/\mathrm{T} = 1$
($\mathrm{Re}_\mathrm{D} = 200$, $\mathrm{Fn}=0.2828$, $\mathrm{St} = 0.2, \mathrm{T}=5000$). Top: All $q = 5000$ singular values in descending order. Bottom:  Resulting retained energy for $1 \leq t \leq \mathrm{T}$ (black) and $\mathrm{T} \leq t \leq 2\mathrm{T}$ (orange). 
}
\label{fig:cylinder_sv_energy}
\end{figure}

Finally, accuracy influences of the 5 investigated truncation
ranks $q$ are assessed.
Figure \ref{fig:error_over_accuracy} shows the respective averaged (black), maximum (orange), and minimum (blue) relative errors of the reconstructed drag (left) and lift (right) coefficients.
 The corresponding errors of the full-storage approach are represented by 
horizontal dashed lines. 
%
As expected, reducing the rank increases the error due to the reduced retained matrix energy. The latter is depicted by the black line  in Fig. \ref{fig:cylinder_sv_energy} (top) for the first 40 singular values and underpins that more than 99\% are retained when four or more singular values are considered. As shown in the Figure's bottom part, the first singular values are in the range of $\mathcal{O}(10^4)$ and quickly decrease by about 5 orders of magnitude above $q/\mathrm{T}=0.1$, cf. bottom of Fig. \ref{fig:cylinder_sv_energy}.
Mind that the data has been generated from scratch and inheres a substantial (detrimental) influence of the initial transient, as documented by the difference between the black and the orange curves in Fig. \ref{fig:cylinder_sv_energy}.

%

\begin{figure}[!ht]
\centering
\iftoggle{tikzExternal}{
\input{./tikz/2D/Cylinder/cylinder_memory_cd_cl_errors_vortext_street.tikz}
}{
\includegraphics{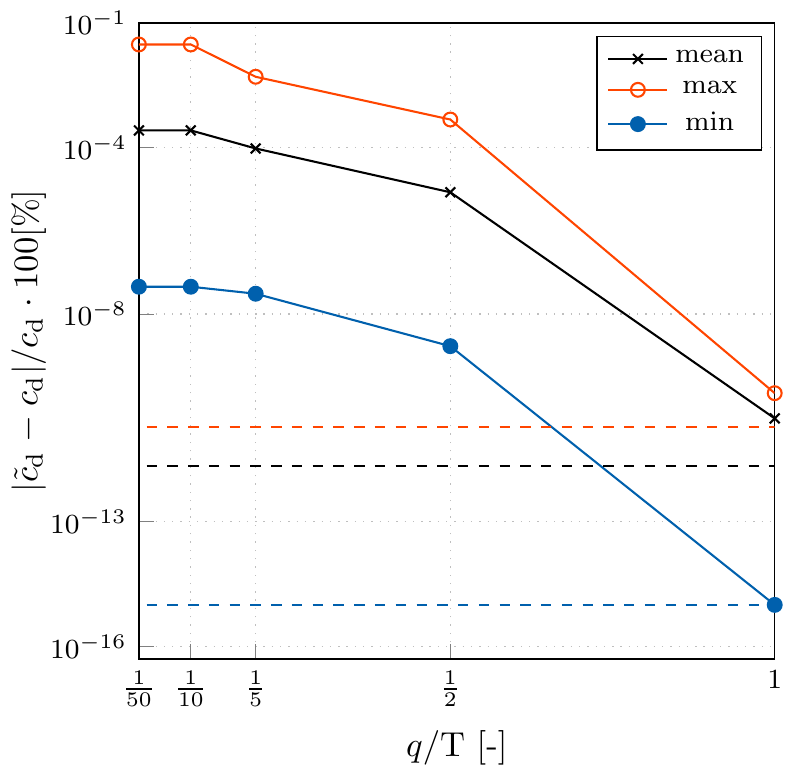}
\includegraphics{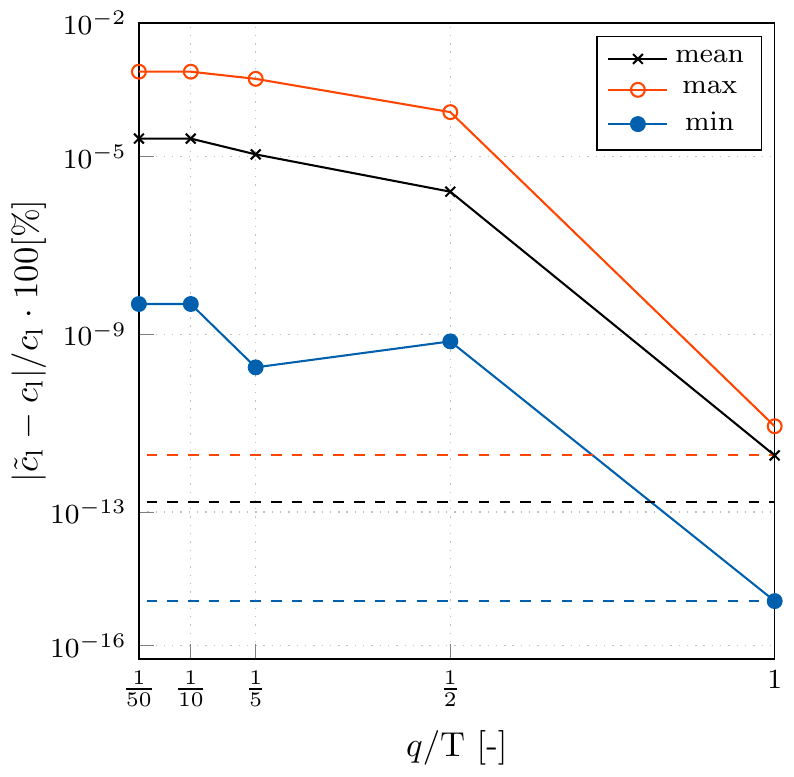}
}
\caption{Single phase cylinder flow for  $q/\mathrm{T} = [0.02, 0.1, 0.2, 0.5, 1]$ ($\mathrm{Re}_\mathrm{D} = 200$, $\mathrm{Fn}=0.2828$, $\mathrm{St} = 0.2, \mathrm{T}=5000$). Left [Right]:  Mean (black), maximum (orange), minimum (blue) relative error of the reconstructed drag [lift] coefficient $\tilde{c}_\mathrm{d}$ [$\tilde{c}_\mathrm{l}$] for itSVD reductions of the restart simulation from $\mathrm{T} \leq t \leq 2\mathrm{T}$.}
\label{fig:error_over_accuracy_vortext_street}
\end{figure}

The orange curves represent the analogue data compiled for time steps $5000 \leq t \leq 10 000$ using a restart mechanism. 
In this case, no initial transient occurs, and 
the retained energy converges after about seven modes, cf. Fig. \ref{fig:cylinder_sv_energy} (bottom). 
The singular values 
are one order of magnitude below the corresponding values obtained from reconstructing the simulation results with  initial transient effects.
The error analyses of Fig. \ref{fig:error_over_accuracy} is repeated for the fully periodic case as depicted in Fig. \ref{fig:error_over_accuracy_vortext_street}. 
Lift values are arguably much less prone to the homogeneous flow initialization, and the error behavior of the lift coefficient is very similar.
In contrast, the drag coefficient's error is of a similar magnitude 
only for large reconstruction values. 
%

\paragraph{To conclude in between} A consistent parallel implementation can be crucial, especially for low-rank approximations. The rank linearly influences the disk usage and nonlinearly influences the accuracy. However, accuracy requirements depend on the computed physics.
Accordingly, the retained energy provides a reasonable quality measure. For efficiency reasons, the bunch measure should be chosen sufficiently large, whereby the maximum allocatable memory has to be considered when designing the domain decomposition. Finally, introducing adaptive elements allows automation of the itSVD process. These aspects will be investigated in the following section.

\section{Adaptivity \& Realizability}
\label{sec:practical_issues}
\begin{figure}[!ht]
\centering
\iftoggle{tikzExternal}{
\input{./tikz/2D/Duncan/duncan_sketch.tikz}
}{
\includegraphics{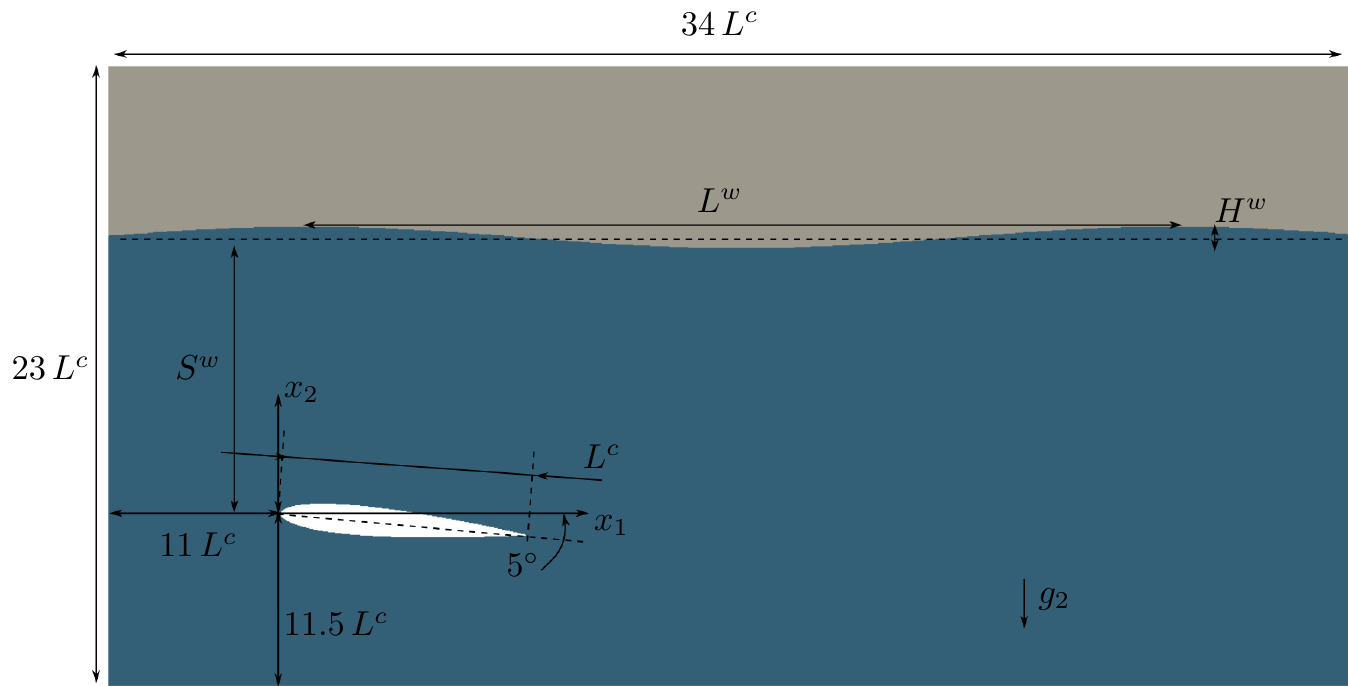}
}
\caption{Illustration of the initial configuration for the two-phase flow around a 2D hydrofoil  ($\mathrm{Re}_\mathrm{L} = \SI{750 000}{}$, $\mathrm{Fn}=0.62$, $L^w / L^c = 8$, $H^w / L^c = 0.1$, $T^w (V_1/L^c) = 4$, $\mathrm{T} = 10000$).}
\label{fig:duncan_sketch}
\end{figure}

 The section reports on adaptive implementations of the parallel itSVD 
which aim at determining an adequate rank and bunch size. Moreover, the realizability of the reconstructed fields is assessed. To this end, a time-dependent, 2D turbulent two-phase flow 
around a submerged NACA0012 hydrofoil
 at $5^\circ$ incidence is investigated. The case refers to experiments conducted by \cite{duncan1981experimental, duncan1983breaking} and is illustrated in Fig. \ref{fig:duncan_sketch}. In contrast to the experiments, the computations are not performed in (steady) calm water conditions but on periodic waves, and the Reynolds number is increased to comply with fully turbulent conditions. 
%
The chord $L^c$ to submergence $S^w$ ratio at the foil's leading edge reads $L^c / S^w = 7 / 9$. Simulations  are conducted for $\mathrm{Re} = \rho^\mathrm{b} \, V_1 \, L^c / \mu^\mathrm{b} = \SI{750 000}{}$ and $\mathrm{Fn} = V_1 / \sqrt{G \, S^w} = \SI{0.62}{}$, based on the gravitational acceleration $G = |g_2|$, the inflow 
velocity $V_1 = |v_1|$ as well as the density $\rho^\mathrm{b}$ and dynamic viscosity $\mu^\mathrm{b}$ of the water phase. The   length, height and period of the approaching wave  read $L^w / L^c = 4$, $H^w / L^c = 1 / 10$ and $T^w (V_1/L^c) = 4$, respectively. 
The submerged hydrofoil is expected to induce a wave field of length $\lambda^c / S^w = 2 \, \pi \, \mathrm{Fn}^2 = 2.416$ downstream of the foil.

\begin{figure}[!ht]
\centering
\subfigure[]{
\includegraphics[height=0.35\textwidth]{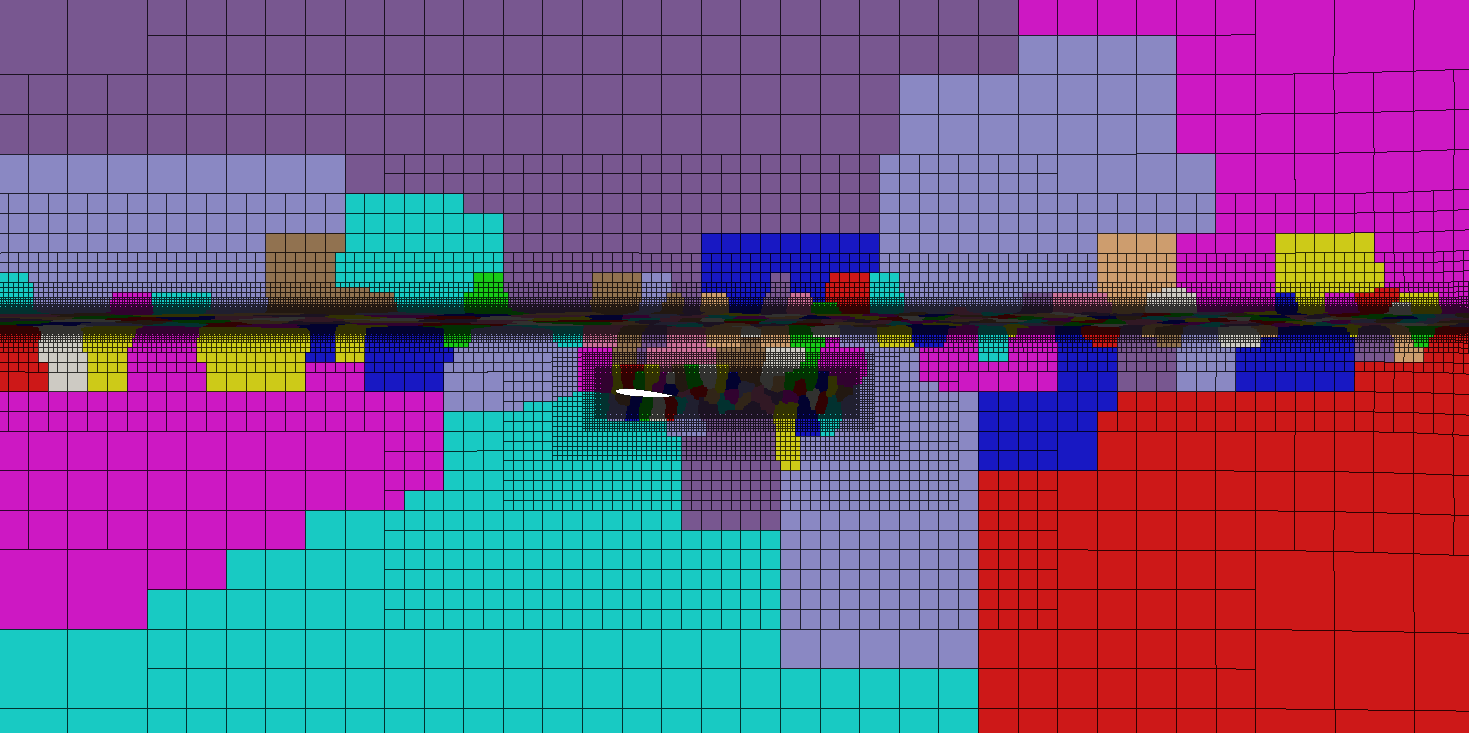}
}
\subfigure[]{
\includegraphics[height=0.35\textwidth]{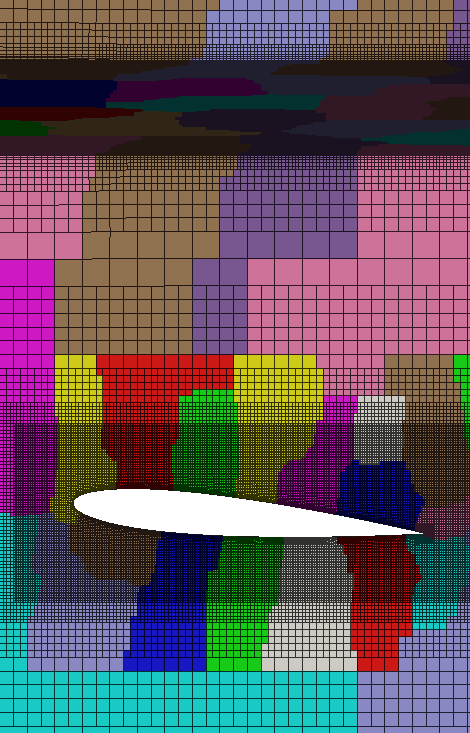}}
\caption{Unstructured grid employed for the two-phase flow around a 2D hydrofoil ($\mathrm{Re}_\mathrm{L} = \SI{750 000}{}$, $\mathrm{Fn}=0.62$, $L^w / L^c = 8$, $H^w / L^c = 0.1$, $T^w (V_1/L^c) = 4$, $\mathrm{T} = 10000$).  Decomposition into 288  partitions (left) and enlargement of the hydrofoil region (right).}
\label{fig:duncan_grid_partitioning}
\end{figure}

The computational grid consists of approximately \SI{530000}{} control volumes and is fractioned into $\mathrm{P} = 288$ partitions, cf. Fig. \ref{fig:duncan_grid_partitioning}. It extends $34 L^c$ in the horizontal ($x_1$)  direction and $23 L^c$ in the vertical direction ($x_2$).  The free surface is refined with cells of size $\Delta x_1 / \lambda^c = 1 / 400$ and $\Delta x_2 / H^w = 1 / 50$, and 
the refinement zone is located  between $- \SI{2.75}{} \, (H^w/2) \leq x_2 - S^w \leq 4 (H^w/2) $, where the origin is placed at the nose of the hydrofoil. The  horizontal  resolution coarsens towards the outlet to meet with the temporally constant hydrostatic outlet conditions. The nose of the hydrofoil is located  $11 L^c$ and $11.5 L^c$ away from the inlet and the bottom,  respectively. The fully turbulent simulations employ wall functions, where the wall normal thickness of the first grid layer reads $y^+ \approx \mathcal{O}(50)$.
At the inlet, 
a linear wave field is superimposed to the horizontal bulk flow together with 
a turbulence intensity of $I = 5$\% 
and a normalized eddy viscosity of $\mu^\mathrm{t} / \mu = 1$.
Slip walls are employed along the top/bottom boundaries, and a hydrostatic pressure boundary is specified along the outlet. 
%
Compared to the single-phase flow, a  smaller time step is utilized to comply with the stability conditions of the  compressive approximation schemes, cf. \cite{manzke2013sub, kuhl2021adjoint}, and the flow field advances in time 
with a time step $\Delta t \, V_1 / L^c = \SI{0.01}{}$ ($\Delta t /T^W= 0.0025$).

\subsection{Adaptive Estimation of the Truncation Rank}
\label{subsec:adaptivity}
The number of retained singular values is crucial to both, the accuracy and the computational cost of the itSVD. An adaptive rank is therefore desirable and will be assessed using  $\mathrm{T}=\SI{10 000}{}$ time steps in this subsection.

 The flow field consists of $\mathrm{S} = 6$ state variables.
 The phase concentration  is dimensionless and normalized. 
In line with Eqn. (\ref{equ:SVD_start}), the normalization of the other field quantities employs $\varphi^{v} = V_1$, $\varphi^{p} = \rho \, G \, L^\mathrm{c}$, $\varphi^{k} = u_\tau^2$ and $\varphi^{\varepsilon} = u_\tau^4$ for the velocity, pressure, turbulent kinetic energy and its dissipation, respectively. 
The friction velocity $u_\tau$ follows from empirical flat plate relations, i.e. $u_\tau = \sqrt{\tau_\mathrm{w} / \rho}$ with $\tau_\mathrm{w} = c_\mathrm{f} \rho V_1^2/2$ and $c_\mathrm{f} = 0.026 / \mathrm{Re}^{(1/7)}$.
%
The discretization effort increases compared to the previous laminar single-phase example. Accordingly, the global matrix length reads $\mathrm{N} = \mathrm{Y} \, \mathrm{P} = \mathrm{(S \, L) \, P}  \approx \SI{3 180 000}{}$
 and the global state matrix size reads $\matr{Y} \in \mathbb{R}^{\SI{3 180 000}{} \times \SI{10 000}{}}$.

\begin{figure}[!ht]
\centering
\iftoggle{tikzExternal}{
\input{./tikz/2D/Duncan/duncan_adaptivity_fixxed_rank.tikz}
}{
\includegraphics{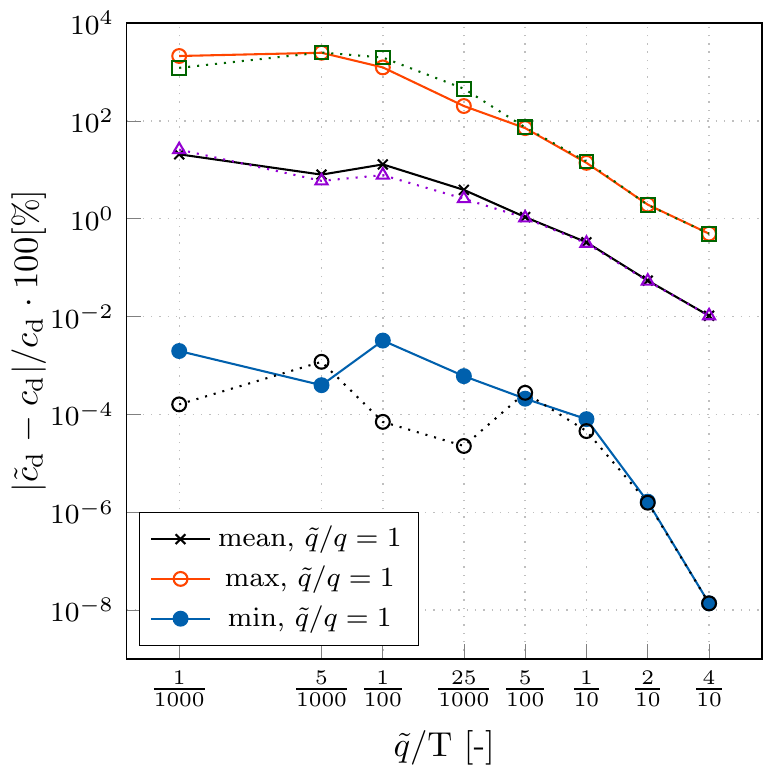}
\includegraphics{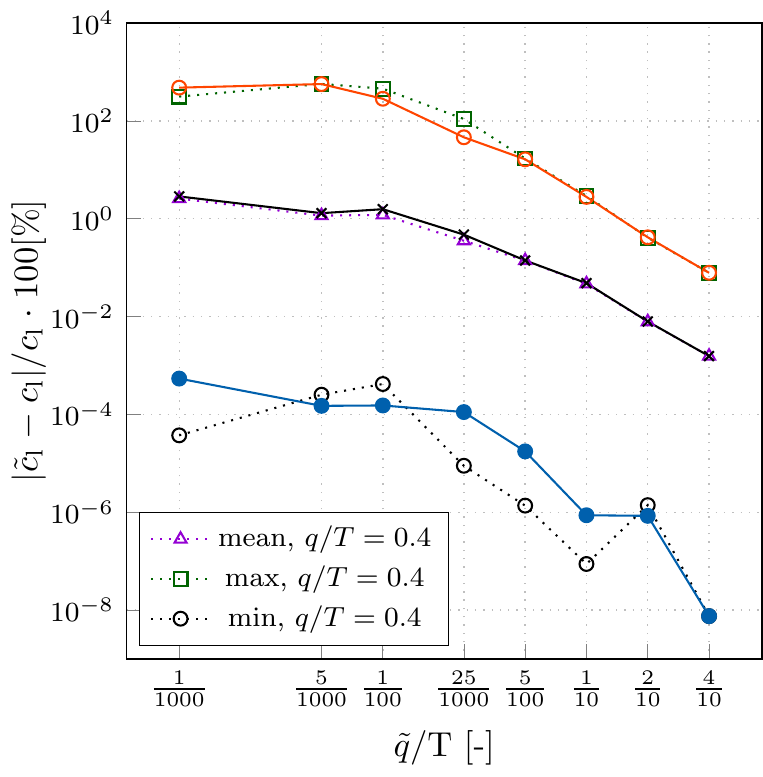}
}
\caption{
Mean (black), 
 maximum (orange), minimum (blue) 
relative error of the reconstructed drag (left) and lift (right) for the two-phase flow around a 2D hydrofoil
with $\tilde{q}/\mathrm{T} = [0.001, 0.005, 0.01, 0.025, 0.05, 0.1, 0.2, 0.4]$ ($\mathrm{Re}_\mathrm{L} = \SI{750 000}{}$, $\mathrm{Fn}=0.62$, $L^w / L^c = 8$, $H^w / L^c = 0.1$, $T^w (V_1/L^c) = 4$).
Solid lines represent $\tilde{q}/q = q/\mathrm{T}$ investigations. Dashed lines mark results for the $\tilde{q}/\mathrm{T}  \leq q/\mathrm{T}$, $q/\mathrm{T} = 0.4$ studies.}
\label{fig:adaptive_study_fixxed_rank}
\end{figure}

Firstly, 8 non-adaptive itSVD constructions featuring different  ranks $q/\mathrm{T} = [0.001, 0.005, 0.01,$ $ 0.025, 0.05, 0.1,$ $ 0.2, 0.4]$ 
are performed and serve as a benchmark.
Due to the increased complexity 
and the previous findings, 
only 0.1\% -- 40\% of all possible singular values are determined.
%
Assigning the evaluation rank  to the corresponding construction rank, i.e.  $\tilde{q} / q = 1$, the resulting mean (black), maximum (orange), and minimum (blue) errors of the drag [lift] coefficients 
predicted from the reconstruction are shown as marked solid lines 
in Fig. \ref{fig:adaptive_study_fixxed_rank} left [right]. Similar to the cylinder studies of Sec. \ref{sec:verification_validation}, the mean errors lay inside a corridor spanned by their minimum and maximum counterparts
and decrease by about 3 orders of magnitude when increasing the rank. 
%
Secondly, to further investigate the interaction of construction ($q$) and evaluation ($\tilde q$) ranks, the latter is varied while the former is assigned to its maximum, i.e., $\tilde{q}/\mathrm{T} \leq q/\mathrm{T}$ with $q/\mathrm{T} = 0.4$. The resulting mean (purple), maximum (green), and minimum (black) errors are added to the previous results as dotted lines. The errors deviate slightly from those of the $\tilde{q}/q=1$ studies, 
and the agreement is best for the averaged results. 
%

\begin{figure}[!ht]
\centering
\iftoggle{tikzExternal}{
\input{./tikz/2D/Duncan/duncan_adaptivity_energy.tikz}
}{
\includegraphics{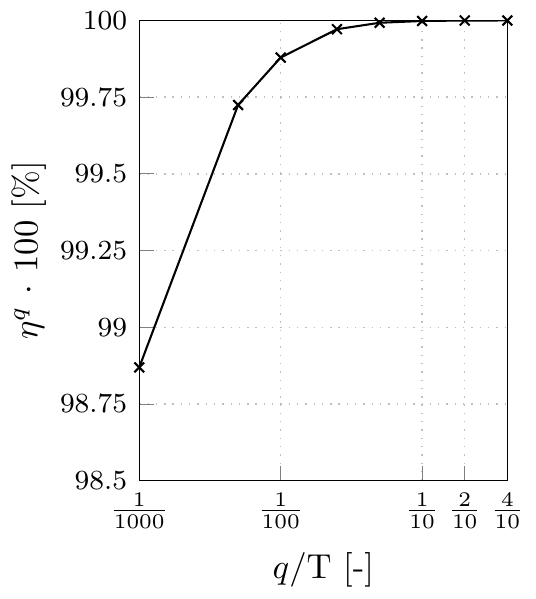}
\includegraphics{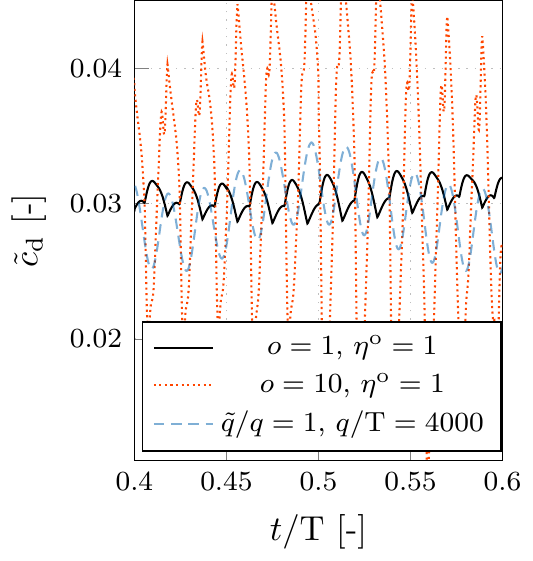}
\includegraphics{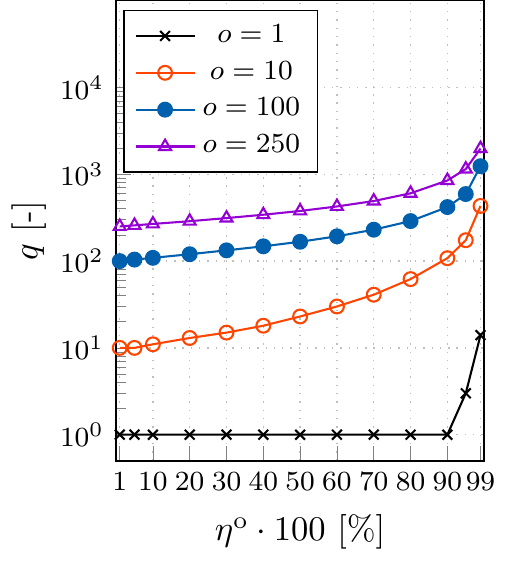}
}
\caption{Two phase hydrofoil flow ($\mathrm{Re}_\mathrm{L} = \SI{750 000}{}$, $\mathrm{Fn}=0.62$, $L^w / L^c = 8$, $H^w / L^c = 0.1$, $T^w (V_1/L^c) = 4$). Left: retained energy $\eta^q$ over construction rank $q/\mathrm{T}$. Center: Drag coefficient obtained from different reconstructions. Right: Estimated required construction rank over the retained matrix energy $\eta^\mathrm{o}$ for  4 minimal ranks.}
\label{fig:adaptive_study_energy}
\end{figure}

\begin{figure}[!ht]
\centering
\iftoggle{tikzExternal}{
\input{./tikz/2D/Duncan/duncan_adaptivity_adaptive_rank.tikz}
}{
\includegraphics{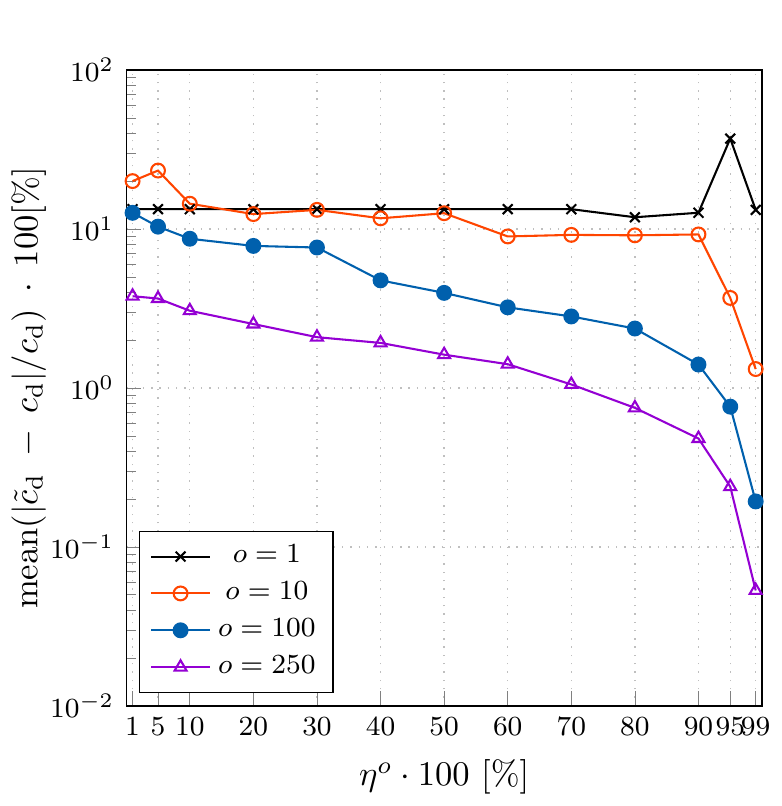}
\includegraphics{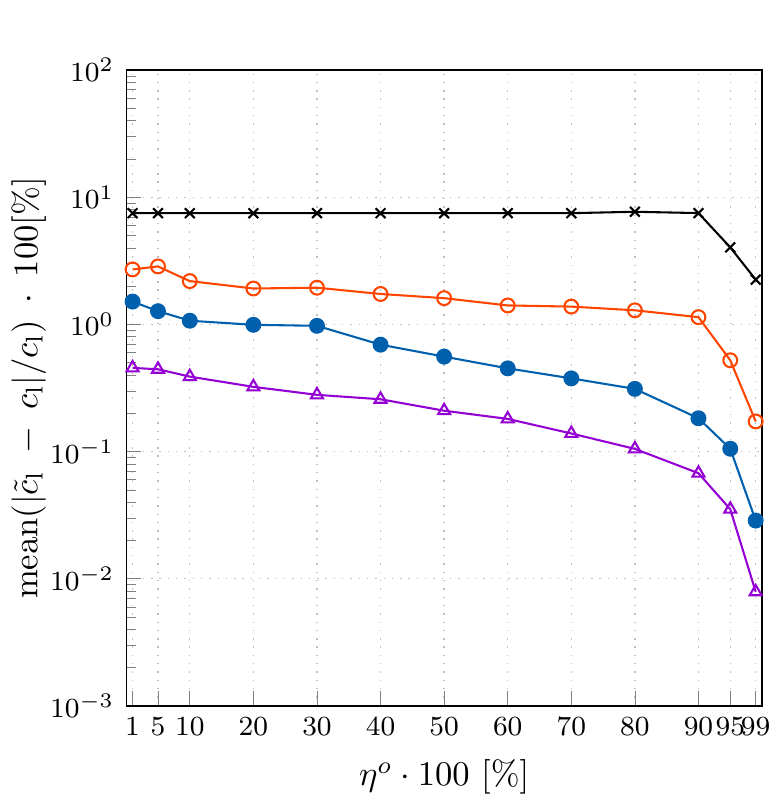}
}
\caption{Mean relative error of the reconstructed drag (left) and lift (right) for the two-phase flow around a 2D hydrofoil ($\mathrm{Re}_\mathrm{L} = \SI{750 000}{}$, $\mathrm{Fn}=0.62$, $L^w / L^c = 8$, $H^w / L^c = 0.1$, $T^w (V_1/L^c) = 4$) 
using an adaptive itSVD with different minimal construction ranks $\eta^o \cdot 100 = [1, 5, 10, 20, 30, 40, 50, 60, 70, 80, 90, 95, 99]$ for $o=1$ (black), $o=10$ (orange), $o=100$ (blue), $o=250$ (purple).}
\label{fig:adaptive_study_adaptive_rank}
\end{figure}

Figure \ref{fig:adaptive_study_energy} (left) shows the variation of the retained energy with the construction rank $q/\mathrm{T}$ as described by Eqn. (\ref{equ:retained_energy}). The figure  indicates the dominance of the first ten modes that  contain about 99\% of the total energy. Considering further modes --and thus also increasing $\tilde{q}/\mathrm{T}$ for fixed $q/\mathrm{T} = 0.4$-- is deemed to have a minor influence on the overall result. 
%
In addition to prescribing a fixed construction rank, four studies are conducted with adaptive ranks and $\tilde q/q=1$.
They differ in the minimum construction rank $o/\mathrm{T} \cdot 100 = [0.01, 0.1, 1, 2.5]$ and are each used in combination with 
a range of 13 different prescribed retained matrix energy levels $\eta^\mathrm{o} \cdot 100 = [1, 5, 10, 20, 30, 40, 50, 60, 70, 80, 90, 95, 99]$, cf. Eqn. (\ref{equ:adaptive_energy}) and Alg. \ref{alg:SVD_adaptive}. Results of these
studies are shown in Fig. \ref{fig:adaptive_study_adaptive_rank} for $o=1$ (black), $o=10$ (orange), $o=100$ (blue), and $o=250$ (purple). Displayed data refers to the mean lift and drag coefficient errors.
Increasing the minimum construction rank generally reduces the  error and yields qualitatively similar behavior for  drag and lift. Increasing the retained energy also reduces the error in almost all scenarios except for the case $o=1$, cf. Eqns. (\ref{equ:retained_energy}, \ref{equ:adaptive_energy}). For this classical starting point of an adaptive approach, the error remains constant over almost the entire energy spectrum, 
which is attributed to the already mentioned extreme energetic contribution of the first mode(s), cf. Fig. \ref{fig:adaptive_study_energy} (left). Accordingly, this case only enters an adaptive process for very demanding requirements on the retained energy, which is confirmed in the right graph of Fig. \ref{fig:adaptive_study_energy}. Therein, 
the resulting construction ranks are measured for each specified retained energy amount. Less strict requirements 
result in ranks that are close to the prescribed minimum  ($o=1$ black, $o=10$ orange, $o=100$ blue, and $o=250$ purple),
which only increases for $\eta^{o = 1} \geq 95\%$ from $q=1$ to $q=4$.
%
Similar observations arise for the drag coefficient in Fig. \ref{fig:adaptive_study_adaptive_rank} (left), where the errors at small retained energy levels $\eta^o \cdot 100 \leq 10\%$  are partly lower for $o=1$ than for $o=10$.
This is attributed to the chosen error evaluation, which is confined to the global mean deviation. For illustration purposes, Fig. \ref{fig:adaptive_study_energy} (center) compares a high-quality drag reconstruction (blue) with
two reconstructions of moderate effort ($\eta^o \cdot 100 = 1\%$) using  $o = 1$ (black) and $o = 10$ (orange).
For $o = 1$, the almost constant mode results in significantly smaller deviations. For $o = 10$, at least nine further modes are considered, that trigger higher deviations and mean errors. For the flows considered herein, such phenomena are restricted to the first $\mathcal{O}(10^1)$ modes, and a significant increase in the construction rank leads to a continuous reduction of averaged approximation errors.

It is concluded that the adaptive itSVD procedure significantly improves from the specification of a minimum construction rank.
However, the identification of the minimum rank $o$ 
  depends on an educated guess, e.g. based on the heuristics presented in Sec. \ref{sec:starting_point}. 

\subsection{Estimation of the Bunch Size}
The bunch size $b$ should be chosen as small as necessary and as large as possible. Ideally it will approach the maximum number of time steps in a one-shot approach, i.e., $b \to \mathrm{T}$. An estimation of a reasonable bunch size $b$ follows from assessing the --hardware specific-- available memory.
Approximations of the required  memory are based on the dominant row size $\mathrm{N}$, cf. Eqn. (\ref{equ:SVD_start}). 
Hence, an itSVD update with incrementally increased bunch size $b$ can be emulated at the beginning of the itSVD construction until the maximum allocatable memory limit is reached. Note, that in addition to the bunch matrix $\matr{B}$, further global matrices, some of identical length (e.g. $\matr{P}$, $\matr{Q}$ in Alg. \ref{alg:SVD_construction}), must be allocated during the update process, which the bunch size estimator should take into account. For instance, the 2D flow of this section with $\mathrm{N}\approx \SI{3 180 000}{}$ and $b=\SI{500}{}$ results in $(\mathrm{N} \times \mathrm{b} \times 8) / 10^9 \approx 12.72$ gigabytes when stored (possibly distributed) in a 8Byte double-precision context.
Finally, a safety factor is imposed on the estimated maximum admissible $b_{\mathrm{max}}$ value  
to guarantee a stable simulation in the present SIMD concept, which is set to to $\approx75\%$ (i.e., $b=\mathcal{O}(0.75\, b_{\mathrm{max}})$) throughout this paper. 

\subsection{Realizability of Reconstructed Fields}
\begin{figure}[!ht]
\centering
\iftoggle{tikzExternal}{
\input{./tikz/2D/Duncan/duncan_realizability.tikz}
}{
\includegraphics{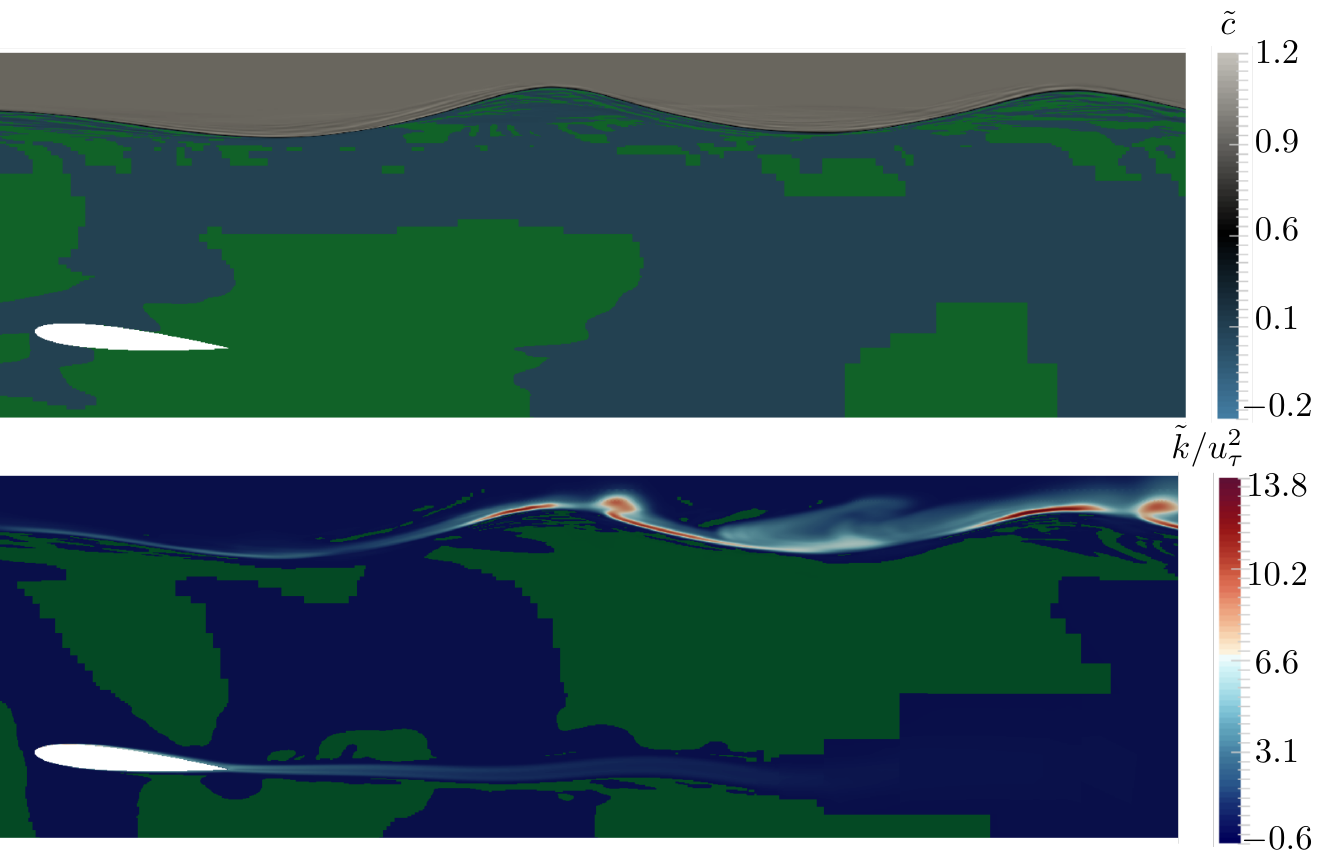}
}
\caption{Reconstructed two-phase flow around a 2D hydrofoil at time instant $t/\mathrm{T} = 0.5$ ($\mathrm{Re}_\mathrm{L} = \SI{750 000}{}$, $\mathrm{Fn}=0.62$, $L^w / L^c = 8$, $H^w / L^c = 0.1$, $T^w V_1/L^c = 4$). Concentration field (top) and normalized turbulent kinetic energy (bottom), whereby green areas indicate regions of nonphysical negative values.}
\label{fig:duncan_realizability}
\end{figure}

Low-rank approximations 
construct a linear combination out of the snapshot dependent reduced basis. They inherit properties of the underlying snapshots, e.g., a solenoidal velocity field is usually well maintained for snapshots that agree with $\partial v_k / \partial x_k = 0$.
However, 
physical constraints on threshold values,
such as bounded volume concentration $c \in [0,1]$ or inherent positive turbulent kinetic energy $k \geq 0$, are not necessarily respected.
Figure \ref{fig:duncan_realizability} shows the wave pattern based on the reconstructed concentration field $\tilde{c}$ (top) and the corresponding normalized turbulent kinetic energy $\tilde{k}/u_\tau^2$ (TKE, bottom) at time instant $t/\mathrm{T} = 0.5$ 
using the adaptive itSVD case with $o=10$ and $\eta^o=90\%$ ($q = 108 = \tilde{q}$).
Regions of nonphysical negative values are highlighted in transparent green. 

Slightly negative TKE values occur in "laminar" regions, but 
no contamination of spurious void regimes is observed in the high shear rate zones close to the airfoil, its wake and the free surface.
Similarly, unrealizable concentration values also appear to be favorably "organized", i.e., too low values are embedded in the water phase and too high values occur inside the air phase. The latter might be much more delicate in violent flows, in particular since exceeding the realizable concentration interval induces utterly wrong fluid properties, e.g., negative fluid densities.
%

Requirements are incorporated 
to restore the realizability of the reconstructed fields. 
Negative energy or concentration values are typically deliberately suppressed or clipped, viz.
\begin{align}
    \tilde{c} &= \mathrm{max}(\alpha, \mathrm{min}(\tilde{c}, 1)),
    \qquad \qquad \mathrm{and} \qquad \qquad
    \tilde{k} = \mathrm{max}(\alpha, \tilde{k}),
    \quad
    \tilde{\varepsilon} = \mathrm{max}(\alpha, \tilde{\varepsilon})\, ,
\end{align}
where $\alpha = 10^{-16}$ refers to the smallest representable number to avoid numerical conflicts. 
%
Note that only minor differences of the integral drag and lift coefficients were observed for the realizable and non-realizable field reconstructions.

\section{Application}
\label{sec:application}
Three issues of practical relevance are studied in this application case: First, again, the influence of the model reduction on global, integral data like the vessel's total resistance and a local quantity, here the wave elevation. Furthermore, the computational overhead due to the itSVD construction during the simulation is analyzed to identify potential bottlenecks within the presented algorithms.

The 3D application refers to the fully turbulent flow around an unappended Kriso container ship (KCS) hull in harmonic head waves. The investigated 1:31.6 scale model 
 offers a large amount of numerical and experimental data obtained in calm water conditions, e.g. \cite{banks2010free, larsson2013numerical, kroger2018adjoint},  and head waves \cite{carrica2011computations, simonsen2013efd, shen2015dynamic}. The distance between the aft and front perpendiculars of the hull model serves as a reference length $L = L^\mathrm{pp}$. Additional reference properties refer to the gravitational acceleration $G = |g_2|$,  
 , the inflow velocity magnitude $V_1 = |v_1|$, and the kinematic viscosity of the water $\nu^\mathrm{b} = \mu^\mathrm{b} / \rho^\mathrm{b}$. Computations were performed at  Reynolds- and Froude-numbers of $\mathrm{Re} = V_1 L / \nu^\mathrm{b} = 1.4 \cdot 10^7$ and $Fn = V_1 / \sqrt{G L} = 0.26$. The ship's motion and propulsion are suppressed during the simulation, and the initial draught refers to $d^\mathrm{h}/L^\mathrm{pp} = 0.04696$.
The length, height and period of the head waves refer to 
$L^\mathrm{w} / L^\mathrm{pp} = 1.3$, $H^\mathrm{w} / L^\mathrm{pp} = 0.075$, and $T^w (V_1/L^\mathrm{pp}) = 1.3$. In addition to the approaching waves, we expect a wave field of length $\lambda^\mathrm{w} / L^\mathrm{pp} = 2 \, \pi \, \mathrm{Fn}^2 = 0.4247$ to be induced by the hull.

\begin{figure}[!ht]
\centering
\subfigure[]{
\includegraphics[width=0.485\textwidth]{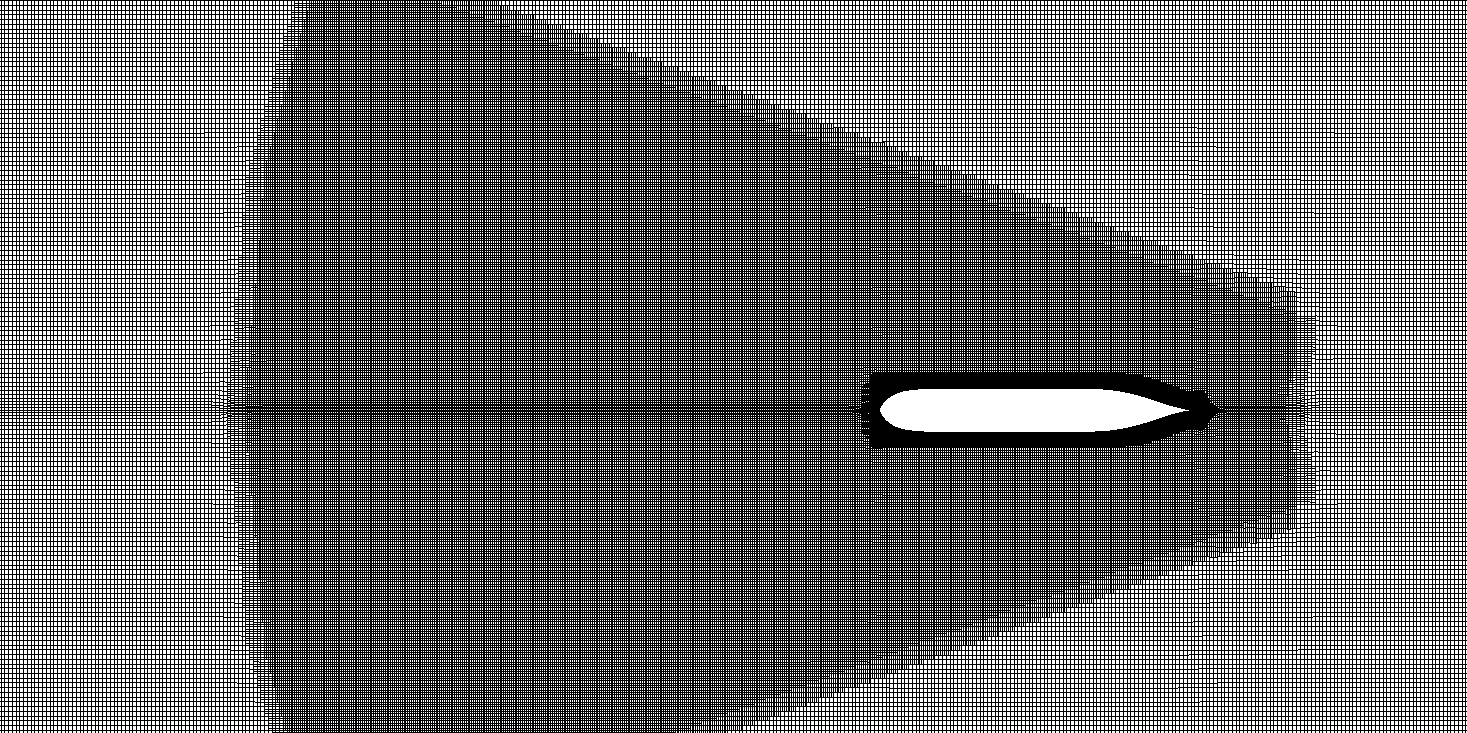}
}
\subfigure[]{
\includegraphics[width=0.485\textwidth]{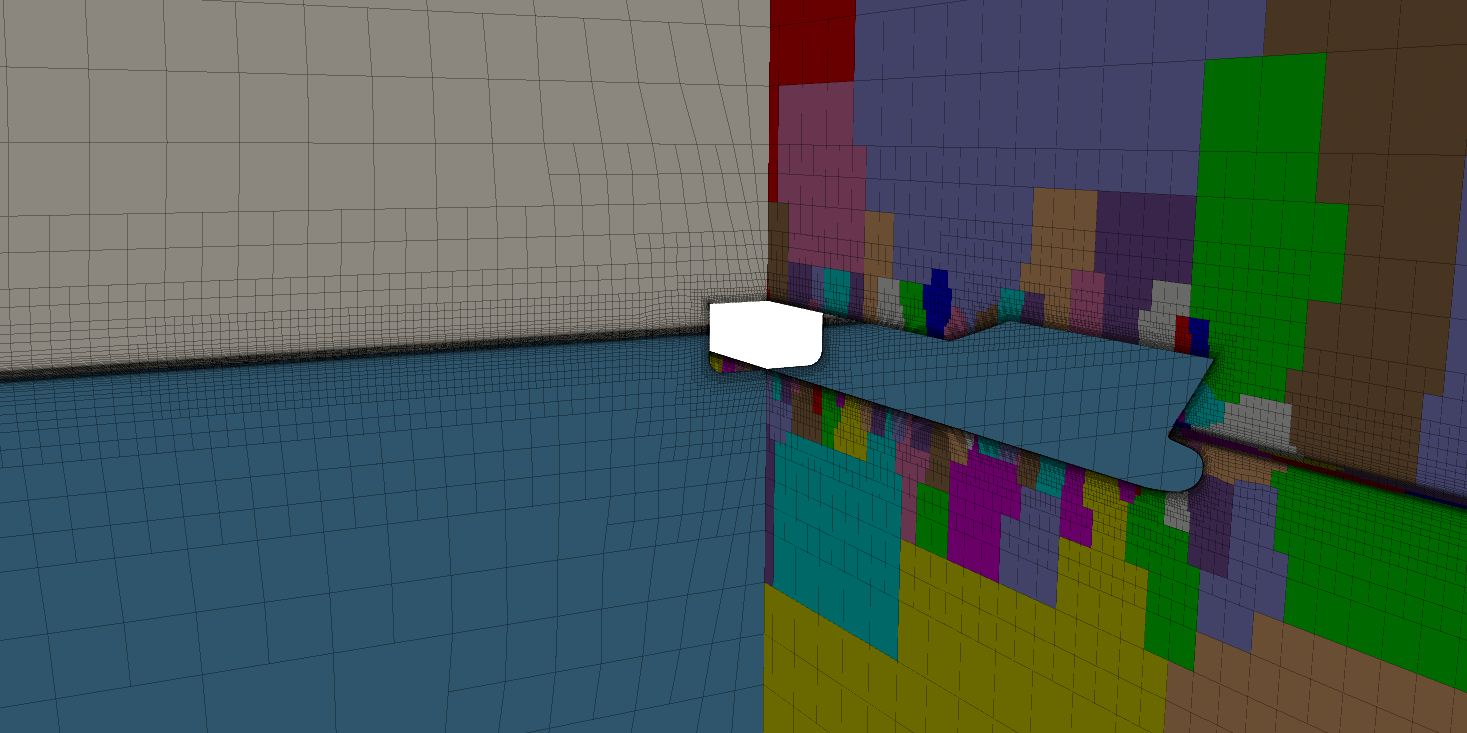}}
\caption{Employed unstructured grid for the Kriso container ship simulation ($\mathrm{Re}_\mathrm{L} = \SI{14 000 000}{}$, $\mathrm{Fn}=0.26$, $d^\mathrm{h}/L^\mathrm{pp} = 0.04696$, $L^\mathrm{w} / L^\mathrm{pp} = 1.3$, $H^\mathrm{w} / L^\mathrm{pp} = 0.075$, $T^w (V_1/L^\mathrm{pp}) = 1.3$). Cutouts of  (a)  the calm water plane 
and (b) 3D view with  indicated $P=2880$ partitioning.}
\label{fig:kcs_grid_partitioning}
\end{figure}

The employed numerical grid consists of approximately 30 million unstructured computational cells and is depicted in Fig. \ref{fig:kcs_grid_partitioning}. The domains extend over $8\, L^\mathrm{pp}$, $6\, L^\mathrm{pp}$, and $5\, L^\mathrm{pp}$ in horizontal ($x_1$), lateral ($x_2$), and vertical ($x_3$) direction. The complete flow is resolved, i.e., no symmetry conditions are imposed, and the inlet (port) [lower] boundary is located at $x_1 / L^\mathrm{pp} = 4$ ($x_2 / L^\mathrm{pp} = 3$) [$x_3 / L^\mathrm{pp} = 3$]. The hull  surface is discretized with approximately \SI{595000}{} surface elements. The wall-normal resolution refers to a dimensionless wall distance of $y^+ = \mathcal{O}(50)$ and justifies the use of wall functions. The vertical resolution of the free surface region is constant throughout the domain to resolve the prescribed wave amplitude of $2 \, H^\mathrm{w}$ by twenty cells, i.e., $\Delta x_2 / H^\mathrm{w} = 1 / 20$. The overall tangential resolution of the free surface is resolved by  $\Delta x_1 / L^\mathrm{w} = 1 / 40$. It is refined within a Kelvin-Wedge to capture the hull induced wave pattern with roughly 50 cells, i.e., $\Delta x_1 / \lambda^\mathrm{w} = 1 / 50$. 
At the inlet, a horizontal bulk velocity and linear wave theory are superimposed together with the corresponding concentration field. In line with the hydrofoil study in Sec. \ref{sec:practical_issues}, turbulent inflow quantities follow from a turbulence intensity of $I = 5$\% 
and $\nu_t/\nu=1$. A hydrostatic pressure boundary is specified along the outlet, and slip walls are employed along the lateral and vertical boundaries. 
%
%
The flow is integrated in time with a time step size of $\Delta t \, V_1 / \lambda^\mathrm{w} = \SI{0.0015}{}$.

The normalization of the $\mathrm{S} = 7$ state variables  employs $\varphi^{v} = V_1$, $\varphi^{p} = \rho \, G \, L^\mathrm{pp}$, $\varphi^{k} = u_\tau^2$, and $\varphi^{\varepsilon} = u_\tau^4$ for the velocities, pressure, turbulent kinetic energy, and its dissipation, respectively. The friction velocity $u_\tau$ again follows from a turbulent flat plate formulae, i.e., $u_\tau = \sqrt{\tau_\mathrm{w} / \rho}$ with $\tau_\mathrm{w} = c_\mathrm{f} \rho V_1^2/2$ and $c_\mathrm{f} = 0.026 / \mathrm{Re}^{(1/7)}$.
%
The  discretization effort drastically increases compared to the previous 2D studies. The length of system matrix reads $\mathrm{N} = \mathrm{Y} \, \mathrm{P} = \mathrm{(S \, L) \, P}  \approx \SI{210 000 000}{}$ which is  distributed on $P = 2880$ partitions.   
Six wave periods, corresponding to $\mathrm{T} = 10000$ time steps, are compressed by the itSVD strategy, and 
the global state matrix reads $\matr{Y} \in \mathbb{R}^{\SI{210 000 000}{} \times \SI{10 000}{}}$.

To avoid initial transient effects, the flow is simulated for $\mathrm{T}=50000$ time steps, and the itSVD is constructed during the final interval $0.8<t/\mathrm{T}<1.0$ by 10000 time steps. A minimum of $o=100$ singular values, i.e. 1\% of the final interval, is used to initiate the  rank  adaptation in combination with a  requirement to retain $\eta^o = 90\%$ of the initially missed matrix energy, as outlined by Eqn. (\ref{equ:adaptive_energy}).
 For the underlying hardware ($\approx \SI{320}{GB}$/96 CPU), the admissible bunch size is estimated to $b_{\textrm{max}} \leq 620$ and thus 
 yields $b = 500$.

\begin{figure}[!ht]
\centering
\iftoggle{tikzExternal}{
\input{./tikz/3D/KCS/kcs__adaptivity.tikz}
}{
\includegraphics{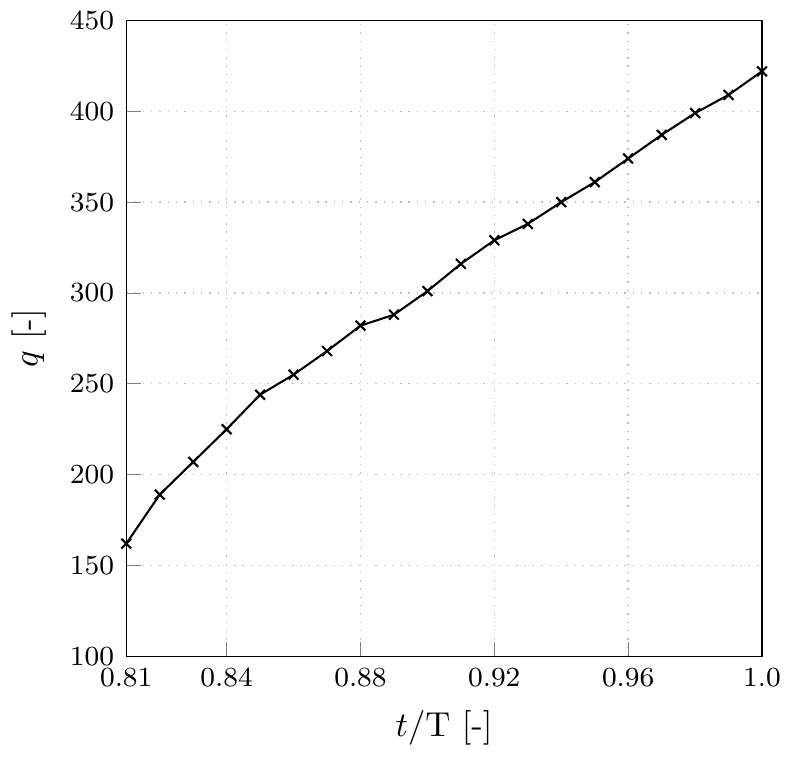}
\includegraphics{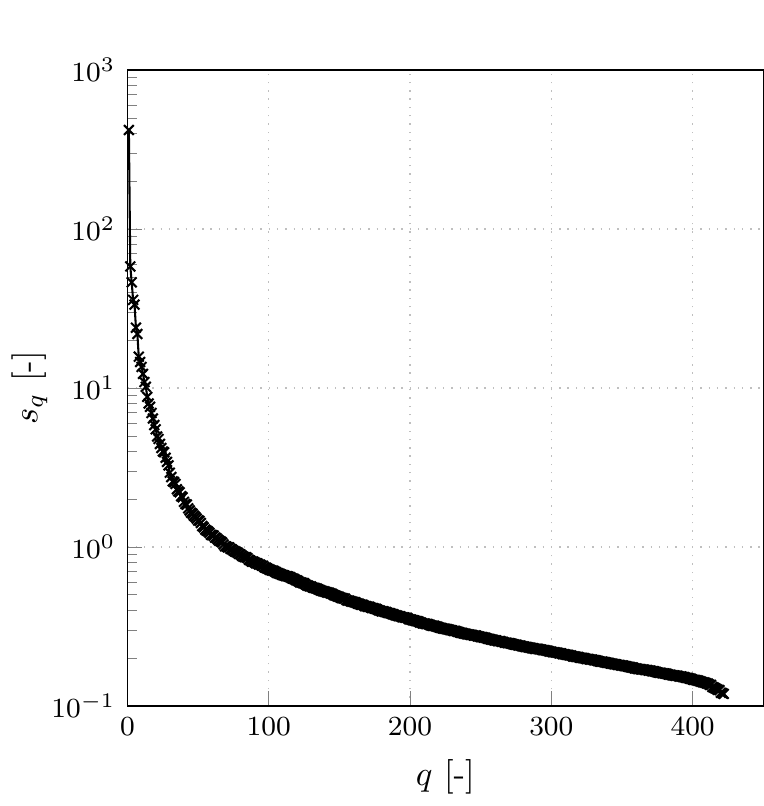}
}
\caption{Kriso container ship case ($\mathrm{Re}_\mathrm{L} = \SI{14 000 000}{}$, $\mathrm{Fn}=0.26$, $d^\mathrm{h}/L^\mathrm{pp} = 0.04696$, $L^\mathrm{w} / L^\mathrm{pp} = 1.3$, $H^\mathrm{w} / L^\mathrm{pp} = 0.075$, $T^w (V_1/L^\mathrm{pp}) = 1.3$).  Left: Evolution of adaptively estimated rank $q$ to retain $\eta^\mathrm{o} = 90\%$ of the matrix energy starting from a minimal rank $o=100$. Right: Distribution of all $q=422$ final singular values at $t/\mathrm{T} = 1$.}
\label{fig:kcs_adaptivity}
\end{figure}

The evolution of the number of considered singular values within the relevant time interval is shown in Fig. \ref{fig:kcs_adaptivity} (left). The itSVD updates become apparent by adjusting every $b=500$th time step and indicate a continuous increase of the considered singular values. Approximately 10-15 singular values are added per update, and the final rank reads $q=422$ at $t/\mathrm{T}=1$. The distribution of the considered singular values is shown in Fig. \ref{fig:kcs_adaptivity} (right), where, again, a dominance of the first 10-30 singular values arises, which 
exceed $\mathcal{O}(10^1)$ and are about two orders of magnitude above the smallest singular values considered.

\begin{figure}[!ht]
\centering
\iftoggle{tikzExternal}{
\input{./tikz/3D/KCS/kcs_compare_forces.tikz}
}{
\includegraphics{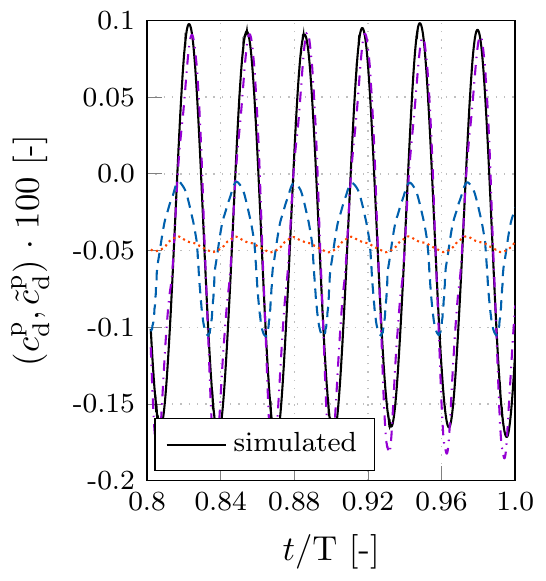}
\includegraphics{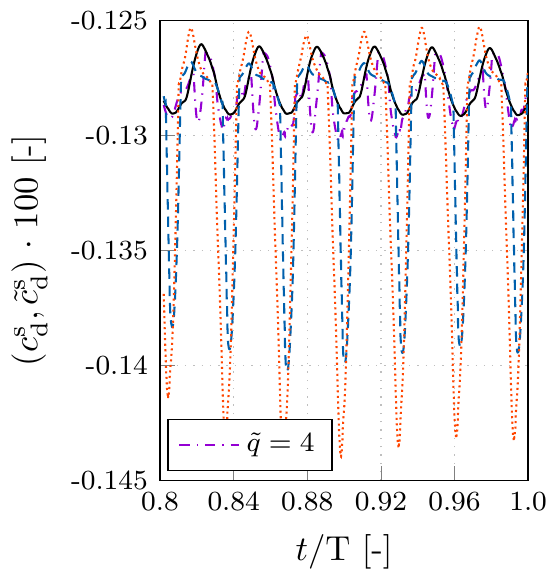}
\includegraphics{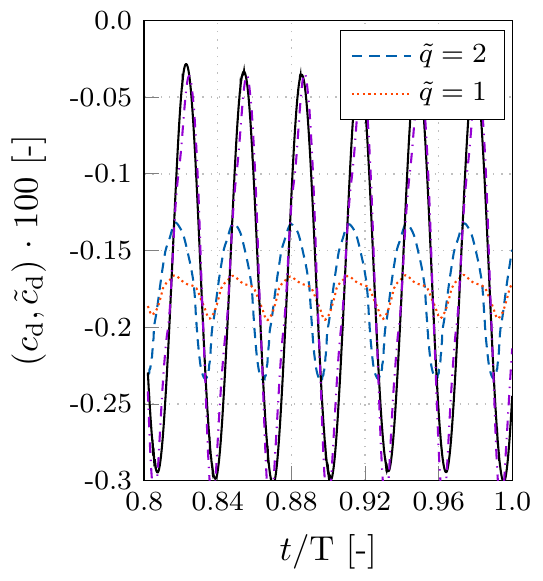}
}
\caption{Kriso container ship case ($\mathrm{Re}_\mathrm{L} = \SI{14 000 000}{}$, $\mathrm{Fn}=0.26$, $d^\mathrm{h}/L^\mathrm{pp} = 0.04696$, $L^\mathrm{w} / L^\mathrm{pp} = 1.3$, $H^\mathrm{w} / L^\mathrm{pp} = 0.075$, $T^w (V_1/L^\mathrm{pp}) = 1.3$).  Simulated (black) and reconstructed drag based on $\tilde{q}=1$ (orange), $\tilde{q}=2$ (blue), and $\tilde{q}=4$ (purple) 
singular values
for the pressure (left) and frictional drag (center), as well as the total resistance (right).}
\label{fig:kcs_compare_forces}
\end{figure}

Based on the adaptively generated itSVD, six evaluation studies are performed that vary the number of singular values used for reconstruction, i.e., $\tilde{q} = [1, 2, 4, 8, 16, 422(=q)]$. A comparison of integral results is shown in Fig. \ref{fig:kcs_compare_forces}, which presents the simulated  drag (black) with companions values obtained from the reconstructed fields using evaluation with $\tilde{q}=1$ (orange), $\tilde{q}=2$ (blue), and $\tilde{q}=4$ (purple).
The figure distinguishes between pressure, friction and  total drag.
Frictional forces always increase the resistance. 
Although the pressure resistance periodically becomes positive, the total resistance experiences no sign change.
For the two smallest evaluation ranks $\tilde{q} = [1,2]$, a visible deviation of the reconstructed contributions from the simulated values becomes apparent in all graphs. In particular the pressure force and the amplitudes of the frictional force are much better recovered for $\tilde{q} =4$.

%
\begin{figure}[!ht]
\centering
\iftoggle{tikzExternal}{
\input{./tikz/3D/KCS/kcs_compare_forces_errors.tikz}
}{
\includegraphics{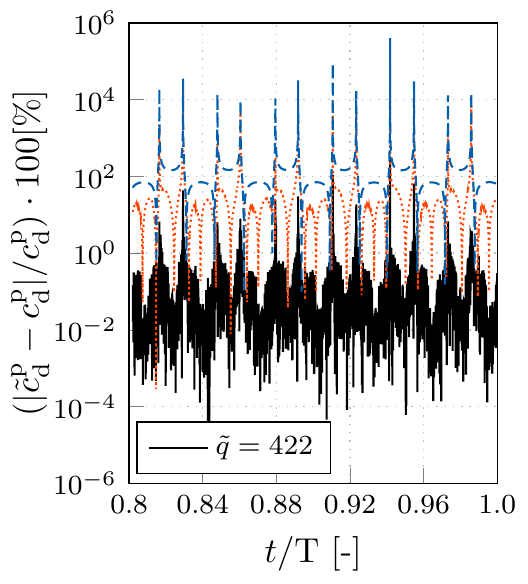}
\includegraphics{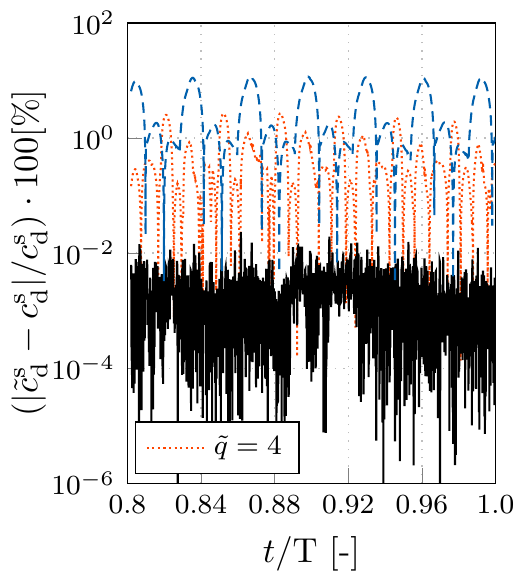}
\includegraphics{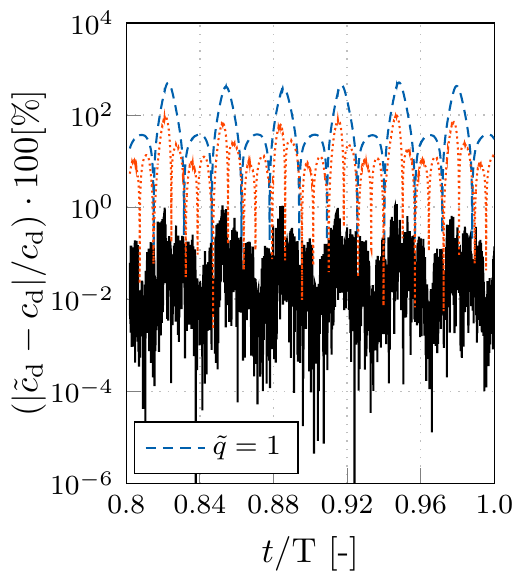}
}
\caption{Kriso container ship case ($\mathrm{Re}_\mathrm{L} = \SI{14 000 000}{}$, $\mathrm{Fn}=0.26$, $d^\mathrm{h}/L^\mathrm{pp} = 0.04696$, $L^\mathrm{w} / L^\mathrm{pp} = 1.3$, $H^\mathrm{w} / L^\mathrm{pp} = 0.075$, $T^w (V_1/L^\mathrm{pp}) = 1.3$). Relative error of the reconstructed $\tilde{c}_\mathrm{d}$ drag coefficient 
for the pressure (left) and frictional drag (center), as well as the total resistance (right) using lower and higher rank approximations.}
\label{fig:kcs_compare_forces_error}
\end{figure}

\begin{figure}[!ht]
\centering
\subfigure[]{
\includegraphics[width=0.475\textwidth]{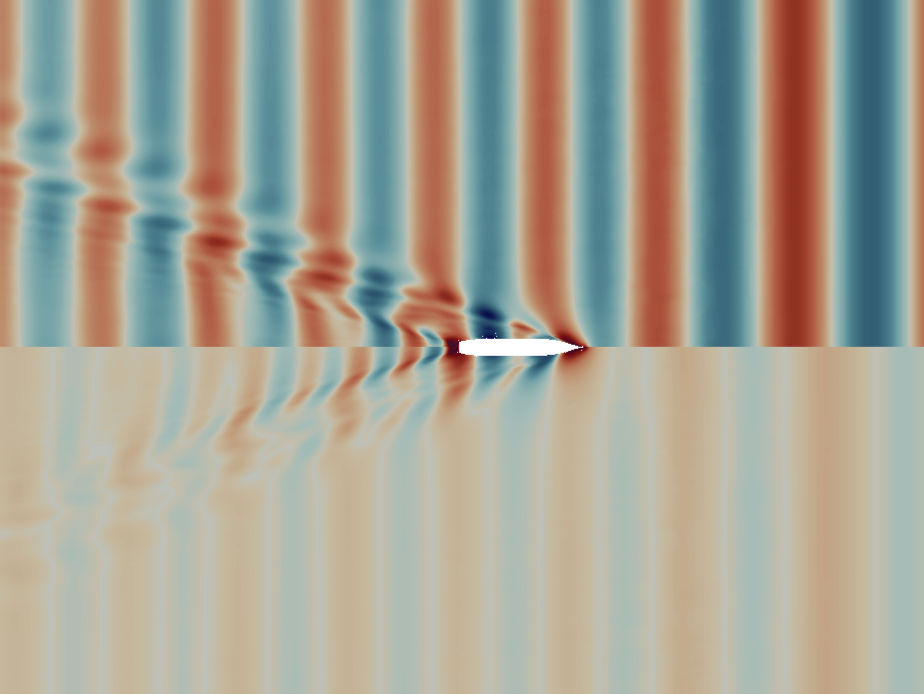}
}
\subfigure[]{
\includegraphics[width=0.475\textwidth]{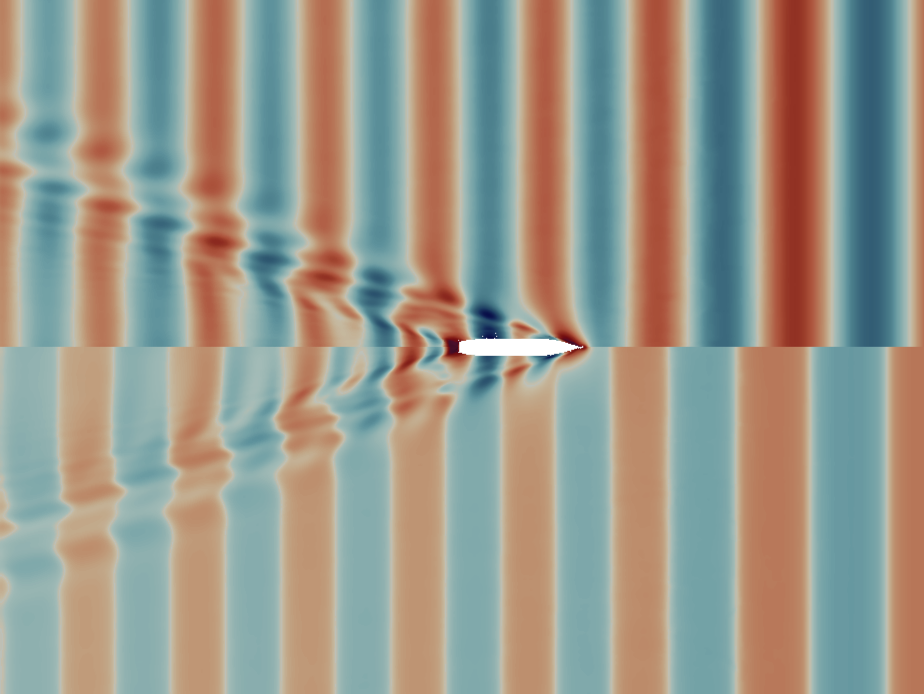}
}
\subfigure[]{
\includegraphics[width=0.475\textwidth]{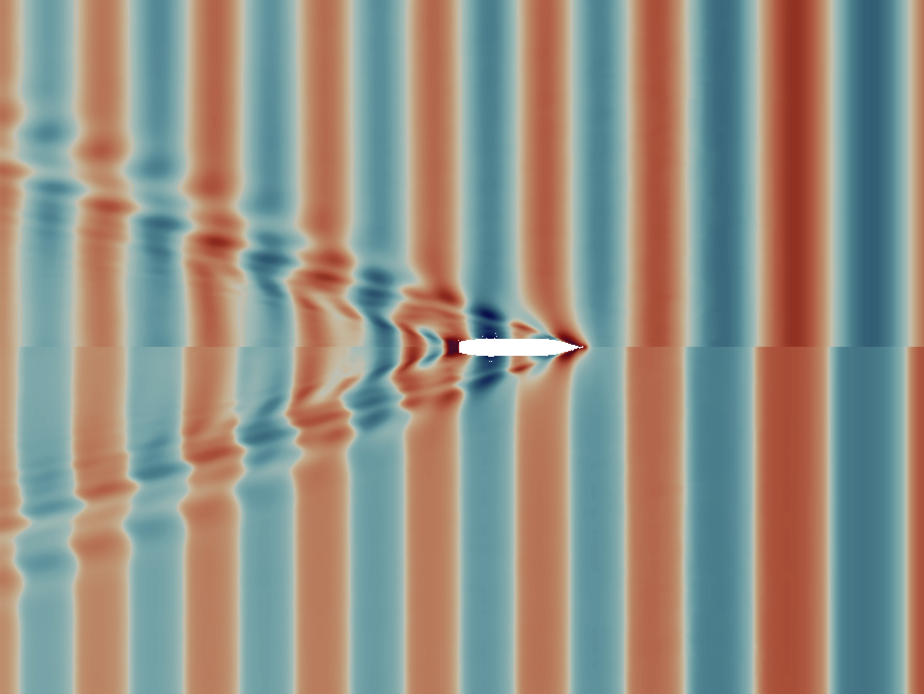}
}
\subfigure[]{
\includegraphics[width=0.475\textwidth]{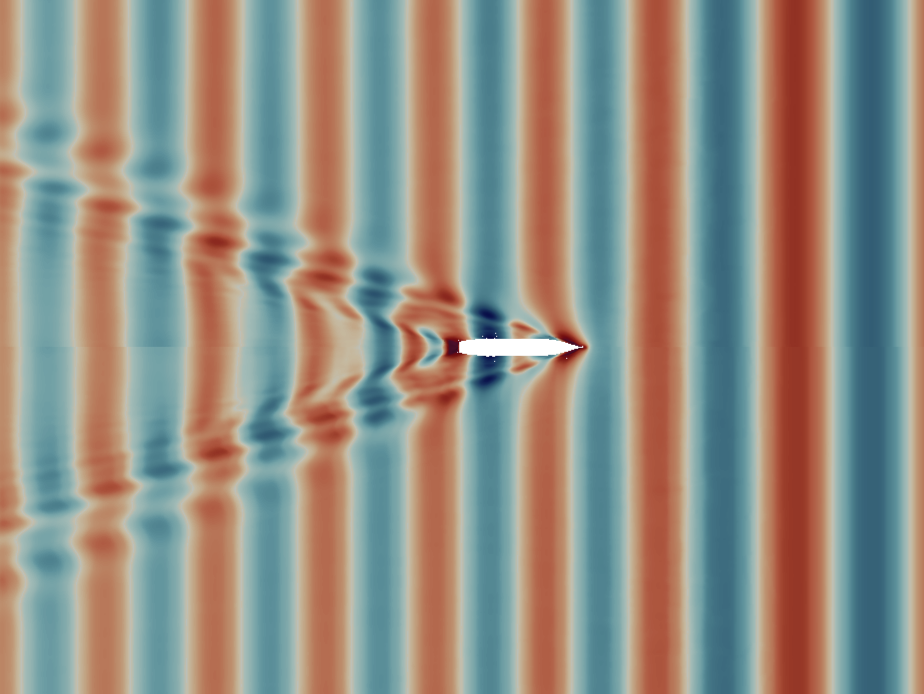}
}
\subfigure[]{
\includegraphics[width=0.475\textwidth]{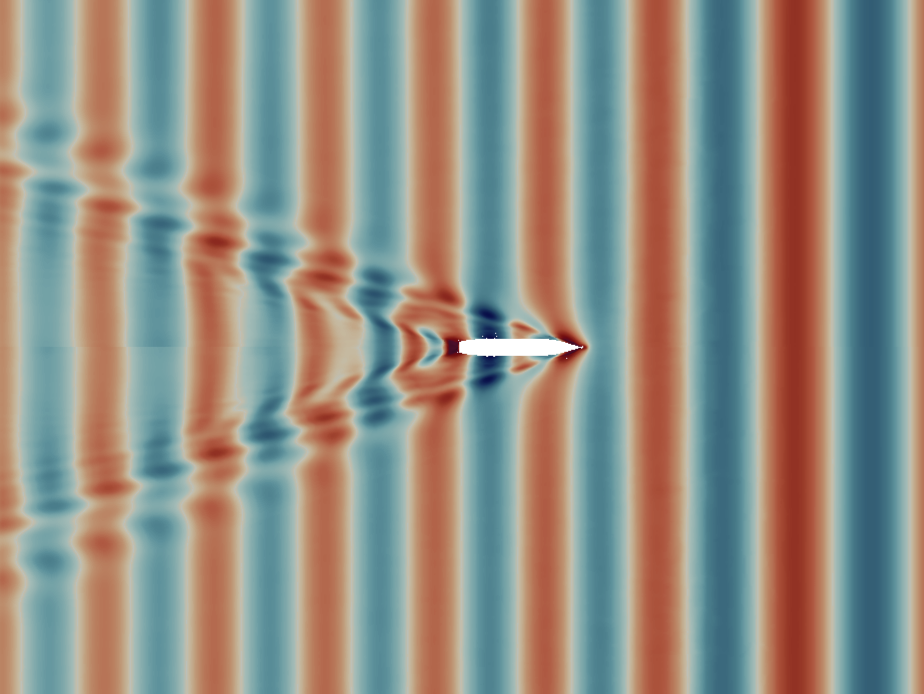}
}
\subfigure[]{
\includegraphics[width=0.475\textwidth]{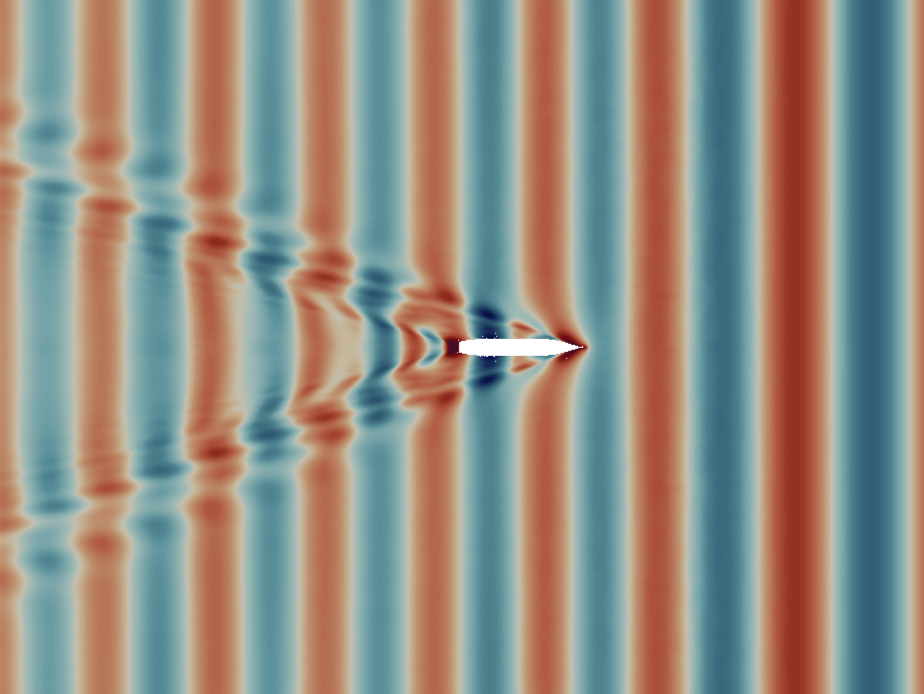}
}
\caption{Kriso container ship case ($\mathrm{Re}_\mathrm{L} = \SI{14 000 000}{}$, $\mathrm{Fn}=0.26$, $d^\mathrm{h}/L^\mathrm{pp} = 0.04696$, $L^\mathrm{w} / L^\mathrm{pp} = 1.3$, $H^\mathrm{w} / L^\mathrm{pp} = 0.075$, $T^w (V_1/L^\mathrm{pp}) = 1.3$): Simulated (upper) as well as reconstructed (lower) wave elevation at $t/\mathrm{T} = 1.0$ for $\tilde{q} = 1$ (a), $\tilde{q} = 2$ (b), $\tilde{q} = 4$ (c), $\tilde{q} = 8$ (d), $\tilde{q} = 16$ (e), and $\tilde{q} = 422 = q$ (f), respectively.}
\label{fig:kcs_fo_elevation}
\end{figure}

\begin{figure}[!ht]
\centering
\iftoggle{tikzExternal}{
\input{./tikz/3D/KCS/kcs_wave_cuts_errors.tikz}
}{
\includegraphics{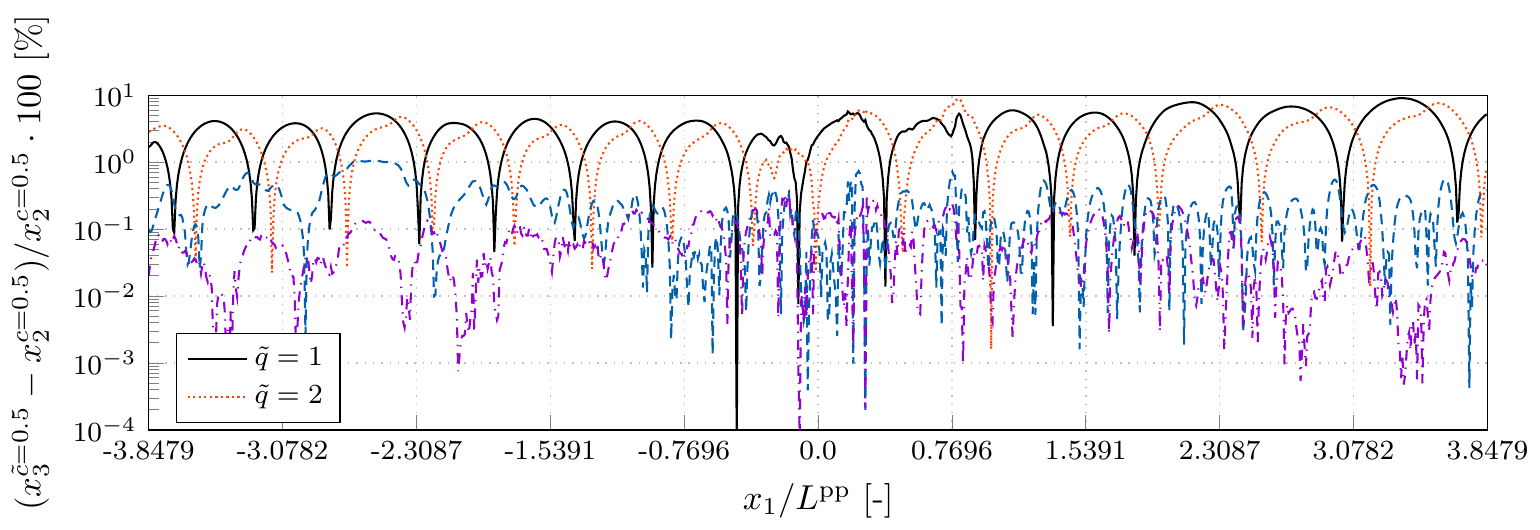}
\includegraphics{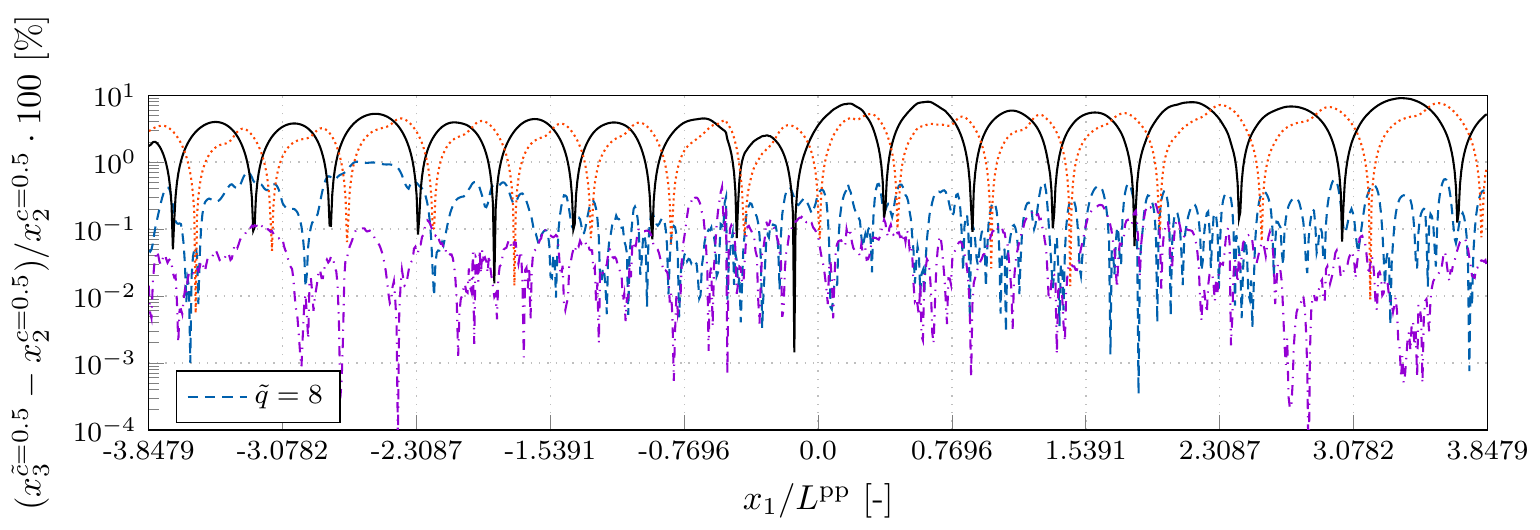}
\includegraphics{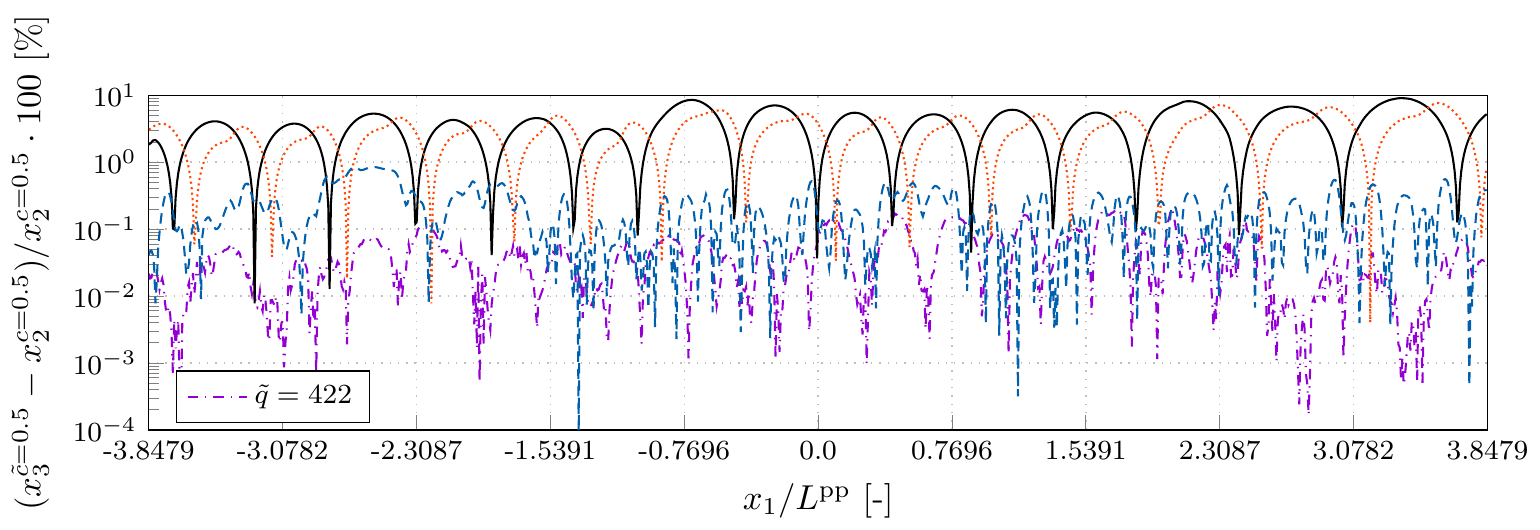}
}
\caption{Kriso container ship case ($\mathrm{Re}_\mathrm{L} = \SI{14 000 000}{}$, $\mathrm{Fn}=0.26$, $d^\mathrm{h}/L^\mathrm{pp} = 0.04696$, $L^\mathrm{w} / L^\mathrm{pp} = 1.3$, $H^\mathrm{w} / L^\mathrm{pp} = 0.075$, $T^w (V_1/L^\mathrm{pp}) = 1.3$): Errors $(x_2^{\tilde{c}=0.5}-x_2^{c=0.5})/x_2^{c=0.5} \cdot 100$ for the wave elevations at $x_2/L^\mathrm{pp} = 0.0741$ (top), $x_2/L^\mathrm{pp} = 0.1509$ (center), $x_2/L^\mathrm{pp} = 0.4224$ (bottom) for four reconstructed scenarios from Fig. \ref{fig:kcs_fo_elevation} at $t/\mathrm{T}=1$.}
\label{fig:kcs_wave_cuts_errors}
\end{figure}

Figure \ref{fig:kcs_compare_forces_error} depicts 
 the error of the reconstructed drag coefficients $|\tilde{c}_\mathrm{d} - c_\mathrm{d}|/c_\mathrm{d} \cdot 100 \%$ 
 over the itSVD relevant time horizon. 
 Again the pressure, the frictional and the total drag contributions are distinguished. 
 All contributions reveal a reduction of errors from approximately  $\mathcal{O}(10^2)$ for $\tilde{q}=1$ to $\mathcal{O}(10^{-2})$ for $\tilde{q}=422$.
Figure \ref{fig:kcs_fo_elevation} compares the wave elevation fields at the final time instant $t/\mathrm{T}=1$.
Each graph compares the simulated data (top) with the respective reconstruction (bottom) for $\tilde{q}=1$ (a), $\tilde{q}=2$ (b), $\tilde{q}=4$ (c), $\tilde{q}=8$ (d), $\tilde{q}=16$ (e), $\tilde{q}=422$ (f). For the very low  rank reconstructions $\tilde{q} = [1,2]$, wave amplitudes are significantly reduced and a noticeable phase shift appears.
 The latter also becomes apparent in three wave cut's at $x_2/L^\mathrm{pp} = 0.0741$ (top), $x_2/L^\mathrm{pp} = 0.1509$ (center), $x_2/L^\mathrm{pp} = 0.4224$ (bottom) depicted by Fig. \ref{fig:kcs_wave_cuts_errors} ($\tilde q=1$, red) and their corresponding reconstruction errors in Fig. \ref{fig:kcs_wave_cuts_errors}. For $\tilde{q} = 4$, the reconstruction of the wave field significantly improves, which is in line with acceptable agreements observed for the drag reconstruction in  Figs. \ref{fig:kcs_compare_forces} and \ref{fig:kcs_compare_forces_error}. Using $\tilde{q} = 16$  singular values, only minor differences are displayed in the wake. These disparities vanish for $\tilde q=422$ in Fig. \ref{fig:kcs_fo_elevation} and 
  wave cut errors reduce to $\mathcal{O}(10^{-2})$ in Fig. \ref{fig:kcs_wave_cuts_errors} (purple).
\begin{figure}[!ht]
\centering
\iftoggle{tikzExternal}{
\input{./tikz/3D/KCS/kcs_wave_cuts.tikz}
}{
\includegraphics{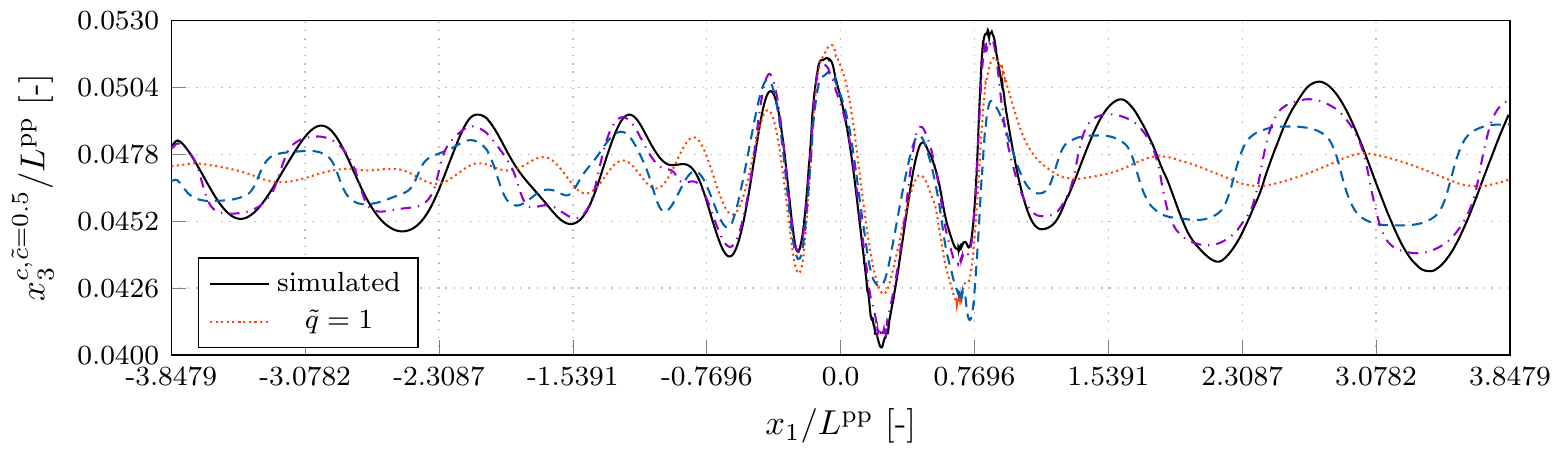}
\includegraphics{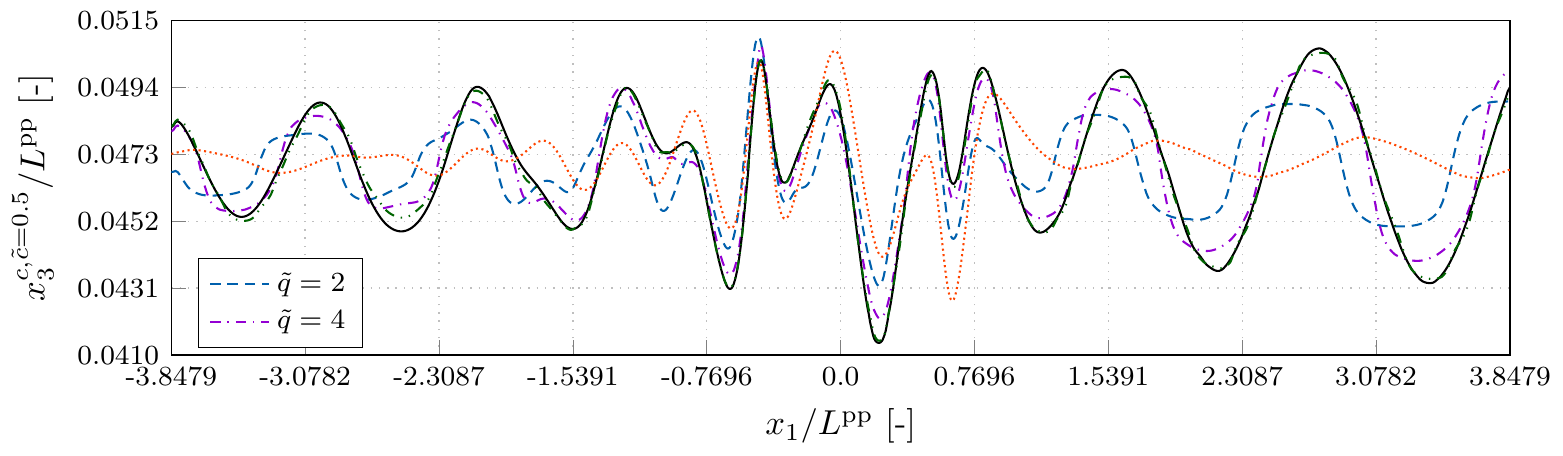}
\includegraphics{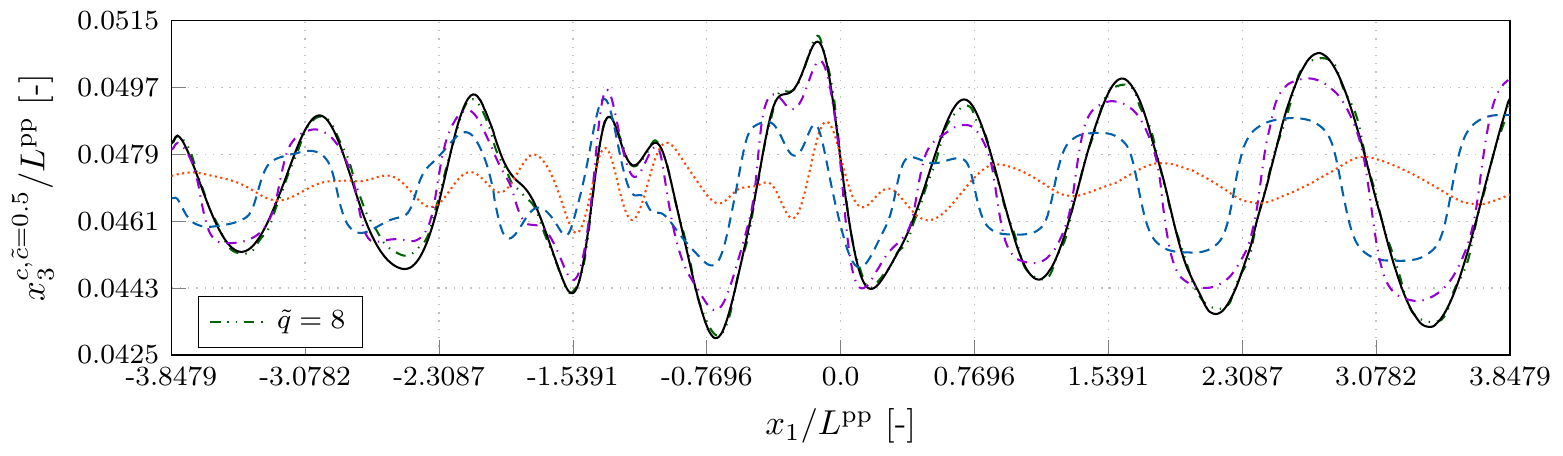}
}
\caption{Kriso container ship case ($\mathrm{Re}_\mathrm{L} = \SI{14 000 000}{}$, $\mathrm{Fn}=0.26$, $d^\mathrm{h}/L^\mathrm{pp} = 0.04696$, $L^\mathrm{w} / L^\mathrm{pp} = 1.3$, $H^\mathrm{w} / L^\mathrm{pp} = 0.075$, $T^w (V_1/L^\mathrm{pp}) = 1.3$): Wave cuts at $x_2/L^\mathrm{pp} = 0.0741$ (top), $x_2/L^\mathrm{pp} = 0.1509$ (center), $x_2/L^\mathrm{pp} = 0.4224$ (bottom) for the simulated (black) and four reconstructed scenarios from Fig. \ref{fig:kcs_fo_elevation} at $t/\mathrm{T}=1$}.
\label{fig:kcs_wave_cuts}
\end{figure}
 \subsection{Overheads}
To adequately reconstruct both local data  
and integral resistance quantities, 
approximately 40-80 singular values are necessary (cf. Figs \ref{fig:kcs_compare_forces}-\ref{fig:kcs_wave_cuts_errors}) which reduces the storage effort for the total $\mathrm{T}=10000$ time steps below one percent.
%
The present 
itSVD  is 
embedded in the time integration loop and thus induces additional run-time overheads. The latter can be extracted for each line in Alg. \ref{alg:SVD_construction}. For the presented adaptive ship flow itSVD with $q=422$ singular values to cover $\mathrm{T}=10000$ time steps with a bunch size of $b = 500$, Tab. \ref{tab:timing} provides an estimate of the expected additional expenses. Using $q/\mathrm{T} \approx 25$ and $b/q \approx 1$ the computational surplus should be 
in the order of $\mathcal O (10\%)$ and 
increase by approximately a factor of 5 when reducing the bunch size to $b\approx 15$.

\begin{figure}[!ht]
\centering
\iftoggle{tikzExternal}{
\input{./tikz/3D/KCS/kcs_timing_lessData.tikz}
}{
\includegraphics{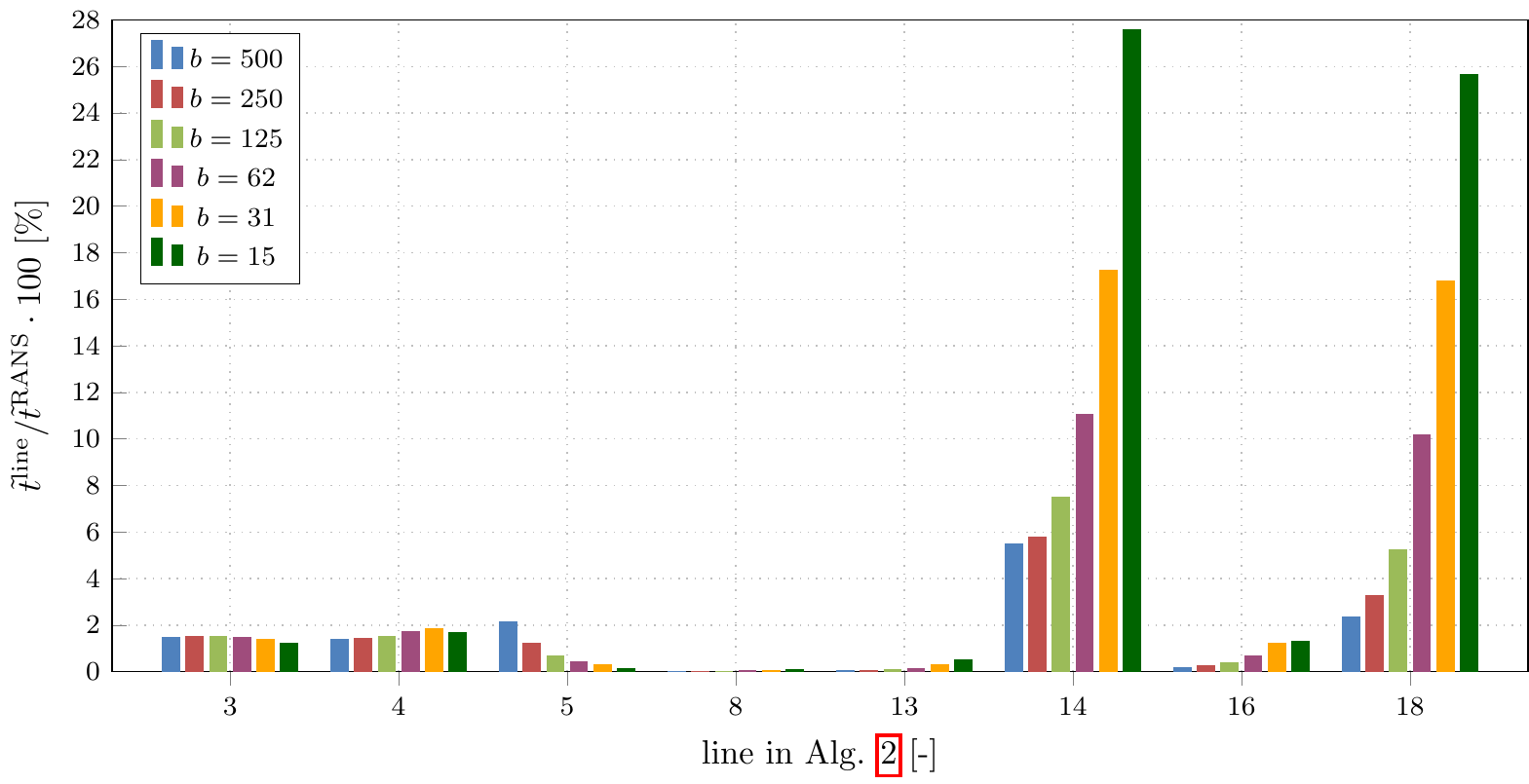}
}
\caption{Kriso container ship case ($\mathrm{Re}_\mathrm{L} = \SI{14 000 000}{}$, $\mathrm{Fn}=0.26$, $d^\mathrm{h}/L^\mathrm{pp} = 0.04696$, $L^\mathrm{w} / L^\mathrm{pp} = 1.3$, $H^\mathrm{w} / L^\mathrm{pp} = 0.075$, $T^w (V_1/L^\mathrm{pp}) = 1.3$): Summed itSVD construction time per line of Alg. \ref{alg:SVD_construction} $\tilde{t}^\mathrm{line}$, normalized with the RANS simulation wall clock time $\tilde{t}^\mathrm{RANS}$.}
\label{fig:kcs_timing_alg}
\end{figure}

 Measured overheads
are presented in Fig. \ref{fig:kcs_timing_alg} as blue bars, based on the relative portion of the total's simulation run-time $\tilde{t}^\mathrm{line}/\tilde{t}^\mathrm{NS} \cdot 100$, where $\tilde{t}^\mathrm{RANS}$ refers to the time integration effort of the RANS system only. The total overhead of 13.07\% is deemed acceptable. 
A considerable amount of the total overhead 
follows from the global QR decomposition (row 14, 5.49\%), which underlines the active discussion of its necessity in \cite{brand2006fast, fareed2018incremental, zhang2022answer, li2022enhanced}.
The second QR decomposition in line 5 contributes with 2.14\% and the two subspace rotations in line 16 and 18 take additional 2.5\%. Surprisingly, the extra effort of the local SVD determination in line 8 is comparably low with $\approx$0.02\%. Further significant efforts arise in connection with the global --thus, with N-scaling-- matrix-vector operations in lines 3 and 4. The effort of all remaining algorithmic lines is below 0.01\% and thus ommited in Fig. \ref{fig:kcs_timing_alg}.
From an efficiency point of view, the proposed itSVD algorithm mainly benefits from bunching the individual snapshots, cf. Sec. \ref{sec:verification_validation}. To underline this, the itSVD construction of the ship flow is repeated with successively halved bunch sizes of $b=[250, 125, 62, 31, 15]$. A noticeable, nonlinear increase in the relative computation time arises for smaller bunch sizes, ranging from 13.58\% ($b=250$) over 16.94\% ($b=125$) and 25.75\% ($b=62$) to 39.27\% ($b=31$) as well as 60.28\% ($b=15$) for the smallest bunch size. 
The experienced increase of overheads when reducing the bunching size confirms the findings of the 2D cylinder flow example in Tab. \ref{tab:timing}.   
The resulting simulation overheads are added to Fig. \ref{fig:kcs_timing_alg} for $b=250$, $b=125$, $b=62$, $b=31$ and $b=15$ in red, light green, purple,  orange and dark green, respectively. As already shown for the cylinder study in Tab. \ref{tab:memory}, a bunch-width reduced to the limit case of $b = 1$ would drive the itSVD construction time --even for this intense reduction of $q/\mathrm{T} \cdot 100 = 4.22\%$-- in regions of the effort of the actual time integration.
Finally, the data is fitted to an exponential function of shape $\tilde{t}(b) = r \, e^{s\, b} + t \, e^{u \, b}$ with $r = 73.79$, $s = -0.03681$, $t = 17.45$, and $u = -0.0006676$ using the Matlab$^\copyright$ Curve Fitting Toolbox, cf. \cite{matlabcurvefitting}, that reveals an extrapolated overhead of $\tilde{t}(b=1) \approx 89\%$.
\begin{figure}[!ht]
\centering
\iftoggle{tikzExternal}{
\input{./tikz/3D/KCS/kcs_timing_total.tikz}
}{
\includegraphics{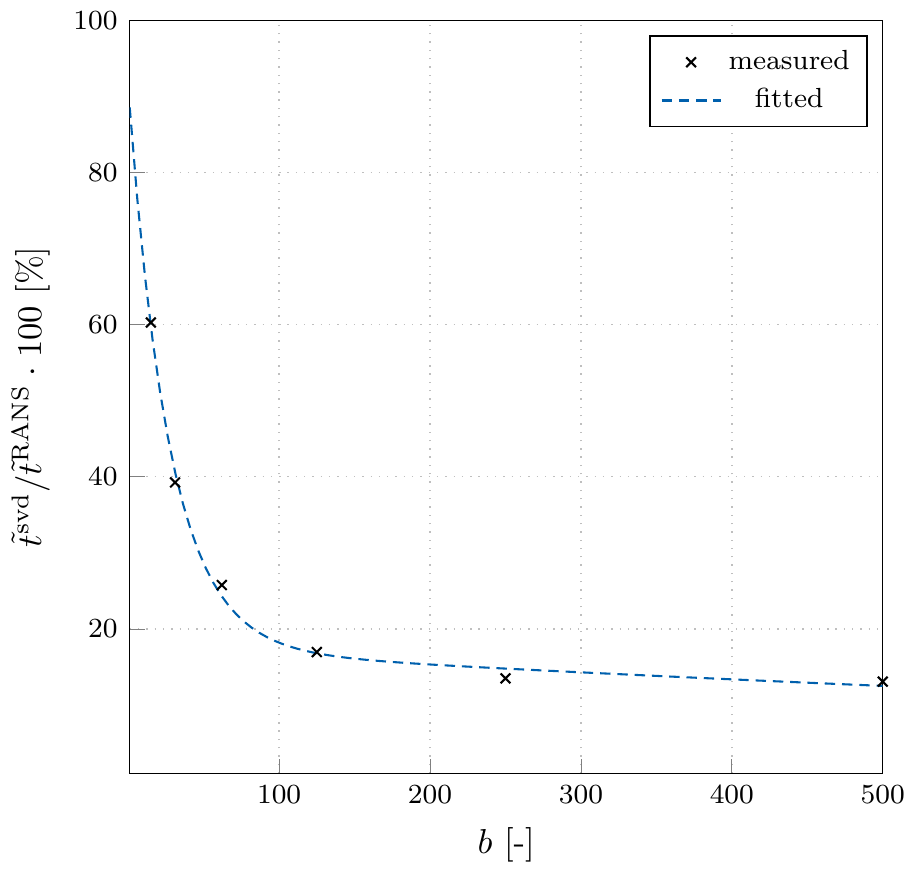}
}
\caption{Kriso container ship case ($\mathrm{Re}_\mathrm{L} = \SI{14 000 000}{}$, $\mathrm{Fn}=0.26$, $d^\mathrm{h}/L^\mathrm{pp} = 0.04696$, $L^\mathrm{w} / L^\mathrm{pp} = 1.3$, $H^\mathrm{w} / L^\mathrm{pp} = 0.075$, $T^w (V_1/L^\mathrm{pp}) = 1.3$): Measured itSVD construction time (summed lines of Alg. \ref{alg:SVD_construction}, cf. Fig. \ref{fig:kcs_timing_alg}) $\tilde{t}^\mathrm{svd}$, normalized with the simulation wall clock time needed to integrate the RANS system $\tilde{t}^\mathrm{RANS}$. Additionally, the efforts are fitted to an exponential function.
}
\label{fig:kcs_timing_fitted}
\end{figure}

\section{Conclusion \& Outlook}
\label{sec:conclusion_outlook}
The paper reports on a parallelized incremental SVD approach to efficiently compress, i.e., reduce, time-evolving PDE results obtained on large spatially partitioned grids on the fly during run time. Such data reduction might be of considerable interest to avoid heavy data transfer rates in subsequent data processing steps or serve as the workhorse for deeper analysis under the umbrella of Data Science.
The procedure is independent of the domain decomposition and has been tested for up to 2900 processes. It delivers the accuracy of formerly introduced SVD approaches and can be coupled to a heuristic quality measure based on the exact energy content to adjust the attainable accuracy. Following the motivation, the most critical aspect refers to the associated storage and run-time overheads. Starting from $\mathcal{O}(1\% \mathrm{T})$ singular values, where $\mathrm{T}$ refers to the temporal data size, a computational surplus of around $\mathcal{O}(10\%)$ in wall-clock time was achieved with fairly small storage overheads of $\mathcal{O}(5\%)$. Reducing the run-time overheads is particularly supported by bunching the incremental SVD updates as suggested herein, which complies with massively parallel applications.
Future directions will look at preserving the reconstructed data's realizability while addressing the procedure's algorithmic efficiency. The employed orthogonalization steps are maybe not necessary for each itSVD update and can be controlled using appropriate non-orthogonality measures. Furthermore, the method could be extended to spatially non-invariant grids, e.g., adaptive refinable/coarsable grids, and alternative preconditioning concepts, e.g., via the local cell volume, could be investigated. Likewise, different parameter handling(s) for the respective fields are conceivable.


\section*{CRediT Authorship Contribution Statement}
\textbf{N.K.}: Funding acquisition, Conceptualization, Formal analysis, Investigation, Methodology, Software, Validation, Visualization, Writing – original draft, Writing – review \& editing.
\textbf{H.F.}: Methodology, Software, Formal analysis, Writing - review \& editing.
\textbf{M.H.}: Funding acquisition, Methodology, Writing - review \& editing. 
\textbf{T.R.}: Funding acquisition, Project administration, Resources, Methodology, Writing – original draft, Writing – review \& editing.

\section*{Declaration of Competing Interest}
The authors declare that they have no known competing ﬁnancial interests or personal relationships that could have appeared to inﬂuence the work reported in this paper.

\section*{Acknowledgements}
This paper is a collaborative contribution to the projects M6 and T4 of the Collaborative Research Centre TRR181, '‘Energy Transfers in Atmosphere and Ocean'’ funded by German Research Foundation (DFG) under Grant Number 274762653, which is acknowledged by N.K., M.H. and T.R.
The second author (H.F.) acknowledges the funding of the DFG within the framework of the International Research Training Group GRK 2657 ''Computational Mechanics Techniques in High Dimensions''  under Grant Number 433082294. 
The authors gratefully acknowledge the computing time granted by the Resource Allocation Board and provided on the supercomputer Lise and Emmy at NHR@ZIB and NHR@Göttingen as part of the NHR infrastructure. The calculations for this research were conducted with computing resources under the projects
hhi00033 (''Hydrodynamic Drag Minimization of Ships'')
and
hhi00037 (''Energy Fluxes at the Air-Sea Interface'').

%
%

\newpage
\begin{appendix}


\section{Enhanced itSVD Algorithm}
\begin{algorithm}[!ht]
\SetKwProg{Fn}{Subroutine}{:}{}
\Fn{\FUpdate{}}{
    $u = \mathrm{size}(\matr{U},2)$, $v = \mathrm{size}(\matr{V},1)$, $c=u+b$, $d=v+b$, $l = \mathrm{min}(q,c)$ \tcp*{Required sizes}
    $\matr{M} = \matr{U}^\mathrm{T} \matr{B}$, parallel{\_}sum($\matr{M}$)\tcp*{1) $\matr{M} \in \mathbb{R}^{u \times b}$ 2) Local to Global Operation. cf. Fig. \ref{fig:global_inner_product}}
    $\matr{P} = \matr{B} - \matr{U} \matr{M}$\tcp*{$\matr{P} \in \mathbb{R}^{\mathrm{Y} \times b}$ around Eqn. \ref{equ:matrix_K_general}}
    $\begin{bmatrix}\matr{Q}_\mathrm{P}, \matr{R}_\mathrm{P} \end{bmatrix} = \mathrm{parallel{\_}gram{\_}schmidt}(\matr{P})$\tcp*{Global QR Decomposition}
    \colorbox{lightgray}{$\matr{Q} = \begin{bmatrix}
    \matr{U} & \matr{Q}_\mathrm{P}
    \end{bmatrix}$}\tcp*{Prepare $\matr{U}$ Update via $\matr{Q} \in \mathbb{R}^{\mathrm{Y} \times l}$, cf. Eqn. \ref{equ:update_U_truncated}}
    \colorbox{lightgray}{$\begin{bmatrix}\matr{Q}_\mathrm{Q}, \matr{R}_\mathrm{Q}\end{bmatrix} = \mathrm{parallel{\_}gram{\_}schmidt}(\matr{Q})$}\tcp*{Global QR Decomposition}
    \If{procID=0}
    {
        \colorbox{lightgray}{$\matr{K} = \matr{R}_\mathrm{Q} \begin{bmatrix}
        \matr{S} & \matr{M} \\
        \matr{0} & \matr{R}_\mathrm{P}
        \end{bmatrix}$}\tcp*{Prepare the SVD input matrix $\matr{K} \in \mathbb{R}^{c \times c}$, cf. Eqn. \ref{equ:matrix_K_column}}
        $\begin{bmatrix}
        \matr{U}^\prime, \vect{s}^\prime, \matr{V}^\prime
        \end{bmatrix} = \mathrm{svd}(\matr{K})$\tcp*{Local SVD yields $\matr{U}^\prime, \matr{V}^\prime \in \mathbb{R}^{c \times c}$ and $\vect{s}^\prime \in \mathbb{R}^{c}$ }
        \If{adaptive}
        {
            q = adaptive{\_}truncation($\vect{s}^\prime$)\tcp*{Adapt the truncation rank}
        }
        parallel{\_}bcast($\matr{U}^\prime, \vect{s}^\prime, \matr{V}^\prime, q$)\tcp*{Globalize the results}
    }
    $\matr{R} = \begin{bmatrix} 
    \matr{V} & \matr{0} \\
    \matr{0} & \matr{I}
    \end{bmatrix}$\tcp*{Prepare $\matr{V}$ Update via $\matr{R} \in \mathbb{R}^{d \times l}$, cf. Eqn. \ref{equ:update_V_truncated}}
    de-allocate($\matr{U},\vect{s},\matr{V}$), allocate($\matr{U}(\mathrm{Y},l),\vect{s}(l),\matr{V}(d,l))$\tcp*{Increase the $\matr{U}$, $\matr{S}$, and $\matr{V}$ sizes}
    $\matr{V}$ = $\matr{R}$ $\matr{V}^\prime$(:,$1$:$l$)\tcp*{Update $\matr{V}$, cf. Eqn. \ref{equ:update_V_truncated}}
    $\vect{s}$ = $\vect{s}^\prime$(1:l)\tcp*{Update $\vect{s}$, cf. Eqn. \ref{equ:update_S_truncated}}
    $\matr{U}$ = $\matr{Q}_Q$ $\matr{U}^\prime($:$,1$:$l)$\tcp*{Update $\matr{U}$, cf. Eqn. \ref{equ:update_U_truncated}}
}

\caption{Improved Incremental rank $b$ update of a Spatially Parallel / Temporal Serial truncated Singular Value Decomposition, where enhancements of Alg. \ref{alg:SVD_construction} are highlighted.}
\label{alg:SVD_construction_improved}
\end{algorithm}

\end{appendix}

\end{document}